\documentclass[pdflatex,sn-mathphys-ay]{sn-jnl}

\usepackage{graphicx}%
\usepackage{multirow}%
\usepackage{amsmath,amssymb,amsfonts}%
\usepackage{amsthm}%
\usepackage{mathrsfs}%
\usepackage[title]{appendix}%
\usepackage{xcolor}%
\usepackage{textcomp}%
\usepackage{manyfoot}%
\usepackage{booktabs}%
\usepackage{algorithm}%
\usepackage{algorithmicx}%
\usepackage{algpseudocode}%
\usepackage{listings}%
\usepackage{url}
\usepackage{hyperref}
\usepackage{arydshln}
\usepackage[table]{xcolor}
\usepackage{colortbl}
\usepackage{longtable,array}
\usepackage{makecell}
\usepackage{enumitem}
\usepackage{adjustbox}
\usepackage{pdflscape}
\usepackage{tablefootnote}
\usepackage{eurosym}
\usepackage{pgfplots}
\usepackage{pdflscape}
\usepackage[T1]{fontenc}
\usepackage{anyfontsize}

\newcommand{\rqbox}[1]{%
  \noindent
  \setlength{\fboxsep}{10pt}%
  \fbox{%
    \parbox{\dimexpr\linewidth-2\fboxsep-2\fboxrule\relax}{%
      \textbf{Summary:} #1%
    }%
  }%
  \vspace{1em}%
}

\begin{document}

\newenvironment{ResearchQuestionOne}[1]
    {
    \vspace{3pt}
    {\narrower \noindent \textit{\textbf{RQ1:} #1}\par}
    \vspace{3pt}
    }

\newenvironment{ResearchQuestionOneOne}[1]
    {
    \vspace{3pt}
    {\narrower \textit{\textbf{RQ1.1:} #1}\par}
    \vspace{3pt}
    }

\newenvironment{ResearchQuestionOneTwo}[1]
    {
    \vspace{3pt}
    {\narrower \textit{\textbf{RQ1.2:} #1}\par}
    \vspace{3pt}
    }

\newenvironment{ResearchQuestionTwo}[1]
    {
    \vspace{3pt}
    {\narrower \noindent \textit{\textbf{RQ2:} #1}\par}
    \vspace{3pt}
    }

\newenvironment{ResearchQuestionThree}[1]
    {
    \vspace{3pt}
    {\narrower \noindent \textit{\textbf{RQ3:} #1}\par}
    \vspace{3pt}
    }

\newenvironment{ResearchQuestionThreeOne}[1]
    {
    \vspace{3pt}
    {\narrower \textit{\textbf{RQ3.1:} #1}\par}
    \vspace{3pt}
    }

\newenvironment{ResearchQuestionThreeTwo}[1]
    {
    \vspace{3pt}
    {\narrower \textit{\textbf{RQ3.2:} #1}\par}
    \vspace{3pt}
    }

\newenvironment{ResearchQuestionFour}[1]
    {
    \vspace{3pt}
    {\narrower \noindent \textit{\textbf{RQ4:} #1}\par}
    \vspace{3pt}
    }

\def\RqOne{How do LLMs perform in paper screening?}
\def\RqOneDotOne{How do LLMs perform in paper screening on the full SESR-Eval dataset?}
\def\RqOneDotTwo{How do LLMs perform in paper screening on the sampled SESR-Eval-Mini dataset?}
\def\rqtwo{How does computing the overall screening decision from criterion-level decisions affect LLM screening performance on the sampled data?}
\def\rqthree{To what extent do LLMs agree with each other?}
\def\rqthreeone{To what extent do LLMs agree with each other in their overall screening decisions?}
\def\rqthreetwo{To what extent do LLMs agree with each other in individual screening criteria?}
\def\rqfour{To what extent does refining the inclusion and exclusion criteria affect LLM screening performance?}

\title[Article Title]{Has LLM Screening Performance Stalled in Software Engineering Systematic Reviews?}


\author*[1]{\fnm{Aleksi} \sur{Huotala} }\email{aleksi.huotala@helsinki.fi}
\author[2]{\fnm{Miikka} \sur{Kuutila}}\email{miikka.kuutila@lut.fi}
\author[1]{\fnm{Mika} \sur{M\"antyl\"a}}\email{mika.mantyla@helsinki.fi}

\affil*[1]{\orgname{University of Helsinki}, \orgaddress{\city{Helsinki}, \country{Finland}}}
\affil[2]{\orgname{LUT University}, \orgaddress{\city{Lahti}, \country{Finland}}}

\abstract{
Screening in systematic reviews (SRs) is manual and time-consuming. Prior work has explored large language models (LLMs) for automating this step, but LLMs are evolving rapidly, so earlier performance claims may no longer accurately reflect their screening performance.  
We used an existing software engineering SR screening benchmark (SESR-Eval) as our data. We also power-sampled a new, smaller dataset (SESR-Eval-Mini) that allows evaluation at lower costs. Using this data, we evaluated eight new LLMs for screening performance. Additionally, we tested different prompts, analyzed LLM agreement in screening decisions and criteria, and examined the effect of refining the inclusion  and exclusion criteria on screening performance.
The eight new LLMs performed marginally better than the seven old ones: avg. MCC across secondary studies rose from 0.347 to 0.365. Differences between secondary studies are still bigger than between LLMs. Computing the overall screening decision from criterion-level decisions degraded screening performance only slightly. LLMs generally agree with each other in their corresponding screening decisions (mean Gwet's AC1 = 0.830), though certain inclusion and exclusion criteria showed larger disagreement than others. Refining the inclusion and exclusion criteria slightly improved recall and made decisions easier for some LLMs, but overall impacts of criteria refinement were modest.
LLMs are not yet ready to replace humans in paper screening and the advantages new, more costly models bring, appear to be very limited. Agent-based approaches, prompt engineering, and further criteria refinement are three potential future research avenues. 
}


\maketitle

\section{Introduction}

Evidence-Based Software Engineering (EBSE) highlights systematic reviews (SRs) as a key method for gathering and synthesizing evidence in software engineering (SE)
~\citep{kitchenhamEvidenceBasedSoftwareEngineering}. However, conducting SRs is labor-intensive and time-consuming~\citep{Felizardo_Deizepe_Coutinho_Gomes_Meireles_Gerosa_Steinmacher_2025}. According to \citet{marshallSystematicReviewAutomation2019}, a single SR can take an average of 67 weeks to complete, with paper screening representing one of the most demanding phases. Prior work has shown that humans spend approximately 89 seconds screening each individual paper~\citep{huotalaPromiseChallengesUsing2024a}, which makes paper screening a substantial bottleneck in the SR process as thousands of papers may require screening.

Automating or supporting paper screening has therefore attracted considerable attention. Earlier approaches have used traditional Machine Learning (ML) and Natural Language Processing (NLP) techniques \citep{hamelEvaluationDistillerSRsMachine2020}, while recently, Large Language Models (LLMs) have been used because of their ability to reason over textual input~\citep{huotalaPromiseChallengesUsing2024a,huotalaSESREvalDatasetEvaluating2025,felizardoChatGPTApplicationSystematic2024,liEvaluatingEffectivenessLarge2024,delgado-chavesTransformingLiteratureScreening2025,shameemExploratoryStudyEffectiveness2026,thodeExploringUseLLMs2025, tranSensitivitySpecificityAvoidable2023,matsuiHumanComparableSensitivityLarge2024a,guoAutomatedPaperScreening2024b,serrettiBenchmarkingLocalSchemaConstrained2026,tradStreamliningSystematicReviews2025,adamPromptEngineeringLarge2026}. Existing studies show promise for paper screening using LLMs, as LLMs achieve performance comparable to that of non-expert human screeners~\citep{huotalaPromiseChallengesUsing2024a,huotalaSESREvalDatasetEvaluating2025,felizardoChatGPTApplicationSystematic2024}. Yet, against expert human screeners, LLMs still make a considerable number of mistakes stemming from sources such as different interpretations of key terms \citep{mantyla2026understanding}. Guidelines for LLM usage have been proposed to mitigate their errors \citep{madeyskiLLM4SCREENLITRecommendationsAssessing2026}.  Human-in-the-loop systems have also been developed to support researchers in applying LLMs and other automation methods to paper screening~\citep{huotalaFSE2026,deizepe2026ailitreview, afonso2026profolaf,ouzzani2016rayyan}. 

This paper extends our prior work \citep{huotalaSESREvalDatasetEvaluating2025}, which introduced the SESR-Eval dataset for monitoring artificial intelligence (AI) performance in paper screening. Nine LLMs were benchmarked, and most of them performed similarly. However, the differences in screening accuracy between the secondary studies were greater than differences between LLMs, but LLMs were fairly cheap to use per secondary study. While the benchmark dataset can be used to benchmark LLMs, LLMs were still not recommended for automating paper screening process.

Next we explain the difference between this paper and \citet{huotalaSESREvalDatasetEvaluating2025}.
First, the evaluated LLMs covered only a single generation of models, prompt, and an output schema, which leaves it unclear whether newer LLMs, different user prompts or output schemas would further improve screening performance and make paper screening feasible with LLMs. Claims of earlier generation models may no longer accurately reflect the capabilities of newer models in paper screening, and benchmarking additional LLMs for paper screening benefit researchers and practitioners in choosing a cost-efficient LLM with an acceptable performance for the task (RQ1 -- RQ2). Benchmarking LLMs on a full dataset (N = 34,528) may not be financially sensible, as LLMs are generally expensive to run~\citep{Felizardo_Deizepe_Coutinho_Gomes_Meireles_Gerosa_Steinmacher_2025}, and token costs of LLMs have increased due to test-time inference and additional model capabilities, such as thinking models (RQ1.2). Computing the overall decision separately from criterion-level decisions can possibly reduce hallucinations, falsely judged screening decisions, and allows for additional flexibility (RQ2).

Second, the agreement in overall screening decisions and across inclusion and exclusion criteria was not analyzed in the paper, even though it could reveal whether LLM screening decisions are uniform across different models, or sensitive to the selected model (RQ3.1 -- RQ3.2).

Third,  the modification of inclusion and exclusion criteria was not studied in the paper, even when certain criteria cannot possibly be evaluated purely by title and abstract. Improving poorly formatted, unclear, or ambiguous criteria may increase the likelihood of a correct overall or criterion screening decision, which directly motivates RQ4.

We therefore address the following research questions:

\ResearchQuestionOne{\RqOne}
\ResearchQuestionOneOne{\RqOneDotOne}
\ResearchQuestionOneTwo{\RqOneDotTwo}
\ResearchQuestionTwo{\rqtwo}
\ResearchQuestionThree{\rqthree}
\ResearchQuestionThreeOne{\rqthreeone}
\ResearchQuestionThreeTwo{\rqthreetwo}
\ResearchQuestionFour{\rqfour}

For RQ1, we benchmark eight new state-of-the-art LLMs against seven older ones. On the full dataset (RQ1.1), we benchmark two of the new LLMs, and all eight new LLMs on a smaller 5\% sample, SESR-Eval-Mini (RQ1.2). We then investigate how prompt design and output structuring affect screening performance in RQ2. For RQ3, we analyze agreement in overall screening decisions across LLMs (RQ3.1) and also across individual inclusion and exclusion criteria (RQ3.2). Finally, for RQ4, we examine whether refining the inclusion and exclusion criteria improves screening performance.

The remainder of the paper is structured as follows. Section~\ref{sec:related_work} reviews related work. Section~\ref{sec:background} provides the necessary background. Section~\ref{sec:methods} describes the methodology, and Section~\ref{sec:results} presents the results. Section~\ref{sec:discussion} discusses the findings, and Section~\ref{sec:conclusions} concludes the paper.

\section{Related Work}

\label{sec:related_work}

\begin{table*}[htbp]
\small
\centering
\caption{Related work of paper screening with LLMs.}
\label{tab:related_work_llm}
\begin{tabular}{lllr}
\toprule
\textbf{Study} & \textbf{Year} & \textbf{Discipline} & \textbf{LLMs} \\
\midrule
\citet{huotalaPromiseChallengesUsing2024a} & 2024 & Software Engineering & 2 \\
\citet{huotalaSESREvalDatasetEvaluating2025} & 2025 & Software Engineering & 9  \\
\citet{felizardoChatGPTApplicationSystematic2024} & 2024 & Software Engineering & 1  \\
\citet{liEvaluatingEffectivenessLarge2024} & 2024 & Software Engineering & 9  \\
\citet{delgado-chavesTransformingLiteratureScreening2025} & 2025 & Software Engineering & 18  \\
\citet{shameemExploratoryStudyEffectiveness2026} & 2026 & Software Engineering & 1  \\
\citet{thodeExploringUseLLMs2025} & 2025 & Software Engineering & 5  \\
\citet{dennstadtTitleAbstractScreening2024} & 2024 & Biomedical & 4  \\
\citet{tranSensitivitySpecificityAvoidable2023} & 2023 & Medical & 1  \\
\citet{matsuiHumanComparableSensitivityLarge2024a} & 2024 & Medical & 2  \\
\citet{guoAutomatedPaperScreening2024b} & 2024 & Medical & 2  \\
\citet{serrettiBenchmarkingLocalSchemaConstrained2026} & 2026 & Medical & 1 \\
\citet{tradStreamliningSystematicReviews2025} & 2025 & Medical & 1 \\
\citet{adamPromptEngineeringLarge2026} & 2026 & Medical & 3  \\
\citet{jozwiakREFLECTIVETIABCosteffectivePrompt2026} & 2026 & Health Technology Assessment & 9  \\
\citet{qiuPromptBasedTutorialLarge2026} & 2026 & Health Technology Assessment & 1  \\
\citet{sujauAcceleratingDiseaseModel2025} & 2025 & Veterinary Medicine & 2  \\
\citet{behrouzianLinguisticPolarityDecision2026} & 2026 & Artificial Intelligence & 2 \\
\bottomrule
\end{tabular}
\end{table*}

Table \ref{tab:related_work_llm} presents related work on using LLMs to automate paper screening in SRs. The remainder of the section presents the current state of LLM-assisted paper screening in multiple disciplines / domains.

LLM-assisted paper screening has been studied in various disciplines, including SE~\citep{huotalaPromiseChallengesUsing2024a,huotalaSESREvalDatasetEvaluating2025,felizardoChatGPTApplicationSystematic2024,liEvaluatingEffectivenessLarge2024,delgado-chavesTransformingLiteratureScreening2025,shameemExploratoryStudyEffectiveness2026,thodeExploringUseLLMs2025}, medicine~\citep{tranSensitivitySpecificityAvoidable2023,matsuiHumanComparableSensitivityLarge2024a,guoAutomatedPaperScreening2024b,serrettiBenchmarkingLocalSchemaConstrained2026,tradStreamliningSystematicReviews2025,adamPromptEngineeringLarge2026}, biomedicine~\citet{dennstadtTitleAbstractScreening2024}, health technology assessment~\citep{jozwiakREFLECTIVETIABCosteffectivePrompt2026,qiuPromptBasedTutorialLarge2026}, veterinary medicine~\citep{sujauAcceleratingDiseaseModel2025}, and artificial intelligence~\citep{behrouzianLinguisticPolarityDecision2026}. Next, we first go over software engineering studies, followed by studies from other disciplines, and conclude with a summary on the related work. 

\subsection{Software Engineering}
\citet{huotalaPromiseChallengesUsing2024a} found that GPT-3.5 and GPT-4 LLMs perform as well as Master's students in paper screening. By simplifying abstracts they were faster to read, but simplification did not increase human performance. GPT-3.5 and GPT-4 LLMs had a precision and recall of (0.650, 0.176) and (0.500, 0.417), respectively. Including task examples in the screening task (one-shot (OS) and few-shot (FS)) was observed to improve screening performance, but the authors must have a list of high-quality seed papers to start with. The usage of prompt optimization was recommended.

\citet{huotalaSESREvalDatasetEvaluating2025} created the SESR-Eval benchmark dataset for benchmarking LLMs in paper screening, and nine LLMs were benchmarked in paper screening using the dataset. Most LLMs performed similarly in paper screening, and the performance differences between secondary studies were greater than for LLMs. GPT-4.1 mini performed the best, but small models, such as Ministral 8B and GPT-4.1 Nano, had poor screening performance. The cost of using LLMs to screen papers was relatively low. The authors concluded that LLMs are not yet recommended for automating paper screening, since accuracy varied widely across secondary studies, and as no LLM managed a high recall with reasonable precision.

\citet{felizardoChatGPTApplicationSystematic2024} evaluated the accuracy of GPT-4 (ChatGPT) in paper screening for two SRs. They replicated the first selection phase of two SRs with the help of ChatGPT, where the LLMs' screening results were compared to those of humans. The results were that the overall accuracy in paper screening was 80.7\%, and that in both SRs the LLMs made false judgments regarding included and excluded papers. For the first SR, there were 16 false negatives (FN) and 17 false positives (FP). For the second SR, there were 35 FNs and 27 FPs. The recommendation was not to rely completely on LLMs for paper screening, but LLMs can be a valuable tool with SRs as they can mitigate bias and inaccuracies in the included evidence.

\citet{liEvaluatingEffectivenessLarge2024} selected benchmark datasets from the SYNERGY dataset. Using that data, they benchmarked nine LLMs in terms of overall accuracy, sensitivity, and specificity. The results of the LLMs were then compared to gold-standard human decisions. The performance of LLMs varied significantly, but GPT-4 demonstrated a good balance of sensitivity and specificity and the accuracy was consistently in the high 90\%s. This indicates a high potential for LLMs in paper screening tasks, but they are not yet ready to fully replace humans in the task.

\citet{delgado-chavesTransformingLiteratureScreening2025} benchmarked 18 LLMs in paper screening. They picked three existing SRs, from which about 6,500 primary studies were evaluated by 18 different LLMs. From the primary studies, 92\%, 88\%, and 40\% of the papers were correctly classified by the LLMs, respectively. There was a significant variation between secondary studies, and the structure of the inclusion and exclusion criteria significantly affected the performance of LLMs in the task. The best-performing LLMs in terms of Matthews Correlation Coefficient (MCC) were GPT-4o (MCC=0.349) and GPT-4o mini (MCC=0.342). The authors emphasized that extra caution should be taken when articulating inclusion and exclusion criteria. A workload reduction between 33\% and 93\% was observed during paper screening.

\citet{shameemExploratoryStudyEffectiveness2026} followed a mixed-methods research to benchmark LLMs on multiple phases of SRs - including formulating research questions, generating search strings, quality assessment, developing taxonomies, and paper screening. First, they selected four SRs as "benchmark studies" (BS), and GPT-4o was used to screen the relevant papers. For all benchmark studies (BS1, BS2, BS3, BS4), GPT-4o correctly classified 64.5\%, 68\%, 58\%, and 72\% of papers correctly, respectively. Recall was 55.17\%, 50\%, 100\%, and 72\%, for the respective benchmark studies. Precision was 76.1\%, 31.2\%, 55.3\%, and 72\%, respectively. However, there were certain domain-specific challenges (e.g., domain-specific literature) that humans would have identified, but the LLM failed to do so.

\citet{thodeExploringUseLLMs2025} benchmarked five LLMs in paper screening, and used the SR's labeled studies as the ground truth. They designed three different prompts for the task. In addition, they created a dual-LLM system to test how a two-reviewer screening workflow would work with LLMs. As a result, LLMs exhibited a high recall (up to 98\%) but low precision (27\%). The dual-LLM system managed an average recall of 99\% with a precision of 27\%. The best-performing LLM was Meta Llama 2 with an average recall of 98\%, while GPT-4 Turbo had the highest average precision (72\%).

\subsection{Other Disciplines}

\citet{dennstadtTitleAbstractScreening2024} used LLMs to screen papers based on the predefined inclusion and exclusion criteria and the paper's title and abstract. The performance of LLMs in paper screening varied depending on the LLM being used and on the dataset. FlanT5 was the best-performing model with sensitivity (recall / true positive rate) and specificity (true negative rate) of 94.48\%/31.78\%. OpenHermes-NeuralChat, Mixtral and Platypus 2 were trailing quite a bit behind, with sensitivity/specificity of 97.58\%/19.12\%, 93\%/75.19\% and 97.58\%/38.34\% for the three models, respectively. Even small changes to the user prompt or the Likert scale from 1-5 to 1-10 significantly impacted the screening performance of LLMs.

\citet{tranSensitivitySpecificityAvoidable2023} evaluated the GPT-3.5 Turbo LLM for paper screening by first developing prompts that aim to mimic the process of screening by human researchers. The LLM was tested on 22,666 papers. The LLM in the paper screening task had a sensitivity of 97.1\%, but with poor specificity (37.7\%). A workload reduction of 65\% was observed. While the tested LLM had a good sensitivity, human intervention was recommended because of the poor specificity. 

\citet{matsuiHumanComparableSensitivityLarge2024a} evaluated paper screening with GPT-3.5 and GPT-4 LLMs in two SRs containing 1,381 and 3,146 papers, respectively. Both of the LLMs had a fairly high paper screening throughput - about 110 records per minute. The total screening task was completed in two hours. In the first SR, GPT-3.5 had a sensitivity/specificity of 0.900/0.709 versus GPT-4's 0.806/0.996. For the second SR, the performance was fairly similar, with sensitivity/specificity of GPT-3.5's 0.958/0.116 versus GPT-4's 0.875/0.855. A deeper analysis found that GPT-3.5 excluded papers because of insufficient domain knowledge, while GPT-4 misinterpreted inclusion criteria.

\citet{guoAutomatedPaperScreening2024b} benchmarked the GPT-4 LLM in paper screening. They screened over 24,000 title-abstract pairs across six clinical reviews. GPT-4 had an accuracy of 91.0\%, F1 score of 0.60, sensitivity of excluded papers of 0.91, and a sensitivity of included papers of 0.76. The agreement between models was high (kappa=0.96) and the LLMs could give justifications for their incorrect classifications. The authors concluded that LLMs have the potential to improve SR efficiency and lead to more accurate and reliable conclusions in medical research.

\citet{serrettiBenchmarkingLocalSchemaConstrained2026} developed a fully local LLM pipeline using the open-weight gpt-oss-20b LLM to benchmark title/abstract screening for three SRs. The performance (F1 score) of the LLM in paper screening was 98.9\%, 76.0\%, and 83.6\% for  the three SRs, respectively. The findings demonstrate that a fully locally run LLM can support researchers in paper screening.

\citet{tradStreamliningSystematicReviews2025} evaluated the performance of a commercial paper screening tool called Rayyan, and an LLM-based paper screening system in automating the screening process. Out of 14,439 papers, two thousand papers were used to train the paper classifier of Rayyan. The paper screening performance was analyzed using different thresholds for paper inclusion. While Rayyan performed adequately (exclusion rate of 50.7\%-72.1\%), the LLM-based screening system was far superior, achieving an exclusion rate of 99.5\%, and a workload reduction of 95.5\%, as the screening took 25.5 hours. The paper demonstrates that LLM-based screening systems are far superior compared to traditional ML-based systems, such as Rayyan.

\citet{adamPromptEngineeringLarge2026} conducted a large experiment with 515 different prompts against open-source LLMs (Meta Llama and Mistral) to study how the prompt structure affects screening performance in four SRs from the medical domain. The resulting F1, recall and precision were 27.6\%, 83.6\% and 18.5\%, respectively. The paper screening performance was observed to increase significantly when 1) larger-parameter LLMs were used; 2) the prompt included the title of the SR being conducted; and 3) the prompt included the selection criteria. Screening accuracy decreased when abstracts were used in screening rather than paper titles. The best-performing prompt achieved 72.2\% recall and 66.1\% specificity. This translated into a workload reduction of around 34.5--37.5\%. The paper highlighted that there is a precision-recall tradeoff to be made and the prompt structure significantly affected screening performance.

\citet{jozwiakREFLECTIVETIABCosteffectivePrompt2026} benchmarked Gemini 3 Flash and GPT-5.2 LLMs in paper screening and experimented with how prompt optimization could improve paper screening performance. A dataset of 100 papers was constructed from the papers that had the highest inter-model disagreement ratio, and was labeled by experts for the experiment. Prompt optimization improved recall for all LLMs by +3.7\% to +37.1\%. The best-performing LLM was Gemini 3 Flash with an accuracy of 91\%, and F1 score of 81.6\%, while costing 25 times less than GPT-5.2 per abstract.

\citet{qiuPromptBasedTutorialLarge2026} benchmarked LLMs for paper screening by using a stepwise training framework with three studies - one for each phase. The first study was used to optimize the technical parameters of the experiment with 1000 abstracts and 50 full texts. The second and third studies validated this approach. The LLMs achieved 100\% sensitivity and a specificity between 85.6\% and 93\%. Full-text screening had an accuracy between 82.1\% and 96.2\%. 

\citet{sujauAcceleratingDiseaseModel2025} evaluated the effectiveness of LLMs to rank primary studies for climate-sensitive zoonotic disease research. The dataset contained 2905 title-abstract pairs that were related to four zoonotic diseases, and a set of paper selection criteria was constructed. Five researchers screened abstracts using the defined criteria. After that, the papers were screened using GPT-4o LLM with zero-shot (ZS) and Chain-of-Thought (CoT) prompting techniques. A workload reduction of at least 70\% was achieved, while the LLM maintained a recall of 95\%. The authors emphasized that further research is needed on standardized benchmark datasets, but the findings demonstrate that LLMs can substantially reduce the screening effort.

\citet{behrouzianLinguisticPolarityDecision2026} used GPT-5.1 and GPT-3.5 Turbo LLMs to benchmark paper screening against a dataset from the Cochrane review corpus, which contained 1,000 abstracts. The recall of GPT-5.1 and GPT-3.5 Turbo ranged from 0.72 to 0.32 and 0.92 to 0.18, respectively. Excluding papers immediately, rather than labeling papers as unsure, made both models perform better with consistency. This demonstrated that LLMs do not handle uncertainty well in paper screening.

\subsection{Summary}

The common consensus of the related work is that LLMs show significant promise in paper screening, by reliably reducing the workload of practitioners and researchers in paper screening tasks, while LLMs have been observed to perform as well as non-expert humans in the task. Despite these promises, LLMs are not yet recommended for fully replacing humans in paper screening, as the screening performance is greatly affected by the structure and clarity of the inclusion and exclusion criteria, the structure of the prompt, certain domain-specific issues, and the trade-off of selecting a cost-effective LLM instead of a state-of-the-art LLM to reduce costs.

Existing literature predominantly treats paper screening as a binary classification task, as they examine only the overall screening performance, and provide only a minimal criterion-level analysis of how LLMs handle specific inclusion/exclusion criteria or whether the performance varies across criteria types. Furthermore, only a few studies investigate inter-LLM agreement and the impact of dataset sampling, or how refining criteria affects paper screening performance.

\section{Background}

\label{sec:background}

In this section, we introduce the core concepts for this study. First, we describe the paper screening process in SRs. Then, we describe paper screening as a classification task, and how the criterion-level decisions are aggregated into an overall screening decision. Finally, we discuss the agreement between multiple screeners.

\subsection{Inclusion and exclusion criteria}

In SRs, inclusion (IC) and/or exclusion (EC) criteria are applied to research papers' titles and abstracts to determine whether the paper should be included or excluded~\citep{kitchenhamGuidelinesPerformingSystematic2007}. Excluded papers are discarded, and the remaining included papers are further examined in the full-text screening phase~\citep{kitchenhamGuidelinesPerformingSystematic2007}. Each of these criteria map to a specific research question (RQ) of the study, so the selected primary studies have the possibility to contain an answer to the RQ. The SR guidelines state that the criteria should be crafted during the SR protocol development, and may be refined during the literature search process~\citep{kitchenhamGuidelinesPerformingSystematic2007}. Importantly, criteria should be formulated so that they can be reliably assessed and classify studies with high accuracy.

Some inclusion and exclusion criteria may be applied to the paper's metadata, such as language, date of publication, and publisher information. While human researchers can easily determine whether a paper was published in a certain year or period, LLMs can have issues making such decisions without having access to the paper's metadata or, in some cases, the publisher's copyright statement in the abstract \citep{Felizardo_Deizepe_Coutinho_Gomes_Meireles_Gerosa_Steinmacher_2025}. In addition, it is noted that criteria that assess the language of the paper, or are biased towards the knowledge of the authors, can influence the decisions based on the criteria \citep{kitchenhamGuidelinesPerformingSystematic2007, kitchenhamEvidenceBasedSoftwareEngineering}.

The implications of these observations raise important questions regarding how LLMs handle screening based on a set of rules. First, since inclusion and exclusion criteria have differences in how they can be assessed objectively, decisions made by LLMs can be more consistent with certain criteria than with others. This motivates RQ3, where we analyze how LLM screening decisions actually agree across paper selection criteria (RQ3.2). Second, as \citet{kitchenhamGuidelinesPerformingSystematic2007} states that the criteria may be refined during the literature search process, it remains uncharted whether such refinement also improves the performance of LLMs in paper screening. This motivates our RQ4, which investigates the effect of refining the paper selection criteria on LLM screening performance.

\subsection{Title-abstract screening as a classification task}

Title-abstract screening (paper screening) can be considered a binary classification task~\citep{duMachineLearningModels2024,vandeschootOpenSourceMachine2021a}, where each paper is either included or excluded based on its title and abstract~\citep{kitchenhamGuidelinesPerformingSystematic2007}.

However, SRs typically include multiple inclusion and exclusion criteria, and the overall inclusion/exclusion decision depends on how individual criterion-level decisions are combined or aggregated. A common pattern is that a paper is excluded if it matches even one exclusion criterion, or if it does not match every inclusion criterion. However, variations of these rule combinations do exist. This complexity of handling multiple criteria raises important questions about inter-screener agreement, and how criterion-level decisions should be aggregated. This also motivates our RQ3.1 and RQ3.2.

When LLMs are applied in paper screening, a common pattern is that the LLM produces an overall decision in binary format. This binary classification task can be expanded to support probabilistic (a floating point between 0.0 and 1.0), or e.g., Likert-scale decisions~\citep{felizardoChatGPTApplicationSystematic2024,huotalaSESREvalDatasetEvaluating2025}. The aggregation behind the overall decision of the LLM can be done in two ways.

In the first way, the LLM produces judgments for each criterion and also the overall decision. This adds a considerable amount of flexibility, but there is an issue. Because most LLMs are decoder-only transformers~\citep{Vaswani_Shazeer_Parmar_Uszkoreit_Jones_Gomez_Kaiser_Polosukhin_2023}, they predict the next token(s) based on the earlier tokens, the overall decision may contradict the LLM's own criterion-level judgments, if e.g., the overall decision is generated before the criterion-level decisions.

In the second way, the LLM produces only the judgments for the criteria, and the overall decision is aggregated manually from the judgments of the criteria. The benefit here is that the decision boundary can be adjusted flexibly, if one wants to e.g., maximize recall~\citep{huotalaSESREvalDatasetEvaluating2025}. However, if a study has a significant number of criteria, a single incorrect criterion judgment is enough to falsely exclude a possibly relevant paper. This trade-off here motivates RQ2. 

\subsection{Agreement between multiple raters}

\begin{figure}[htbp]
	\centering
	\begin{tikzpicture}
		\begin{axis}[
				xlabel={Number of decision makers},
				ylabel={Accuracy of a majority vote},
				xmin=1, xmax=19,
				ymin=0.4, ymax=1.02,
				grid=major,
				legend pos=south east,
				mark size=1.5pt
			]
			
			\addplot[color=green, mark=triangle*] coordinates {
				(1,0.7000) (3,0.7840) (5,0.8369) (7,0.8740) (9,0.9012) (11,0.9218) (13,0.9376) (15,0.9500) (17,0.9597) (19,0.9674)
			};
			\addlegendentry{$p=0.70$}
			
			\addplot[color=red, mark=square*] coordinates {
				(1,0.6000) (3,0.6480) (5,0.6826) (7,0.7102) (9,0.7334) (11,0.7535) (13,0.7712) (15,0.7869) (17,0.8011) (19,0.8139)
			};
			\addlegendentry{$p=0.60$}
			
			\addplot[color=black, dashed] coordinates {(1,0.5) (19,0.5)};
			\addlegendentry{Random guess}
		\end{axis}
	\end{tikzpicture}
	\caption{Accuracy of majority voting as a function of the number of independent decision makers.}
	\label{fig:condorcet-jury-theorem}
\end{figure}
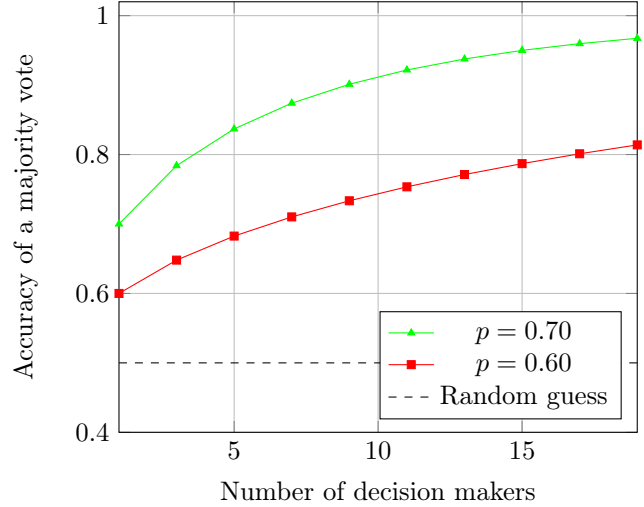

The standard SR methodology recommends that at least two independent reviewers screen papers, to reduce bias and ensure rigor~\citep{kitchenhamGuidelinesPerformingSystematic2007}. The benefit of multiple reviewers can be formalized by the Condorcet Jury Theorem~\citep{listEpistemicDemocracyGeneralizing2001, shteingartMajorityVotingCondorcets2020}, which demonstrates that when individual decision-makers make the correct binary decision with a higher than 50/50 chance, the accuracy of a majority vote increases as the number of independent decision-makers increases (Figure \ref{fig:condorcet-jury-theorem})~\citep{listEpistemicDemocracyGeneralizing2001,shteingartMajorityVotingCondorcets2020}. Empirical evidence supports this principle. A study by \citet{abramowitzWhoReviewsReviewers2025} found that multiple independent reviewers provide more reliable quality assessment than a single reviewer in peer review contexts.

Three conditions of the Condorcet Jury Theorem are important when LLMs screen papers. First, the benefit of multiple raters requires a majority. Second, with two LLMs as decision makers, a majority vote cannot resolve a possible disagreement. Using two LLMs to screen is therefore only useful to flag disagreements, or to include a paper if either of the models includes it. This increases recall, but lowers precision. Third, the LLMs making the decisions must reach their decisions independently. It is known that LLMs share much of their training data, so choosing e.g. models from only a single provider might reduce the collaborative usage of these LLMs, and most likely the errors of the LLMs are correlated. A high agreement between LLMs is therefore ambiguous. This might indicate that the LLMs themselves are reliable, or they make the same mistakes, which makes it pointless to use multiple LLMs for screening.

A comprehensive evaluation of LLM reliability requires examining both inter-model agreement and the consistency of decisions across different LLMs and criteria. Such analysis is essential for assessing whether LLMs can reliably replace or augment human screeners in the paper screening phase of SRs. For these reasons, LLM agreement is analyzed between LLMs (RQ3.1) and between inclusion and exclusion criteria (RQ3.2).

\section{Methodology}
\label{sec:methods}

\begin{figure}
\resizebox{\textwidth}{!}{%
\includegraphics[]{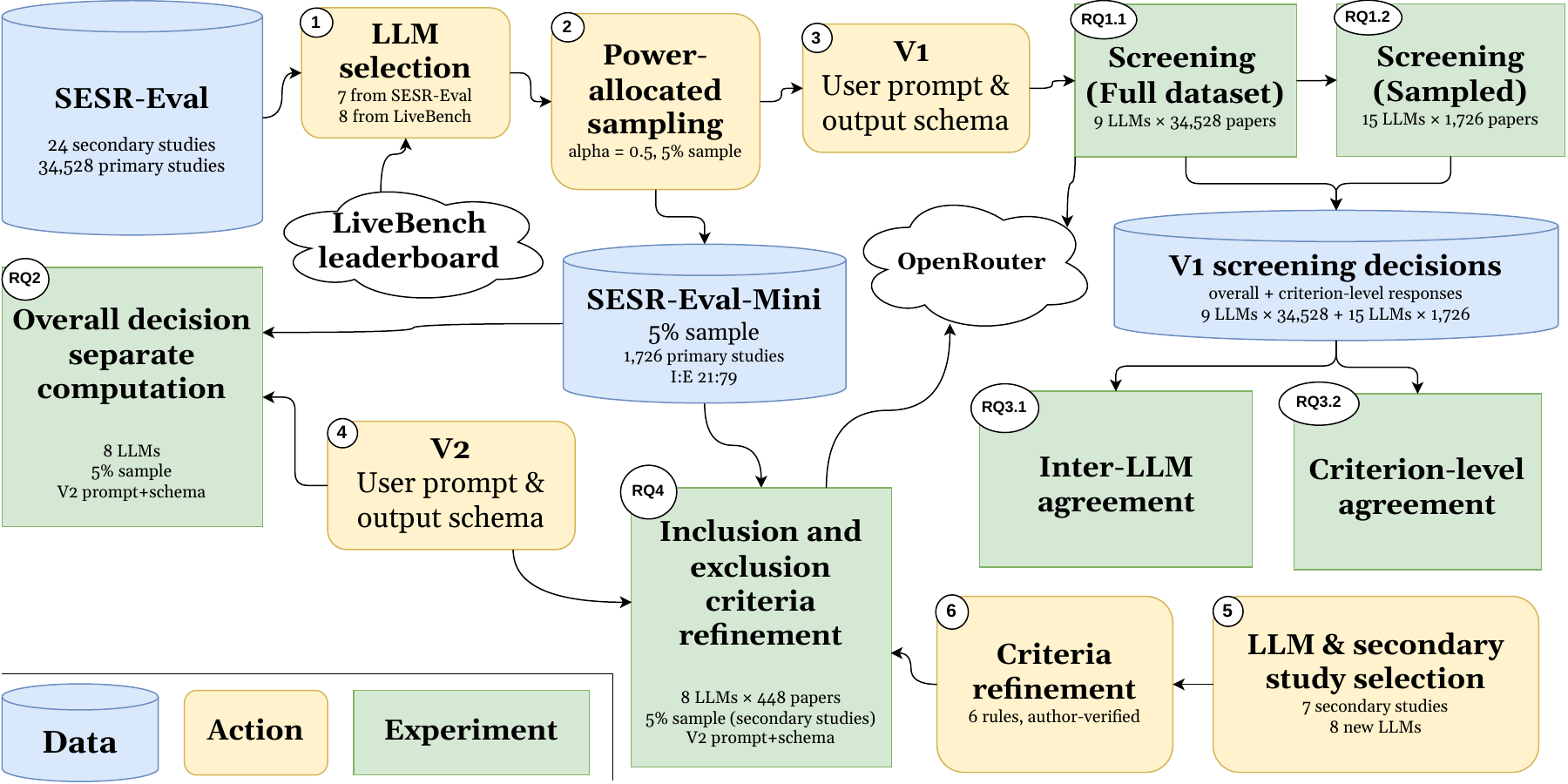}
}%
\caption{The study methodology, including numbered actions and the conducted experiments.}
\label{fig:methodology}
\end{figure}

Next, we cover the methodology used in the study. We followed the SIGSOFT empirical standards to conduct this research~\citep{ralph2020empirical}. The high-level overview of the methodology can be seen in Figure~\ref{fig:methodology}. First, in Subsection~\ref{sec:performance_selecting_llms}, we describe how LLMs were selected for the experiments \textcircled{1}. Then, we describe how the data sampling for the SESR-Eval-Mini dataset was conducted in Subsection~\ref{sec:data_selection_llms} \textcircled{2}. After this, we give a detailed overview of how we benchmarked the LLMs (Subsection~\ref{sec:performance_benchmarking_llms}), how the computation of the overall screening decision was separated (Subsection~\ref{sec:separating_computation}), the setup behind the LLM agreement between each other and criteria (Subsection~\ref{sec:llm_agreement_between_each_other}), and the criteria modification process (Subsection~\ref{sec:criteria_modification}). In Subsection~\ref{sec:statistical_methods}, we cover the statistical methods used.

\subsection{LLM selection (all RQs)}
\label{sec:performance_selecting_llms}

\begin{table*}[htbp]
	\centering
	\caption{Selected models with their reasoning effort setting, costs per 1M tokens (in U.S. dollars), and Pydantic AI output type for OpenRouter. SESR-Eval models are picked from \citet{huotalaSESREvalDatasetEvaluating2025}. \\ \\ \textsuperscript{1} Pricing as of September 2026 \\ \textsuperscript{2} Pricing as of April 2025}
	\label{tab:companies_and_models}
	\resizebox{\textwidth}{!}{%
		\begin{tabular}{llllrrl}%
			\toprule
			                                                    & \textbf{Publisher}   & \textbf{Model}    & \textbf{\makecell[l]{Reasoning \\ effort}} & \textbf{\makecell[l]{Input (\$) / \\ 1M tokens}} & \textbf{\makecell[l]{Output (\$) / \\ 1M tokens}} & \textbf{\makecell[l]{Output \\ type}} \\
			\midrule
			\multirow{8}{*}{\rotatebox{90}{\textbf{LiveBench}}} & \textbf{OpenAI}      & GPT-5.5           & Low    & \$5.00 \textsuperscript{1}  & \$30.00 \textsuperscript{1} & Native   \\
			                                                    & \textbf{Anthropic}   & Claude Opus 4.7   & Low    & \$5.00 \textsuperscript{1}  & \$25.00 \textsuperscript{1} & Tool     \\
			                                                    & \textbf{Google}      & Gemini 3.1 Pro    & Low    & \$2.00 \textsuperscript{1}  & \$12.00 \textsuperscript{1} & Tool     \\
			                                                    & \textbf{Z.AI}        & GLM 5.1           & Low    & \$1.05 \textsuperscript{1}  & \$3.50 \textsuperscript{1}  & Tool     \\
			                                                    & \textbf{Moonshot AI} & Kimi K2.6         & Low    & \$0.74 \textsuperscript{1}  & \$3.49 \textsuperscript{1}  & Tool     \\
			                                                    & \textbf{SpaceXAI}    & Grok 4.20 Beta    & Low    & \$1.25 \textsuperscript{1}  & \$2.50 \textsuperscript{1}  & Tool     \\
			                                                    & \textbf{Alibaba}     & Qwen 3.6 Plus     & Medium & \$0.32 \textsuperscript{1}  & \$3.20 \textsuperscript{1}  & Prompted \\
			                                                    & \textbf{DeepSeek}    & V4 Pro            & High   & \$0.435 \textsuperscript{1} & \$0.87 \textsuperscript{1}  & Tool     \\
			\midrule
			\multirow{7}{*}{\rotatebox{90}{\textbf{SESR-Eval}}} & \textbf{OpenAI}      & o3-mini           & High   & \$1.10 \textsuperscript{2}  & \$4.40 \textsuperscript{2}  & Native   \\
			                                                    & \textbf{OpenAI}      & GPT-4o            & N/A    & \$2.50 \textsuperscript{2}  & \$10.00 \textsuperscript{2} & Native   \\
			                                                    & \textbf{OpenAI}      & GPT-4.1           & N/A    & \$2.00 \textsuperscript{2}  & \$8.00 \textsuperscript{2}  & Native   \\
			                                                    & \textbf{OpenAI}      & GPT-4.1 mini      & N/A    & \$0.40 \textsuperscript{2}  & \$1.60 \textsuperscript{2}  & Native   \\
			                                                    & \textbf{Anthropic}   & Claude 3.7 Sonnet & N/A    & \$3.00 \textsuperscript{2}  & \$10.00 \textsuperscript{2} & Native   \\
			                                                    & \textbf{DeepSeek}    & DeepSeek R1       & N/A    & \$0.54 \textsuperscript{2}  & \$2.18 \textsuperscript{2}  & Native   \\
			                                                    & \textbf{Meta}        & Llama 4 Maverick  & N/A    & \$0.19 \textsuperscript{2}  & \$0.85 \textsuperscript{2}  & Native   \\
			\bottomrule
		\end{tabular}%
	}%
\end{table*}

For the LLM selection \textcircled{1}, we first picked existing LLMs that were benchmarked with the SESR-Eval dataset~\citep{huotalaSESREvalDatasetEvaluating2025}, excluding Ministral 8B and GPT-4.1 Nano, as they exhibited poor screening performance compared to the rest of the models in the work. Selecting the existing LLMs gives us a good baseline that we can use to directly compare the newer models against.

In addition to the existing LLMs, we selected some new LLMs from the LiveBench leaderboard by \citet{livebench} (version 2026-01-08~\footnote{https://web.archive.org/web/20260504142848/https://livebench.ai/\#/?highunseenbias=true}), to see how the current state-of-the-art performs in paper screening. LiveBench actively maintains a list of LLMs alongside the performance results in various domains and contexts, including reasoning, coding, mathematics, and data analysis. From the leaderboard, we first picked eight distinct publishers: OpenAI~\footnote{https://openai.com}, Anthropic~\footnote{https://www.anthropic.com}, Google~\footnote{https://ai.google}, Z.AI~\footnote{https://z.ai}, Moonshot AI~\footnote{https://www.moonshot.ai}, SpaceXAI~\footnote{https://x.ai}, Alibaba~\footnote{https://qwen.ai}, DeepSeek~\footnote{https://deepseek.com} -- not in any particular order. Next, for each publisher, we selected their LLM with the highest global average performance score to minimize the risk of picking LLMs that fare better in some domain or context. As demonstrated by \citet{huotalaSESREvalDatasetEvaluating2025}, selecting the most expensive model does not guarantee an increase in screening performance. 

While most new LLMs were set to a 'low' reasoning effort, the reasoning effort of DeepSeek V4 Pro and Qwen 3.6 Plus were set to 'high' and 'medium', since they were the first LLMs we tested, we mirrored the reasoning effort setting of the existing o3-mini LLM. As we previously demonstrated that "thinking" models give negligible improvements in screening performance when reasoning effort is increased, and that it increases token output significantly~\citep{huotalaSESREvalDatasetEvaluating2025}, we then decided to configure the rest of the models to a 'low' reasoning effort.

In total, we selected 15 LLMs for the experiments (Table \ref{tab:companies_and_models}) -- seven of which were previously benchmarked in the SESR-Eval paper, and eight new LLMs, which represent the current state-of-the-art.

\subsection{Data sampling -- SESR-Eval-Mini (all RQs)}

\label{sec:data_selection_llms}

\begin{table}[htbp]
	\centering
	\caption{Number of papers per secondary study, and the number of papers in the 5\% power-allocated sample, \(\alpha = 0.5\). Included:Excluded (I:E) is the percentage split of included to excluded papers. Secondary studies sourced from \citet{huotalaSESREvalDatasetEvaluating2025}.}
	\label{tab:sample_sizes}
	\begin{tabular}{lrrrr}
		\toprule
		& \multicolumn{2}{c}{\textbf{Full}} & \multicolumn{2}{c}{\textbf{Sampled}} \\
		\cmidrule(lr){2-3} \cmidrule(lr){4-5}
		\textbf{Study}                                           & \textbf{N} & \textbf{I:E} & \textbf{N} & \textbf{I:E} \\
		\midrule
		\citet{casalaroModeldrivenEngineeringMobile2022a}         & 2,260      & 3:97         & 109        & 3:97         \\
		\citet{sharbafConflictManagementTechniques2023a}          & 322        & 53:47        & 41         & 46:54        \\
		\citet{tambonHowCertifyMachine2022a}                      & 1,024      & 18:82        & 73         & 25:75        \\
		\citet{rodriguez-perezPerceivedDiversitySoftware2021a}    & 2,748      & 8:92         & 120        & 6:94         \\
		\citet{bjarnasonEmpiricallyBasedModel2023a}               & 3,703      & 1:99         & 139        & 1:99         \\
		\citet{kuutilaTimePressureSoftware2020b}                  & 4,186      & 5:95         & 148        & 5:95         \\
		\citet{geraldiSoftwareProductLine2020a}                   & 167        & 34:66        & 30         & 27:73        \\
		\citet{zhangTestingVerificationNeuralnetworkbased2020a}   & 104        & 100:0        & 23         & 100:0        \\
		\citet{tebesAnalyzingDocumentingSystematic2020a}          & 495        & 72:28        & 51         & 78:22        \\
		\citet{ebrahimiMobileAppPrivacy2021a}                     & 111        & 50:50        & 24         & 46:54        \\
		\citet{lewowskiHowFarAre2022a}                            & 1,651      & 10:90        & 93         & 10:90        \\
		\citet{vandinterPredictiveMaintenanceUsing2022a}          & 559        & 25:75        & 54         & 19:81        \\
		\citet{stradowskiMachineLearningSoftware2023a}            & 1,194      & 62:38        & 79         & 59:41        \\
		\citet{alonsoSystematicMappingStudy2023b}                 & 222        & 86:14        & 34         & 79:21        \\
		\citet{somersDigitaltwinbasedTestingCyber2023b}           & 405        & 33:67        & 46         & 26:74        \\
		\citet{teixeiraModelingAutomaticCode2017a}                & 318        & 59:41        & 41         & 63:37        \\
		\citet{lenarduzziSystematicLiteratureReview2021a}         & 301        & 14:86        & 40         & 15:85        \\
		\citet{pereiraLearningSoftwareConfiguration2021a}         & 69         & 100:0        & 19         & 100:0        \\
		\citet{barisicMultiparadigmModelingCyber2022a}            & 326        & 47:53        & 41         & 51:49        \\
		\citet{deckersSystematicLiteratureReview2022a}            & 321        & 6:94         & 41         & 7:93         \\
		\citet{raniDecadeCodeComment2023a}                        & 2,079      & 3:97         & 104        & 1:99         \\
		\citet{zakeri-nasrabadiSystematicLiteratureReview2023b}   & 9,695      & 5:95         & 225        & 7:93         \\
		\citet{shevtsovControlTheoreticalSoftwareAdaptation2018a} & 1,487      & 11:89        & 88         & 17:83        \\
		\citet{linOpinionMiningSoftware2022a}                     & 781        & 16:84        & 63         & 14:86        \\
		\midrule
		\textbf{Total}                                           & 34,528     & 12:88        & 1,726      & 21:79        \\
		\bottomrule
	\end{tabular}
\end{table}

After the LLM selection \textcircled{1}, we continued with data sampling \textcircled{2}. We decided to conduct data sampling, as our main goal was to provide a cheaper alternative of SESR-Eval dataset, which is motivated by Table~\ref{tab:companies_and_models} as it shows that newer flagship models have significantly increased input and output token costs. In addition, we did not want to trade better performance for lower costs by choosing smaller LLMs, which are evidently cheaper~\citep{Felizardo_Deizepe_Coutinho_Gomes_Meireles_Gerosa_Steinmacher_2025}. As the data source, we used SESR-Eval by \citet{huotalaSESREvalDatasetEvaluating2025}, which is a gold-standard paper screening dataset. The dataset contains 34,528 pre-labeled primary studies from 24 software engineering secondary studies, which were rigorously sourced from high-reputation SE journals~\citep{huotala2025research}. The labels of the primary studies are the overall screening decisions. 

For the paper sampling strategy, we decided to use power-allocated sampling with $\alpha=0.5$~\citep{bankierPowerAllocationsDetermining1988}. Power-allocated sampling balances the proportional and equal allocation to control for variance across strata~\citep{bankierPowerAllocationsDetermining1988}. In practice, pure random sampling (a power-allocated sample with an $\alpha = 1.0$) at 5\% would give us a sample where the smallest secondary study (with 69 primary studies) would get an average of 3.45 primary studies, which would not allow any sort of secondary study-level inference on the accuracy. On the other hand, with an $\alpha = 0.0$, all secondary studies would get equal weight, which at 5\% would be 72 primary studies for each secondary study ($34,528$ $\times$ $0.05$ / 24). This would distort the overall dataset too much in favor of small studies. The value of $\alpha = 0.5$ is a compromise as it balances between equally allocating across all strata, and allocating proportional to the size of each stratum~\citet{bankierPowerAllocationsDetermining1988}. Table \ref{tab:sample_sizes} lists the secondary studies, their full primary study count and the sampled primary study count. The full dataset used in the experiments contains 34,528 rows, whereas the 5\% sample set contains 1,726 papers. One can notice that the sampled data has a different I:E ratio overall (21:79) than the full dataset (12:88). This is because smaller secondary studies had more inclusions and their proportional weight increases in power-allocated sampling. For RQ4, the subset of studies was further reduced to 448 papers, as we selected only those papers that were part of the seven secondary studies most suitable for criteria improvement \textcircled{5}.

The 5\% sampled dataset is called the SESR-Eval-Mini dataset, which is available in our prior work's \citep{huotalaSESREvalDatasetEvaluating2025} research artifact as an update~\citep{sesr_eval_research_artifact}.

\subsection{Benchmarking the LLMs (all RQs)}

\label{sec:performance_benchmarking_llms}

\begin{table}[htbp]
	\centering
	\caption{Tested models for each research question, with sample counts. F = full (N=34,528), S = 5\% power-allocated sample (N=1,726), R = papers of the 7 secondary studies from 5\% power-allocated sample (N=448). Bottom row indicates the user prompt and output schema version used.}
	\label{tab:test_llms_by_rq} 
	\begin{tabular}{l|l|cc|cc|ccc}
		\toprule 
		                                                    & \textbf{Model}             & \textbf{RQ1.1} & \textbf{RQ1.2} & \textbf{RQ2} & \textbf{RQ3.1} & \textbf{RQ3.2} & \textbf{RQ4} \\
		\midrule 
		\multirow{8}{*}{\rotatebox{90}{\textbf{LiveBench}}} & \textbf{GPT-5.5}           & -              & S              & S            & -              & -              & R            \\
		                                                    & \textbf{Claude Opus 4.7}   & -              & S              & S            & -              & -              & R            \\
		                                                    & \textbf{Gemini 3.1 Pro}    & -              & S              & S            & -              & -              & R            \\
		                                                    & \textbf{GLM 5.1}           & -              & S              & S            & -              & -              & R            \\
		                                                    & \textbf{Kimi K2.6}         & -              & S              & S            & -              & -              & R            \\
		                                                    & \textbf{Grok 4.20 Beta}    & -              & S              & S            & -              & -              & R            \\
		                                                    & \textbf{Qwen 3.6 Plus}     & F              & S              & S            & F, S           & F, S           & R            \\
		                                                    & \textbf{DeepSeek V4 Pro}   & F              & S              & S            & F, S           & F, S           & R            \\
		\midrule
		\multirow{9}{*}{\rotatebox{90}{\textbf{SESR-Eval}}} & \textbf{o3-mini}           & F              & S              & -            & F, S           & F, S           & -            \\
		                                                    & \textbf{GPT-4o}            & F              & S              & -            & F, S           & F, S           & -            \\
		                                                    & \textbf{GPT-4.1}           & F              & S              & -            & F, S           & F, S           & -            \\
		                                                    & \textbf{GPT-4.1 mini}      & F              & S              & -            & F, S           & F, S           & -            \\
		                                                    & \textbf{Claude 3.7 Sonnet} & F              & S              & -            & F, S           & F, S           & -            \\
		                                                    & \textbf{DeepSeek R1}       & F              & S              & -            & F, S           & F, S           & -            \\
		                                                    & \textbf{Llama 4 Maverick}  & F              & S              & -            & F, S           & F, S           & -            \\
        \midrule
         & \textbf{\makecell[l]{User prompt \& \\ Output schema \\ version}}             & \textbf{V1}    & \textbf{V1}    & \textbf{V2}  & \textbf{V1}    & \textbf{V1}    & \textbf{V2}  \\
		\bottomrule
	\end{tabular}
\end{table}

Next, we describe how we benchmarked the LLMs for each research question. Tested models for each RQ are listed in Table~\ref{tab:test_llms_by_rq}.


For screening on the full dataset (RQ1.1), we re-used the screening results of the seven existing LLMs that were gathered in prior work ~\citep{huotalaSESREvalDatasetEvaluating2025}, and benchmarked two new LLMs (see Table~\ref{tab:test_llms_by_rq}). For RQ1.2, we derived the sampled screening results for the seven existing LLMs and two new LLMs from RQ1.1, but ran the benchmarks for the remaining six new LLMs that were not benchmarked in RQ1.1. For RQ2, all eight new LLMs were benchmarked on the 5\% sample. In the case of RQ3.1 and RQ3.2, the same old and new LLMs as for RQ1.1 were used, as RQ3.1 and RQ3.2 are based on the data gathered from RQ1.1. As benchmarking on the full dataset made it possible to derive the results for the sampled data, we decided to also derive the results for the sampled data in RQ3.1 and RQ3.2, respectively. As all new LLMs were benchmarked for RQ2, we decided to also benchmark them on criteria refinement for RQ4. This makes sense, as both RQ2 and RQ4 use the same user prompt and output schema.

One might wonder why certain LLMs were not tested for certain RQs. After we first had benchmarked two new LLMs, Qwen 3.6 Plus and DeepSeek V4 Pro, for RQ1.1, we decided to not run rest of them (GPT-5.5, Claude Opus 4.7, Gemini 3.1 Pro, GLM 5.1, Kimi K2.6, or Grok 4.20 Beta) on the full dataset, as running these LLMs on the full dataset would have inflated costs significantly, and current LLMs are quite expensive to run~\citep{Felizardo_Deizepe_Coutinho_Gomes_Meireles_Gerosa_Steinmacher_2025}.\footnote{Our current cost for this study was \$1,100. Had we tested those models with the full SESR-Eval dataset, we estimated a cost of about \$6,500.} After reviewing prior observations that the performance differences between LLMs are fairly narrow~\citep{huotalaSESREvalDatasetEvaluating2025}, testing only two of the new LLMs for RQ1.1 was justified. As the sampled data represents 5\% of the full dataset with greatly reduced costs, we were able to benchmark all new LLMs for RQ1.2, RQ2, and RQ4.

We used OpenRouter~\citep{openrouter} to access the LLMs, which provides a unified application programming interface (API) to call the LLMs routed through the service. An additional benefit of using OpenRouter was that it supports Pydantic AI out of the box, which is a model-agnostic Python framework. It provides automatic output validation, error handling, and retry support that makes it possible to call LLMs reliably. Our recent work used OpenRouter and Pydantic AI for paper screening and it worked very well for benchmarking multi-vendor LLMs~\citep{huotalaSESREvalDatasetEvaluating2025}. The scripts are available in the paper's research artifact~\citep{zenodo_research_artifact}.

\newcommand{\graycell}{\cellcolor{gray!20}}

The benchmarking was done with a parallelized Python script, which used the Pydantic AI library for calling LLMs. The LLMs were configured with a top\_p value of 0.1, and the following system prompt: \begin{quote}You are an expert research assistant.\end{quote} Recent works by \citet{huotalaPromiseChallengesUsing2024a,huotalaSESREvalDatasetEvaluating2025} and \citet{felizardoChatGPTApplicationSystematic2024} have used the same parameters and system prompt to increase the reproducibility and determinism of LLM outputs without affecting output quality. We did not set a temperature parameter this time, as OpenAI's current recommendation is to adjust either top\_p or temperature, not both~\citep{Chiang_Lee_2023}. The results were output as Comma-Separated Values (CSV) files, which were further processed using \texttt{pandas (3.0.1)} and \texttt{scikit-learn (1.8.0)} Python libraries.

For both RQ1.1 and RQ1.2, we computed accuracy, precision, recall, F1, MCC, and lost evidence.

\subsection{Separating the computation of the overall screening decision (RQ2)}
\label{sec:separating_computation}

\begin{figure}

\centering
\caption{User prompt and output schema versions, the changes made for each version, and the RQs in which they are used. V1 user prompt and output schema are from \citet{huotalaSESREvalDatasetEvaluating2025}.}
\label{fig:versions}
\resizebox{\textwidth}{!}{%
\includegraphics[]{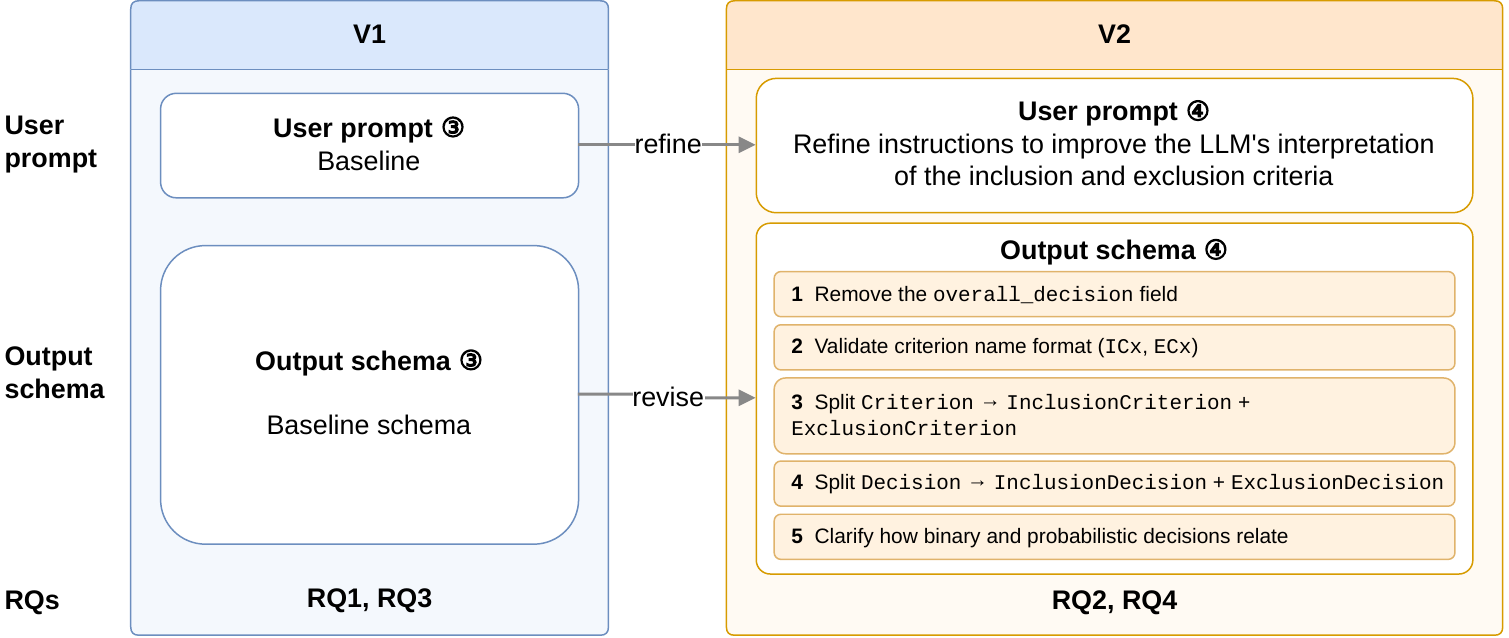}
}%
\end{figure}

Figure \ref{fig:versions} describes the changes made to the user prompt and the LLM output schema, respectively. 

The first version (V1) of the user prompt was the same as that used in our prior work~\citep{huotalaSESREvalDatasetEvaluating2025}. This same user prompt was then used for RQ1.1, RQ1.2, RQ3.1, and RQ3.2, as these RQs were directly compared with the results of our prior work, so no changes were required nor made to the user prompt \textcircled{3}.

For RQ2, the changes made to the user prompt in V2 were tightly coupled with the motivation behind the respective RQ. Removing the overall screening decision field necessitated refining the user prompt, as the screening decision of the LLMs relied on the criterion decisions. Because of this, the instructions given to the LLMs were refined to better guide the LLMs' interpretation of the criteria. Also for RQ4, the V2 version user prompt was used, because it focused solely on refining inclusion and exclusion criteria; the refinements made for V2 version of user prompt aimed to provide an improved guidance for the LLMs and to reduce the ambiguity of applying criteria \textcircled{4}.

No changes were made for the output schema for RQ1.1, RQ1.2, RQ3.1, or RQ3.2, as the results were directly compared with prior work~\citep{huotalaSESREvalDatasetEvaluating2025}. On the other hand, significant changes were implemented to the V2 output schema (RQ2 and RQ4).

First, the overall decision field was removed from the output schema, as the overall decision was computed from the inclusion and exclusion criterion decisions, making the overall decision field redundant.

Second, we added extra validation for the criterion name format as an attempt to improve reliability of the structured responses. The increased number of LLMs tested increases the likelihood of LLMs hallucinating the output, and letting the Pydantic AI library apply stricter output validations is an attempt to mitigate this.

Third, the Python classes for Criterion and Decision were split for inclusion and exclusion criteria, as we wanted to make sure that the LLMs rigorously understand the difference between inclusion and exclusion criterion evaluations. Lastly, we clarified how the binary and probabilistic decisions are related to each other, e.g. the probabilistic decision means the probability of an inclusion criteria being met, and for exclusion criteria, this means the probability of the exclusion criterion being met. This was to mitigate a potential misunderstanding of the LLMs where the LLM could misunderstand the probability of exclusion as "how certain the LLM is about its answer". If the LLM output JavaScript Object Notation (JSON) that did not meet the output schema, Pydantic AI automatically retried the request and instructed the LLM to correct the mistake.

Because the overall decision field was removed for RQ2 and RQ4, the overall decision was determined by the results of the inclusion and exclusion criterion. Each secondary study in the SESR-Eval dataset contained inclusion and exclusion criteria, which we converted into YAML Ain't Markup Language (YAML) files and compared against the original inclusion and exclusion criteria to ensure correctness. The resulting YAML files were consumed by the experiment scripts. Each secondary study had rules, by which the inclusion criteria (IC) and exclusion criteria (EC) are evaluated. This information was also provided to the LLM as a boolean formula, e.g. \begin{quote}(IC1 AND IC2 AND IC3) AND NOT (EC1 OR EC2)\end{quote} Both versions of the user prompts and output structures are included in the research artifact~\citep{zenodo_research_artifact}. For RQ2, we also computed accuracy, precision, recall, F1, MCC, and lost evidence.

\subsection{LLM agreement - Overall decision and criteria (RQ3.1 \& RQ3.2)} \label{sec:llm_agreement_between_each_other}

To study the agreement between LLMs in overall decision RQ3.1 and criteria RQ3.2, we first retrieved the raw responses of the LLMs originally collected for RQ1.1. These raw responses contained the original screening responses, which we parsed as objects. From the objects, we retrieved the \texttt{binary\_decision} field, which indicates the LLMs overall screening decision. For studying the agreement across the inclusion and exclusion criteria in RQ3.2, we extracted the results corresponding to the inclusion and/or exclusion criteria, that were also found from the raw responses originally collected for RQ1.1.

\subsection{Modifying criteria (RQ4)} 
\label{sec:criteria_modification}

\begin{table}[htbp]
\caption{Criteria modification process.}
\label{fig:criteria_modification_process}
\centering
\begin{tabular}{lll}
\toprule
\# & \textbf{Description} & \\
\midrule
1 & Expand abbreviations and acronyms & \ref{subsec:expand_abbrev} \\
2 & Split compound criteria into multiple rules where possible & \ref{subsec:split_compound} \\
3 & Remove criteria not verifiable from title/abstract (e.g., publication year, venue) & \ref{subsec:remove_criteria} \\
4 & Convert negations to affirmative form & \ref{subsec:convert_negations} \\
5 & Preserve original wording --- minimize rewording & \ref{subsec:preserve_wording} \\
6 & If a rule contains vague terms, search the SR for their definition or description & \ref{subsec:vague} \\
\bottomrule
\end{tabular}
\end{table}

Our motivation behind modifying criteria in RQ4 was that improving poorly formatted, unclear, or ambiguous criteria may decrease the likelihood of an incorrect overall or criterion screening decision. As this part was somewhat experimental and as modifying criteria takes a lot of time, we focused on the secondary studies that LLMs had challenges making the correct decision for includes and excludes.

We began by analyzing the criterion-level agreements from Online resource 3~\citep{zenodo_research_artifact} and looking at Table~\ref{tab:rq11_per_study_metrics} to see, which secondary studies had most disagreement \textcircled{5}. We then selected seven secondary studies from Table~\ref{tab:rq11_per_study_metrics} \citep{rodriguez-perezPerceivedDiversitySoftware2021a,bjarnasonEmpiricallyBasedModel2023a,zhangTestingVerificationNeuralnetworkbased2020a,tebesAnalyzingDocumentingSystematic2020a,alonsoSystematicMappingStudy2023b,lenarduzziSystematicLiteratureReview2021a,deckersSystematicLiteratureReview2022a}. For this experiment, we selected all new LLMs from Table~\ref{tab:test_llms_by_rq} \textcircled{5}, as we wanted to make the results fully comparable with RQ2.

The criteria modification process \textcircled{6} was performed by the third author. The second author verified the third author's criteria modification suggestions. After the criteria modification process was finished, we then re-benchmarked the primary studies using eight LLMs (GPT-5.5, Claude Opus 4.7, Gemini 3.1 Pro, GLM 5.1, Kimi K2.6, Grok 4.20 Beta, Qwen 3.6 Plus, and DeepSeek V4 Pro) on the 5\% sampled data, filtered to the selected seven secondary studies (N=448), using the V2 user prompt and output schema. For RQ4, we computed accuracy, precision, recall, F1, MCC, Gwet's First-order Agreement Coefficient (Gwet's AC1), and lost evidence.

Next, we describe each item of the criteria modification process in detail, in Subsections \ref{subsec:expand_abbrev} to \ref{subsec:vague}. The following modifications were made to the criteria (Table \ref{fig:criteria_modification_process}):

\subsubsection{Expand abbreviations and acronyms}
\label{subsec:expand_abbrev}

Abbreviations and acronyms were expanded, as acronyms can be misleading, e.g. if the LLM fails to understand the context of the abbreviation or acronym. For instance, continuous integration (CI) and confidence interval (CI) have matching acronyms, but mean totally different things. This might lead the LLM to misclassify a paper that mentions continuous integration in the abstract, but the LLM interprets the acronym as confidence interval.

\subsubsection{Split compound criteria into multiple rules where possible}
\label{subsec:split_compound}

Long criteria that contain multiple different rules were split into separate, simpler rules so each can be evaluated clearly and independently by the LLM.

\subsubsection{Remove criteria not verifiable from title/abstract (e.g., publication year, venue)} 
\label{subsec:remove_criteria}

If the LLM cannot determine a criterion's outcome by the paper's title/abstract, it can lead to false negatives or false positives. Paper abstracts rarely contain information regarding e.g. publication year or venue.

\subsubsection{Convert negations to affirmative form}
\label{subsec:convert_negations}

We converted negations to affirmative form, so they describe the desired criteria directly. Affirmative rules are easier to interpret and we have witnessed LLMs tripping over negation, especially when present in exclusion rules \citep{mantyla2026understanding}. 

\subsubsection{Preserve original wording --- minimize rewording} 
\label{subsec:preserve_wording}

We kept the criteria as close as possible to their original wording to preserve the intent of the underlying SR. The rewording was limited to changes needed for consistency, clarity, and LLM interpretability, as larger changes to the criterion could alter the meaning of it.

\subsubsection{If a rule contains vague terms, search the SR for their definition or description}
\label{subsec:vague}

We searched the SR for definitions and explanations, or contextual descriptions of vague terms used in the criteria. When available, these definitions were incorporated into the criteria to make them more precise.

\subsection{Statistical methods used}
\label{sec:statistical_methods}

Next, we describe each statistical method we used and motivate their use. For all the metrics (excluding regression model), we used the confusion matrices as a single source of truth, and include them in our research artifact~\citep{zenodo_research_artifact} as recommended by \citet{madeyskiLLM4SCREENLITRecommendationsAssessing2026}. True positives (TP) indicate correctly included studies, while true negatives (TN) indicate correctly excluded studies. On the contrary, false negatives (FN) are studies incorrectly excluded and false positives (FP) are studies incorrectly included.

\subsubsection{Accuracy, Precision, Recall, F1}

In our experiments we reported accuracy, precision, recall, and F1 scores to make the screening results comparable with prior work, and the corresponding metrics are industry-standard metrics for reporting empirical software engineering research. Accuracy was used to measure the general correctness of the LLMs, and precision and recall were used to measure the LLM's correctness in correctly included studies, while F1 presents precision and recall as a harmonic mean~\citep{Manning_Raghavan_Schutze_2008}. Accuracy, precision, recall, and F1 are computed as follows:

\begin{equation}Accuracy = \frac{TP + TN}{n}\end{equation}

Where $n$ is the total number of screened studies (included or excluded).

\begin{equation}Precision = \frac{TP}{TP+FP}\end{equation}
\begin{equation}Recall = \frac{TP}{TP+FN}\end{equation}
\begin{equation}F1 = \frac{2 \times TP}{2 \times TP + FP + FN}\end{equation}

\subsubsection{MCC}

We reported the Matthews Correlation Coefficient (MCC), as it was recommended by \citet{madeyskiLLM4SCREENLITRecommendationsAssessing2026} for ranking, and because the SESR-Eval dataset~\citep{huotalaSESREvalDatasetEvaluating2025} contains secondary studies that sample far more excluded than included papers. The SESR-Eval dataset by \citet{huotalaSESREvalDatasetEvaluating2025} used in our experiments contains 34,528 papers, from which 4,197 were labeled as included, and 30,331 as excluded. This gives an I:E ratio of 12:88, which makes the excluded papers a highly prevalent category. The same situation holds for the 5\% sampled data, which also had a similar level of class imbalance (21:79). MCC uses the whole confusion matrix and behaves adequately with class imbalance, which the F1 score does not. MCC is computed as follows:

\begin{equation}MCC = \frac{TP \times TN - FP \times FN}{\sqrt{(TP + FP)(TP + FN)(TN + FP)(TN + FN)}}\end{equation}

\subsubsection{Lost evidence}

Lost evidence (FN) is the number and percentage of relevant studies that the LLM excludes incorrectly (false negatives). In the context of SRs, missing relevant papers is more costly than including irrelevant ones, as the irrelevant papers are usually caught in the full-text screening phase~\citep{kitchenhamGuidelinesPerformingSystematic2007}. Reporting lost evidence is useful because two LLMs can have a similar F1 score, but the lost evidence of one LLM can be significantly higher than the other. The confusion matrices of each experiment are included in our research artifact~\citep{zenodo_research_artifact}.

\subsubsection{LLM Agreement - Gwet's AC1}
Gwet's AC1 \citep{gwet2008computing} was used instead of Cohen's kappa as our agreement metric. Cohen's kappa is well known to suffer from the "prevalence problem"\footnote{Also known as the prevalence paradox and prevalence effect} making it unsuited for datasets with high category imbalance. In our data, we have secondary studies where nearly all decisions are includes or excludes. Should there be any mistakes in LLMs' screening decisions in those studies, Cohen's kappa would collapse close to zero as it overestimates chance agreement from the base rate. As our data has a prevalence problem, Gwet's AC1 becomes the most suitable metric for it. Gwet's AC1 does not artificially deflate the agreement estimate simply because one category was highly prevalent~\citep{wongpakaranComparisonCohensKappa2013}. Gwet's AC1 is computed as follows:

\begin{equation}AC1 = \frac{p_a - p_e}{1 - p_e}\end{equation}

where $p_a$ is the percentage of observed agreement and $p_e$ is the probability of chance agreement.

\subsubsection{Bootstrap confidence intervals and regression model for statistical significance}
Our main metrics, F1 and MCC, come from machine learning and are not typically accompanied by statistical tests. Because these are composite metrics, traditional statistical tests like the t-test are not applicable. Therefore, to test the statistical significance of our results, we reported the 95\% percentile bootstrap confidence intervals \citep{Efron_Tibshirani_1994} for precision, recall, F1, MCC, lost evidence, and Gwet's AC1. We used $B = 2000$ samples. Because the secondary studies in the SESR-Eval dataset differ significantly in the size and I:E ratio, we resampled papers with replacement within each secondary study, so that every resample preserved the composition of the dataset. The 2.5th and 97.5th percentiles of each metric's bootstrap distribution formed the 95\% confidence interval. For differences between conditions evaluated on the same primary studies, we computed the difference on each resample. This yielded paired confidence intervals, whose difference we considered meaningful when its confidence interval excluded 0.

To compare the screening performance between all LLMs and secondary studies, we chose models benchmarked for RQ1.2 (see Table~\ref{tab:test_llms_by_rq}). Then, we fit a logistic regression model that predicted the correctness of LLMs' screening decisions based on the LLM and the primary study. The logistic regression model was fitted and the odds ratios were calculated using the R programming language and the methods \emph{glm} and \emph{coef}. We chose the model and secondary study with highest overall probability of correct prediction for the reference classes.

\subsection{AI disclosure}

Claude Code was used to support the development of scripts that we used to run the LLMs and to generate and format tables. All outputs generated by the tool were independently reviewed and validated by the authors, who retain full responsibility for the final scripts, tables, and other associated outputs.

\section{Results}

\label{sec:results}

\subsection{RQ1.1: Paper screening on full dataset}

\begin{sidewaystable*}[htbp]
  \centering
  \caption{Screening performance metrics of LLMs across primary studies (RQ1.1). The models above the double line are from \citet{huotalaSESREvalDatasetEvaluating2025}.}
  \label{tab:rq11_model_metrics_ci_primary}
  \begin{adjustbox}{max width=\linewidth, max totalheight=0.34\textwidth}
  \begin{tabular}{lrrrrrr}
    \toprule
    \textbf{Model} & \textbf{Accuracy} & \textbf{Precision} & \textbf{Recall} & \textbf{F1} & \textbf{MCC} & \textbf{Lost Evidence} \\
    \midrule
    \textbf{o3-mini} & 0.86 [0.86, 0.86] & 0.43 [0.42, 0.45] & 0.52 [0.51, 0.53] & 0.47 [0.46, 0.48] & 0.39 [0.38, 0.41] & 2018 (48\%) [1944, 2095] \\
    \textbf{GPT-4o} & 0.84 [0.83, 0.84] & 0.40 [0.39, 0.41] & 0.66 [0.64, 0.67] & 0.50 [0.48, 0.51] & 0.42 [0.41, 0.44] & 1443 (34\%) [1378, 1506] \\
    \textbf{GPT-4.1} & 0.83 [0.83, 0.83] & 0.38 [0.37, 0.39] & 0.63 [0.61, 0.64] & 0.47 [0.46, 0.48] & 0.39 [0.38, 0.41] & 1565 (37\%) [1499, 1633] \\
    \textbf{GPT-4.1 mini} & 0.90 [0.89, 0.90] & 0.60 [0.59, 0.62] & 0.43 [0.42, 0.45] & 0.50 [0.49, 0.52] & 0.46 [0.44, 0.47] & 2375 (57\%) [2294, 2454] \\
    \textbf{Claude 3.7 Sonnet} & 0.89 [0.89, 0.89] & 0.56 [0.54, 0.58] & 0.46 [0.45, 0.48] & 0.51 [0.50, 0.52] & 0.45 [0.44, 0.46] & 2249 (54\%) [2171, 2328] \\
    \textbf{DeepSeek R1} & 0.89 [0.89, 0.89] & 0.56 [0.55, 0.58] & 0.42 [0.41, 0.43] & 0.48 [0.47, 0.49] & 0.43 [0.41, 0.44] & 2433 (58\%) [2348, 2517] \\
    \textbf{Llama 4 Maverick} & 0.87 [0.86, 0.87] & 0.47 [0.45, 0.48] & 0.61 [0.59, 0.62] & 0.53 [0.51, 0.54] & 0.46 [0.44, 0.47] & 1655 (39\%) [1587, 1721] \\
    \midrule
    \textbf{Average} & 0.867 & 0.486 & 0.532 & 0.494 & 0.429 & 1962.6 (47\%) \\
    \addlinespace
    \addlinespace
    \hline
    \hline
    \addlinespace
    \addlinespace
    \textbf{DeepSeek V4 Pro} & 0.87 [0.87, 0.88] & 0.48 [0.47, 0.50] & 0.47 [0.46, 0.49] & 0.48 [0.46, 0.49] & 0.41 [0.39, 0.42] & 2214 (53\%) [2135, 2290] \\
    \textbf{Qwen 3.6 Plus} & 0.86 [0.86, 0.87] & 0.45 [0.44, 0.46] & 0.52 [0.50, 0.53] & 0.48 [0.47, 0.49] & 0.40 [0.39, 0.42] & 2034 (48\%) [1959, 2110] \\
    \addlinespace
    \midrule
    \textbf{Average} & 0.869 & 0.466 & 0.494 & 0.479 & 0.405 & 2124.0 (51\%) \\
    \bottomrule
  \end{tabular}
  \end{adjustbox}
  \bigskip
  \caption{Screening performance metrics of LLMs across secondary studies (RQ1.1). The models above the double line are from \citet{huotalaSESREvalDatasetEvaluating2025}.}
  \label{tab:rq11_model_metrics_ci_secondary}
  \begin{adjustbox}{max width=\linewidth, max totalheight=0.48\textwidth}
  \begin{tabular}{lrrrrrrrr}
    \toprule
    \textbf{Model} & \textbf{Accuracy} & \textbf{Precision} & \textbf{Recall} & \textbf{F1} & \textbf{MCC} & \textbf{Lost Evidence} & \makecell{\textbf{P95 time} \\ \textbf{per paper (s)}} & \makecell{\textbf{Cost per} \\ \textbf{study (USD)}} \\
    \midrule
    \textbf{o3-mini} & 0.73 [0.72, 0.74] & 0.60 [0.58, 0.62] & 0.49 [0.48, 0.51] & \makecell[r]{\rule{0pt}{2.6ex}0.46 [0.44, 0.48]\rlap{$^1$} \\ 0.54 [0.52, 0.56]\rlap{$^2$}\rule[-1.2ex]{0pt}{0pt}} & 0.30 [0.29, 0.32] & 84.1 (51\%) [81.0, 87.3] & 86.1 & 36.7\rlap{$^\dagger$} \\
    \hline
    \textbf{GPT-4o} & 0.73 [0.72, 0.74] & 0.53 [0.51, 0.54] & 0.66 [0.64, 0.67] & \makecell[r]{\rule{0pt}{2.6ex}0.51 [0.49, 0.52]\rlap{$^1$} \\ 0.59 [0.57, 0.60]\rlap{$^2$}\rule[-1.2ex]{0pt}{0pt}} & 0.31 [0.29, 0.32] & 60.1 (34\%) [57.4, 62.8] & 14.1 & 13.4\rlap{$^\dagger$} \\
    \hline
    \textbf{GPT-4.1} & 0.73 [0.72, 0.74] & 0.53 [0.51, 0.54] & 0.60 [0.58, 0.62] & \makecell[r]{\rule{0pt}{2.6ex}0.48 [0.47, 0.50]\rlap{$^1$} \\ 0.56 [0.55, 0.57]\rlap{$^2$}\rule[-1.2ex]{0pt}{0pt}} & 0.29 [0.27, 0.31] & 65.2 (40\%) [62.5, 68.0] & 24.7 & 12.1\rlap{$^\dagger$} \\
    \hline
    \textbf{GPT-4.1 mini} & 0.73 [0.72, 0.74] & 0.62 [0.59, 0.64] & 0.38 [0.36, 0.40] & \makecell[r]{\rule{0pt}{2.6ex}0.41 [0.39, 0.42]\rlap{$^1$} \\ 0.47 [0.45, 0.49]\rlap{$^2$}\rule[-1.2ex]{0pt}{0pt}} & 0.28 [0.26, 0.30] & 99.0 (62\%) [95.6, 102.3] & 10.5 & 2.2\rlap{$^\dagger$} \\
    \hline
    \textbf{Claude 3.7 Sonnet} & 0.73 [0.72, 0.74] & 0.58 [0.56, 0.60] & 0.45 [0.44, 0.47] & \makecell[r]{\rule{0pt}{2.6ex}0.44 [0.43, 0.46]\rlap{$^1$} \\ 0.51 [0.49, 0.52]\rlap{$^2$}\rule[-1.2ex]{0pt}{0pt}} & 0.28 [0.26, 0.30] & 93.7 (55\%) [90.5, 97.0] & 33.7 & 24.0\rlap{$^\dagger$} \\
    \hline
    \textbf{DeepSeek R1} & 0.72 [0.71, 0.73] & 0.58 [0.55, 0.60] & 0.41 [0.39, 0.43] & \makecell[r]{\rule{0pt}{2.6ex}0.41 [0.39, 0.43]\rlap{$^1$} \\ 0.48 [0.46, 0.50]\rlap{$^2$}\rule[-1.2ex]{0pt}{0pt}} & 0.27 [0.25, 0.29] & 101.4 (59\%) [97.8, 104.9] & 42.8 & 3.8\rlap{$^\dagger$} \\
    \hline
    \textbf{Llama 4 Maverick} & 0.74 [0.73, 0.75] & 0.54 [0.52, 0.55] & 0.61 [0.59, 0.62] & \makecell[r]{\rule{0pt}{2.6ex}0.50 [0.48, 0.51]\rlap{$^1$} \\ 0.57 [0.56, 0.58]\rlap{$^2$}\rule[-1.2ex]{0pt}{0pt}} & 0.31 [0.29, 0.33] & 69.0 (39\%) [66.1, 71.7] & 23.0 & 1.4\rlap{$^\dagger$} \\
    \addlinespace
    \midrule
    \textbf{Average} & 0.731 & 0.566 & 0.515 & \makecell[r]{\rule{0pt}{2.6ex}0.459\rlap{$^1$} \\ 0.531\rlap{$^2$}\rule[-1.2ex]{0pt}{0pt}} & 0.294 & 81.8 (49\%) & & \\
    \hline
    \hline
    \addlinespace
    \textbf{DeepSeek V4 Pro} & 0.73 [0.73, 0.74] & 0.62 [0.60, 0.64] & 0.47 [0.45, 0.48] & \makecell[r]{\rule{0pt}{2.6ex}0.45 [0.43, 0.46]\rlap{$^1$} \\ 0.53 [0.52, 0.55]\rlap{$^2$}\rule[-1.2ex]{0pt}{0pt}} & 0.31 [0.29, 0.33] & 92.3 (53\%) [89.0, 95.4] & 103.0 & 12.1 \\
    \hline
    \textbf{Qwen 3.6 Plus} & 0.73 [0.72, 0.74] & 0.59 [0.57, 0.62] & 0.51 [0.49, 0.52] & \makecell[r]{\rule{0pt}{2.6ex}0.46 [0.45, 0.48]\rlap{$^1$} \\ 0.55 [0.53, 0.56]\rlap{$^2$}\rule[-1.2ex]{0pt}{0pt}} & 0.31 [0.29, 0.33] & 84.8 (49\%) [81.6, 87.9] & 94.7 & 11.2 \\
    \addlinespace
    \midrule
    \textbf{Average} & 0.733 & 0.608 & 0.486 & \makecell[r]{\rule{0pt}{2.6ex}0.456\rlap{$^1$} \\ 0.539\rlap{$^2$}\rule[-1.2ex]{0pt}{0pt}} & 0.309 & 88.5 (51\%) & & \\
    \bottomrule
  \end{tabular}
  \end{adjustbox}
  \par\footnotesize $^1$ Mean of the per-secondary-study F1 scores. $^2$ F1 computed from the averaged precision and recall. $^\dagger$Old models: costs previously recorded -- only an estimate.\par
\end{sidewaystable*}

\begin{table*}[htbp]
	\centering
	\caption{Accuracy, precision, recall, F1 score, MCC, and lost evidence for the secondary studies across all LLMs (RQ1.1, full dataset). Lost evidence is the share of relevant papers missed (false negatives), pooled across LLMs. MCC is N/A where it is undefined (all papers, or all predictions, belong to one class); the average MCC is taken over the remaining studies.}
	\label{tab:rq11_per_study_metrics}
	\resizebox{\linewidth}{!}{%
		\begin{tabular}{lcccccc}
			\toprule
			\textbf{Study}                                           & \textbf{Accuracy} & \textbf{Precision} & \textbf{Recall} & \textbf{F1} & \textbf{MCC} & \textbf{\makecell{Lost \\ evidence}} \\
			\midrule
			\citet{casalaroModeldrivenEngineeringMobile2022a}         & 0.93              & 0.27               & 0.74            & 0.39        & 0.42         & 26\%                   \\
			\citet{sharbafConflictManagementTechniques2023a}          & 0.69              & 0.73               & 0.68            & 0.70        & 0.39         & 32\%                   \\
			\citet{tambonHowCertifyMachine2022a}                      & 0.87              & 0.74               & 0.46            & 0.57        & 0.52         & 54\%                   \\
			\citet{rodriguez-perezPerceivedDiversitySoftware2021a}    & 0.91              & 0.41               & 0.18            & 0.25        & 0.23         & 82\%                   \\
			\citet{bjarnasonEmpiricallyBasedModel2023a}               & 0.86              & 0.05               & 0.64            & 0.10        & 0.16         & 36\%                   \\
			\citet{kuutilaTimePressureSoftware2020b}                  & 0.94              & 0.39               & 0.40            & 0.39        & 0.36         & 60\%                   \\
			\citet{geraldiSoftwareProductLine2020a}                   & 0.65              & 0.47               & 0.46            & 0.47        & 0.20         & 54\%                   \\
			\citet{zhangTestingVerificationNeuralnetworkbased2020a}   & 0.26              & 1.00               & 0.26            & 0.41        & N/A          & 74\%                   \\
			\citet{tebesAnalyzingDocumentingSystematic2020a}          & 0.36              & 0.80               & 0.16            & 0.27        & 0.07         & 84\%                   \\
			\citet{ebrahimiMobileAppPrivacy2021a}                     & 0.62              & 0.59               & 0.78            & 0.67        & 0.25         & 22\%                   \\
			\citet{lewowskiHowFarAre2022a}                            & 0.91              & 0.56               & 0.64            & 0.60        & 0.55         & 36\%                   \\
			\citet{vandinterPredictiveMaintenanceUsing2022a}          & 0.70              & 0.45               & 0.72            & 0.55        & 0.37         & 28\%                   \\
			\citet{stradowskiMachineLearningSoftware2023a}            & 0.83              & 0.89               & 0.82            & 0.85        & 0.64         & 18\%                   \\
			\citet{alonsoSystematicMappingStudy2023b}                 & 0.23              & 0.77               & 0.15            & 0.25        & -0.11        & 85\%                   \\
			\citet{somersDigitaltwinbasedTestingCyber2023b}           & 0.71              & 0.59               & 0.45            & 0.51        & 0.32         & 55\%                   \\
			\citet{teixeiraModelingAutomaticCode2017a}                & 0.52              & 0.83               & 0.24            & 0.37        & 0.22         & 76\%                   \\
			\citet{lenarduzziSystematicLiteratureReview2021a}         & 0.73              & 0.17               & 0.24            & 0.20        & 0.04         & 76\%                   \\
			\citet{pereiraLearningSoftwareConfiguration2021a}         & 0.79              & 1.00               & 0.79            & 0.88        & N/A          & 21\%                   \\
			\citet{barisicMultiparadigmModelingCyber2022a}            & 0.67              & 0.78               & 0.41            & 0.53        & 0.35         & 59\%                   \\
			\citet{deckersSystematicLiteratureReview2022a}            & 0.78              & 0.19               & 0.81            & 0.31        & 0.32         & 19\%                   \\
			\citet{raniDecadeCodeComment2023a}                        & 0.96              & 0.32               & 0.27            & 0.29        & 0.27         & 73\%                   \\
			\citet{zakeri-nasrabadiSystematicLiteratureReview2023b}   & 0.89              & 0.29               & 0.68            & 0.41        & 0.40         & 32\%                   \\
			\citet{shevtsovControlTheoreticalSoftwareAdaptation2018a} & 0.85              & 0.36               & 0.55            & 0.44        & 0.36         & 45\%                   \\
			\citet{linOpinionMiningSoftware2022a}                     & 0.89              & 0.63               & 0.70            & 0.67        & 0.60         & 30\%                   \\
			\midrule
			\textbf{Average}                                         & 0.73              & 0.55               & 0.51            & 0.46        & 0.32         & 49\%                   \\
			\bottomrule
		\end{tabular}
	}%
\end{table*}

We started our research by rerunning the full dataset of our prior SESR-Eval paper~\citep{huotalaSESREvalDatasetEvaluating2025} with the models we selected, starting from the cheapest: DeepSeek V4 Pro and Qwen 3.6 Plus, see Table \ref{tab:companies_and_models}. We did not run the other models on the full dataset, as we estimated it would have cost \$6,500; instead, we opted to run them only on the sampled data, see next section.

Table~\ref{tab:rq11_model_metrics_ci_primary} and Table~\ref{tab:rq11_model_metrics_ci_secondary} show the accuracy, precision, recall, F1, MCC, and lost evidence scores of the tested models for primary studies and secondary studies, respectively. We report results for both primary and secondary studies for the same reasons as for our prior work~\citep{huotalaSESREvalDatasetEvaluating2025}, as some secondary studies contain a large number of papers and therefore dominate the primary study scores.

\textbf{Screening performance across primary studies.} Across primary studies (Table~\ref{tab:rq11_model_metrics_ci_primary}) we see that in terms of MCC, the two new LLMs (DeepSeek V4 Pro and Qwen 3.6 Plus) do not exceed the best-performing old LLMs (GPT-4.1 mini and Llama 4 Maverick, MCC=0.46) that were first benchmarked 18 months prior. The same is observed for F1 score, where DeepSeek V4 Pro and Qwen 3.6 Plus have exactly the same F1=0.48, while the old LLM with the best F1 score was Llama 4 Maverick (F1=0.53). In terms of lost evidence for primary studies, both DeepSeek V4 Pro (53\%) and Qwen 3.6 Plus (48\%) lose less evidence than GPT-4.1 mini (57\%), Claude 3.7 Sonnet (54\%) and DeepSeek R1 (58\%), but do not reach the lost evidence levels of GPT-4o (34\%) or GPT-4.1 (37\%).  Accuracy, precision, and recall for the two new LLMs are at about the same level as for the seven old LLMs. On average, the old LLMs perform marginally better than the new LLMs (old MCC = 0.429 vs. new MCC = 0.405).

\textbf{Screening performance across secondary studies.} In Table~\ref{tab:rq11_model_metrics_ci_secondary}, we can see that there is a four-way tie as the best-performing LLM in paper screening over the full dataset in terms of MCC -- the two new LLMs (DeepSeek V4 Pro and Qwen 3.6 Plus) have the same MCC as GPT-4o and Llama 4 Maverick (MCC=0.31). If we look at screening performance between secondary studies (Table~\ref{tab:rq11_per_study_metrics}), the F1 score ranges between 0.10 and 0.88, MCC between -0.11 and 0.64 averaged over all LLMs. Lost evidence varies significantly per-study -- the average is 49\% while it ranges from 18\% to 85\%. For studies \citet{zhangTestingVerificationNeuralnetworkbased2020a} and \citet{pereiraLearningSoftwareConfiguration2021a}, MCC was not defined as they did not contain excluded studies at all (Table~\ref{tab:sample_sizes}). Next, if we look at the more detailed two-dimensional table (Online resource 1) \citep{zenodo_research_artifact}, the performance of DeepSeek V4 Pro and Qwen 3.6 Plus are fairly similar compared to the rest of the LLMs, where F1 for DeepSeek V4 Pro ranges from 0.08 to 0.86, and 0.07 to 0.86 for Qwen 3.6 Plus, respectively. Meanwhile, MCC for DeepSeek V4 Pro ranges from -0.27 to 0.70. For Qwen 3.6 Plus, the MCC ranges between -0.26 and 0.65. In terms of lost evidence for secondary studies (Table~\ref{tab:rq11_model_metrics_ci_secondary}), both DeepSeek V4 Pro and Qwen 3.6 Plus lose 53\% and 49\% evidence on average, respectively. Compared to older LLMs, this is at almost the same level as o3-mini (51\%), and Claude 3.7 Sonnet (55\%). The new LLMs are not even close to GPT-4o, which loses only 34\% of evidence on average. When looking at the average across secondary studies, the new LLMs perform marginally better than the old LLMs (old MCC = 0.294 vs. new MCC = 0.309).

Looking at the cost per secondary study and the time taken to screen one paper (95th percentile), we see that both of the new models (DeepSeek V4 Pro = 103.0s, Qwen 3.6 Plus = 94.7s) are the two slowest models around. The previous slowest model was o3-mini with a screening speed of 86.1s per paper. The fastest LLM to screen one paper is GPT-4.1 mini (10.5s), which is almost ten times faster. In addition, the two models are about as expensive as GPT-4o and GPT-4.1, where DeepSeek V4 Pro costs 12.1 dollars per secondary study, and Qwen 3.6 Plus cost 11.2 dollars per secondary study on average.

\rqbox{
RQ1.1 results show that the two new models (DeepSeek V4 Pro and Qwen 3.6 Plus) tested on full datasets are not clearly more accurate than the older ones. Average MCC is marginally higher across secondary studies (0.309 vs 0.294) but lower across primary studies (0.405 vs 0.429). Differences between secondary studies are still significant. 
}

\subsection{RQ1.2: Paper screening on sampled dataset}

\begin{sidewaystable*}[htbp]
  \centering
  \caption{Screening performance metrics of LLMs across primary studies (RQ1.2). The models above the double line are from \citet{huotalaSESREvalDatasetEvaluating2025}.}
  \label{tab:rq12_model_metrics_ci_primary}
  \begin{adjustbox}{width=\linewidth}
  \begin{tabular}{lrrrrrr}
    \toprule
    \textbf{Model} & \textbf{Accuracy} & \textbf{Precision} & \textbf{Recall} & \textbf{F1} & \textbf{MCC} & \textbf{Lost Evidence} \\
    \midrule
    \textbf{o3-mini} & 0.83 [0.81, 0.84] & 0.60 [0.55, 0.66] & 0.50 [0.45, 0.55] & 0.55 [0.50, 0.59] & 0.44 [0.39, 0.49] & 179 (50\%) [159, 200] \\
    \textbf{GPT-4o} & 0.80 [0.78, 0.81] & 0.51 [0.46, 0.55] & 0.64 [0.60, 0.69] & 0.57 [0.53, 0.60] & 0.44 [0.39, 0.49] & 128 (36\%) [111, 145] \\
    \textbf{GPT-4.1} & 0.80 [0.78, 0.82] & 0.51 [0.47, 0.56] & 0.61 [0.56, 0.65] & 0.55 [0.51, 0.59] & 0.43 [0.38, 0.48] & 141 (39\%) [123, 159] \\
    \textbf{GPT-4.1 mini} & 0.84 [0.83, 0.85] & 0.69 [0.63, 0.75] & 0.41 [0.37, 0.46] & 0.52 [0.47, 0.56] & 0.45 [0.40, 0.50] & 210 (59\%) [189, 231] \\
    \textbf{Claude 3.7 Sonnet} & 0.82 [0.80, 0.84] & 0.59 [0.54, 0.65] & 0.42 [0.38, 0.47] & 0.49 [0.45, 0.54] & 0.40 [0.35, 0.45] & 207 (58\%) [186, 228] \\
    \textbf{DeepSeek R1} & 0.83 [0.81, 0.84] & 0.61 [0.55, 0.67] & 0.44 [0.40, 0.49] & 0.51 [0.47, 0.56] & 0.42 [0.36, 0.47] & 200 (56\%) [179, 222] \\
    \textbf{Llama 4 Maverick} & 0.82 [0.80, 0.83] & 0.56 [0.51, 0.61] & 0.58 [0.53, 0.62] & 0.57 [0.53, 0.61] & 0.45 [0.40, 0.50] & 152 (42\%) [134, 171] \\
    \hline
    \addlinespace
    \textbf{Average} & 0.818 & 0.583 & 0.514 & 0.537 & 0.433 & 173.9 (49\%) \\
    \addlinespace
    \addlinespace
    \hline
    \hline
    \addlinespace
    \addlinespace
    \addlinespace
    \textbf{Claude Opus 4.7} & 0.85 [0.83, 0.86] & 0.75 [0.69, 0.81] & 0.38 [0.34, 0.43] & 0.51 [0.46, 0.55] & 0.46 [0.41, 0.51] & 221 (62\%) [201, 242] \\
    \textbf{DeepSeek V4 Pro} & 0.83 [0.81, 0.84] & 0.61 [0.56, 0.67] & 0.44 [0.40, 0.49] & 0.51 [0.47, 0.55] & 0.42 [0.37, 0.47] & 200 (56\%) [180, 220] \\
    \textbf{Gemini 3.1 Pro} & 0.83 [0.82, 0.85] & 0.67 [0.61, 0.73] & 0.39 [0.35, 0.44] & 0.50 [0.45, 0.54] & 0.43 [0.38, 0.48] & 217 (61\%) [196, 239] \\
    \textbf{GLM 5.1} & 0.83 [0.82, 0.85] & 0.71 [0.65, 0.77] & 0.33 [0.29, 0.38] & 0.45 [0.41, 0.50] & 0.41 [0.36, 0.46] & 239 (67\%) [216, 261] \\
    \textbf{GPT-5.5} & 0.82 [0.80, 0.84] & 0.60 [0.54, 0.66] & 0.41 [0.36, 0.45] & 0.48 [0.44, 0.53] & 0.39 [0.34, 0.44] & 213 (59\%) [192, 234] \\
    \textbf{Grok 4.20 Beta} & 0.82 [0.80, 0.83] & 0.58 [0.53, 0.63] & 0.40 [0.36, 0.45] & 0.48 [0.43, 0.52] & 0.38 [0.33, 0.43] & 214 (60\%) [193, 235] \\
    \textbf{Kimi K2.6} & 0.83 [0.81, 0.84] & 0.62 [0.56, 0.67] & 0.42 [0.37, 0.46] & 0.50 [0.45, 0.54] & 0.41 [0.36, 0.46] & 209 (58\%) [188, 230] \\
    \textbf{Qwen 3.6 Plus} & 0.82 [0.80, 0.83] & 0.57 [0.51, 0.62] & 0.48 [0.44, 0.53] & 0.52 [0.48, 0.56] & 0.41 [0.36, 0.46] & 185 (52\%) [165, 206] \\
    \hline
    \addlinespace
    \textbf{Average} & 0.827 & 0.638 & 0.407 & 0.493 & 0.412 & 212.3 (59\%) \\
    \bottomrule
  \end{tabular}
  \end{adjustbox}
  \par\footnotesize Lost Evidence CI is for the false-negative count.\par
\end{sidewaystable*}

\begin{sidewaystable*}[htbp]
  \centering
  \caption{Screening performance metrics of LLMs across secondary studies (RQ1.2).}
  \label{tab:rq12_model_metrics_ci_secondary}
  \begin{adjustbox}{width=\linewidth}
  \begin{tabular}{lrrrrrrrr}
    \toprule
    \textbf{Model} & \textbf{Accuracy} & \textbf{Precision} & \textbf{Recall} & \textbf{F1} & \textbf{MCC} & \textbf{Lost Evidence} & \makecell{\textbf{P95 time} \\ \textbf{per paper (s)}} & \makecell{\textbf{Cost per} \\ \textbf{sample (USD)}} \\
    \midrule
    \textbf{o3-mini} & 0.75 [0.73, 0.78] & 0.68 [0.59, 0.73] & 0.57 [0.46, 0.61] & \makecell[r]{\rule{0pt}{2.6ex}0.53 [0.44, 0.57]\rlap{$^1$} \\ 0.62 [0.52, 0.65]\rlap{$^2$}\rule[-1.2ex]{0pt}{0pt}} & 0.40 [0.31, 0.44] & 7.5 (43\%) [6.6, 8.3] & 86.1 & 1.82\rlap{$^\dagger$} \\
    \hline
    \textbf{GPT-4o} & 0.72 [0.70, 0.75] & 0.53 [0.48, 0.58] & 0.74 [0.63, 0.77] & \makecell[r]{\rule{0pt}{2.6ex}0.53 [0.47, 0.57]\rlap{$^1$} \\ 0.62 [0.55, 0.65]\rlap{$^2$}\rule[-1.2ex]{0pt}{0pt}} & 0.35 [0.29, 0.40] & 5.3 (26\%) [4.6, 6.0] & 14.1 & 0.67\rlap{$^\dagger$} \\
    \hline
    \textbf{GPT-4.1} & 0.74 [0.71, 0.76] & 0.58 [0.50, 0.62] & 0.69 [0.57, 0.73] & \makecell[r]{\rule{0pt}{2.6ex}0.53 [0.45, 0.57]\rlap{$^1$} \\ 0.63 [0.54, 0.66]\rlap{$^2$}\rule[-1.2ex]{0pt}{0pt}} & 0.36 [0.28, 0.40] & 5.9 (31\%) [5.1, 6.6] & 24.7 & 0.60\rlap{$^\dagger$} \\
    \hline
    \textbf{GPT-4.1 mini} & 0.75 [0.72, 0.77] & 0.70 [0.56, 0.74] & 0.44 [0.35, 0.49] & \makecell[r]{\rule{0pt}{2.6ex}0.45 [0.37, 0.49]\rlap{$^1$} \\ 0.54 [0.44, 0.58]\rlap{$^2$}\rule[-1.2ex]{0pt}{0pt}} & 0.35 [0.27, 0.40] & 8.8 (56\%) [7.9, 9.6] & 10.5 & 0.11\rlap{$^\dagger$} \\
    \hline
    \textbf{Claude 3.7 Sonnet} & 0.73 [0.70, 0.75] & 0.53 [0.45, 0.61] & 0.45 [0.36, 0.49] & \makecell[r]{\rule{0pt}{2.6ex}0.42 [0.35, 0.46]\rlap{$^1$} \\ 0.49 [0.41, 0.53]\rlap{$^2$}\rule[-1.2ex]{0pt}{0pt}} & 0.27 [0.20, 0.32] & 8.6 (55\%) [7.8, 9.5] & 33.7 & 1.20\rlap{$^\dagger$} \\
    \hline
    \textbf{DeepSeek R1} & 0.73 [0.71, 0.76] & 0.59 [0.49, 0.64] & 0.53 [0.41, 0.56] & \makecell[r]{\rule{0pt}{2.6ex}0.47 [0.38, 0.51]\rlap{$^1$} \\ 0.56 [0.46, 0.59]\rlap{$^2$}\rule[-1.2ex]{0pt}{0pt}} & 0.34 [0.25, 0.38] & 8.3 (47\%) [7.5, 9.3] & 42.8 & 0.19\rlap{$^\dagger$} \\
    \hline
    \textbf{Llama 4 Maverick} & 0.74 [0.72, 0.77] & 0.56 [0.51, 0.60] & 0.67 [0.55, 0.70] & \makecell[r]{\rule{0pt}{2.6ex}0.51 [0.46, 0.55]\rlap{$^1$} \\ 0.61 [0.54, 0.64]\rlap{$^2$}\rule[-1.2ex]{0pt}{0pt}} & 0.35 [0.28, 0.40] & 6.3 (33\%) [5.6, 7.1] & 23.0 & 0.07\rlap{$^\dagger$} \\
    \hline
    \addlinespace
    \textbf{Average} & 0.738 & 0.596 & 0.583 & \makecell[r]{\rule{0pt}{2.6ex}0.493\rlap{$^1$} \\ 0.580\rlap{$^2$}\rule[-1.2ex]{0pt}{0pt}} & 0.347 & 7.2 (42\%) & & \\
    \addlinespace
    \addlinespace
    \hline
    \hline
    \addlinespace
    \addlinespace
    \textbf{Claude Opus 4.7} & 0.74 [0.72, 0.77] & 0.79 [0.61, 0.81] & 0.47 [0.35, 0.52] & \makecell[r]{\rule{0pt}{2.6ex}0.49 [0.39, 0.53]\rlap{$^1$} \\ 0.59 [0.46, 0.62]\rlap{$^2$}\rule[-1.2ex]{0pt}{0pt}} & 0.42 [0.32, 0.46] & 9.2 (53\%) [8.4, 10.1] & 15.6 & 2.75 \\
    \hline
    \textbf{DeepSeek V4 Pro} & 0.74 [0.72, 0.76] & 0.66 [0.55, 0.70] & 0.54 [0.42, 0.58] & \makecell[r]{\rule{0pt}{2.6ex}0.50 [0.42, 0.54]\rlap{$^1$} \\ 0.59 [0.49, 0.62]\rlap{$^2$}\rule[-1.2ex]{0pt}{0pt}} & 0.38 [0.29, 0.42] & 8.3 (46\%) [7.5, 9.2] & 103.0 & 0.60 \\
    \hline
    \textbf{Gemini 3.1 Pro} & 0.74 [0.72, 0.76] & 0.65 [0.52, 0.68] & 0.44 [0.33, 0.48] & \makecell[r]{\rule{0pt}{2.6ex}0.45 [0.37, 0.49]\rlap{$^1$} \\ 0.52 [0.41, 0.55]\rlap{$^2$}\rule[-1.2ex]{0pt}{0pt}} & 0.35 [0.26, 0.39] & 9.0 (56\%) [8.2, 10.0] & 14.3 & 0.93 \\
    \hline
    \textbf{GLM 5.1} & 0.73 [0.71, 0.76] & 0.68 [0.51, 0.71] & 0.40 [0.29, 0.44] & \makecell[r]{\rule{0pt}{2.6ex}0.41 [0.32, 0.45]\rlap{$^1$} \\ 0.50 [0.38, 0.54]\rlap{$^2$}\rule[-1.2ex]{0pt}{0pt}} & 0.34 [0.24, 0.38] & 10.0 (60\%) [9.0, 10.9] & 51.2 & 0.50 \\
    \hline
    \textbf{GPT-5.5} & 0.73 [0.71, 0.75] & 0.65 [0.53, 0.70] & 0.47 [0.36, 0.52] & \makecell[r]{\rule{0pt}{2.6ex}0.46 [0.37, 0.49]\rlap{$^1$} \\ 0.54 [0.43, 0.58]\rlap{$^2$}\rule[-1.2ex]{0pt}{0pt}} & 0.35 [0.26, 0.40] & 8.9 (53\%) [8.0, 9.8] & 17.6 & 1.83 \\
    \hline
    \textbf{Grok 4.20 Beta} & 0.73 [0.71, 0.75] & 0.69 [0.58, 0.74] & 0.47 [0.35, 0.51] & \makecell[r]{\rule{0pt}{2.6ex}0.47 [0.38, 0.51]\rlap{$^1$} \\ 0.56 [0.45, 0.59]\rlap{$^2$}\rule[-1.2ex]{0pt}{0pt}} & 0.36 [0.26, 0.40] & 8.9 (53\%) [8.0, 9.8] & 53.0 & 0.61 \\
    \hline
    \textbf{Kimi K2.6} & 0.74 [0.72, 0.76] & 0.63 [0.52, 0.67] & 0.50 [0.38, 0.54] & \makecell[r]{\rule{0pt}{2.6ex}0.47 [0.38, 0.50]\rlap{$^1$} \\ 0.56 [0.45, 0.59]\rlap{$^2$}\rule[-1.2ex]{0pt}{0pt}} & 0.35 [0.26, 0.39] & 8.7 (50\%) [7.8, 9.6] & 225.6 & 0.66 \\
    \hline
    \textbf{Qwen 3.6 Plus} & 0.73 [0.70, 0.75] & 0.63 [0.53, 0.68] & 0.57 [0.45, 0.60] & \makecell[r]{\rule{0pt}{2.6ex}0.50 [0.42, 0.54]\rlap{$^1$} \\ 0.60 [0.50, 0.63]\rlap{$^2$}\rule[-1.2ex]{0pt}{0pt}} & 0.37 [0.28, 0.41] & 7.7 (43\%) [6.9, 8.6] & 94.7 & 0.55 \\
    \hline
    \addlinespace
    \textbf{Average} & 0.736 & 0.672 & 0.481 & \makecell[r]{\rule{0pt}{2.6ex}0.469\rlap{$^1$} \\ 0.558\rlap{$^2$}\rule[-1.2ex]{0pt}{0pt}} & 0.365 & 8.8 (52\%) & & \\
    \bottomrule
  \end{tabular}
  \end{adjustbox}
  \par\footnotesize $^1$ Mean of the per-secondary-study F1 scores. $^2$ F1 computed from the averaged precision and recall. $^\dagger$Old models: costs previously recorded -- only an estimate; Scaled to the sample by the model's token-based cost per paper.\par
\end{sidewaystable*}

\begin{table*}[htbp]
	\centering
	\caption{Accuracy, precision, recall, F1 score, MCC, and lost evidence for the secondary studies across all LLMs (RQ1.2, 5\% sample). Lost evidence is the share of relevant papers missed (false negatives). MCC is N/A where it is undefined (all sampled papers, or all predictions, belong to one class).}
	\label{tab:rq12_per_study_metrics}
	\resizebox{\linewidth}{!}{%
		\begin{tabular}{lcccccc}
			\toprule
			\textbf{Study}                                           & \textbf{Accuracy} & \textbf{Precision} & \textbf{Recall}   & \textbf{F1}       & \textbf{MCC}        & \textbf{\makecell{Lost \\ evidence}} \\
			\midrule
			\citet{casalaroModeldrivenEngineeringMobile2022a}         & 0.94 [0.90, 0.97] & 0.25 [0.00, 0.51]  & 0.67 [0.00, 0.93] & 0.37 [0.00, 0.59] & 0.38 [0.17, 0.59]   & 33\% [0, 47]           \\
			\citet{sharbafConflictManagementTechniques2023a}          & 0.76 [0.66, 0.86] & 0.80 [0.62, 0.92]  & 0.65 [0.48, 0.81] & 0.72 [0.56, 0.83] & 0.52 [0.32, 0.70]   & 35\% [19, 52]          \\
			\citet{tambonHowCertifyMachine2022a}                      & 0.82 [0.75, 0.89] & 0.81 [0.56, 0.95]  & 0.34 [0.19, 0.51] & 0.48 [0.29, 0.64] & 0.45 [0.26, 0.59]   & 66\% [49, 81]          \\
			\citet{rodriguez-perezPerceivedDiversitySoftware2021a}    & 0.94 [0.91, 0.97] & 0.53 [0.17, 0.86]  & 0.23 [0.08, 0.37] & 0.32 [0.11, 0.46] & 0.32 [0.11, 0.48]   & 77\% [63, 92]          \\
			\citet{bjarnasonEmpiricallyBasedModel2023a}               & 0.83 [0.79, 0.88] & 0.04 [0.00, 0.12]  & 0.87 [0.00, 0.87] & 0.07 [0.00, 0.21] & 0.16 [0.14, 0.30]   & 13\% [0, 13]           \\
			\citet{kuutilaTimePressureSoftware2020b}                  & 0.95 [0.92, 0.97] & 0.51 [0.23, 0.75]  & 0.58 [0.29, 0.87] & 0.55 [0.27, 0.73] & 0.52 [0.25, 0.72]   & 42\% [13, 71]          \\
			\citet{geraldiSoftwareProductLine2020a}                   & 0.73 [0.60, 0.85] & 0.50 [0.17, 0.81]  & 0.44 [0.16, 0.73] & 0.47 [0.17, 0.68] & 0.29 [-0.03, 0.58]  & 56\% [27, 84]          \\
			\citet{zhangTestingVerificationNeuralnetworkbased2020a}   & 0.18 [0.08, 0.30] & 1.00 [1.00, 1.00]  & 0.18 [0.08, 0.30] & 0.30 [0.14, 0.46] & N/A                 & 82\% [70, 92]          \\
			\citet{tebesAnalyzingDocumentingSystematic2020a}          & 0.31 [0.20, 0.42] & 0.83 [0.57, 1.00]  & 0.15 [0.08, 0.24] & 0.26 [0.14, 0.37] & 0.05 [-0.18, 0.21]  & 85\% [77, 92]          \\
			\citet{ebrahimiMobileAppPrivacy2021a}                     & 0.62 [0.46, 0.76] & 0.56 [0.32, 0.76]  & 0.80 [0.63, 0.96] & 0.66 [0.44, 0.80] & 0.27 [0.00, 0.52]   & 20\% [4, 37]           \\
			\citet{lewowskiHowFarAre2022a}                            & 0.93 [0.89, 0.96] & 0.65 [0.37, 0.82]  & 0.56 [0.32, 0.81] & 0.60 [0.36, 0.77] & 0.56 [0.33, 0.74]   & 44\% [19, 68]          \\
			\citet{vandinterPredictiveMaintenanceUsing2022a}          & 0.84 [0.77, 0.90] & 0.55 [0.31, 0.75]  & 0.74 [0.55, 0.92] & 0.63 [0.42, 0.77] & 0.54 [0.34, 0.71]   & 26\% [8, 45]           \\
			\citet{stradowskiMachineLearningSoftware2023a}            & 0.85 [0.78, 0.91] & 0.89 [0.81, 0.95]  & 0.85 [0.76, 0.93] & 0.87 [0.80, 0.92] & 0.69 [0.55, 0.81]   & 15\% [7, 24]           \\
			\citet{alonsoSystematicMappingStudy2023b}                 & 0.23 [0.12, 0.35] & 0.58 [0.25, 0.94]  & 0.09 [0.03, 0.17] & 0.15 [0.05, 0.28] & -0.20 [-0.44, 0.06] & 91\% [83, 97]          \\
			\citet{somersDigitaltwinbasedTestingCyber2023b}           & 0.75 [0.66, 0.83] & 0.52 [0.28, 0.73]  & 0.42 [0.27, 0.58] & 0.46 [0.29, 0.61] & 0.30 [0.12, 0.48]   & 58\% [42, 73]          \\
			\citet{teixeiraModelingAutomaticCode2017a}                & 0.51 [0.38, 0.65] & 0.92 [0.79, 0.99]  & 0.25 [0.13, 0.38] & 0.39 [0.22, 0.55] & 0.27 [0.13, 0.41]   & 75\% [62, 87]          \\
			\citet{lenarduzziSystematicLiteratureReview2021a}         & 0.79 [0.69, 0.87] & 0.30 [0.03, 0.57]  & 0.33 [0.04, 0.71] & 0.32 [0.04, 0.57] & 0.19 [-0.09, 0.48]  & 67\% [28, 96]          \\
			\citet{pereiraLearningSoftwareConfiguration2021a}         & 0.71 [0.56, 0.86] & 1.00 [1.00, 1.00]  & 0.71 [0.56, 0.86] & 0.83 [0.72, 0.92] & N/A                 & 29\% [14, 44]          \\
			\citet{barisicMultiparadigmModelingCyber2022a}            & 0.62 [0.49, 0.74] & 0.93 [0.83, 0.99]  & 0.27 [0.17, 0.39] & 0.42 [0.28, 0.55] & 0.35 [0.24, 0.46]   & 73\% [61, 83]          \\
			\citet{deckersSystematicLiteratureReview2022a}            & 0.73 [0.63, 0.83] & 0.18 [0.00, 0.38]  & 0.76 [0.00, 0.87] & 0.29 [0.00, 0.51] & 0.27 [0.11, 0.44]   & 24\% [0, 40]           \\
			\citet{raniDecadeCodeComment2023a}                        & 0.99 [0.98, 1.00] & 0.54 [0.00, 0.83]  & 1.00 [0.00, 1.00] & 0.70 [0.00, 0.91] & 0.73 [0.62, 0.92]   & 0\% [0, 0]             \\
			\citet{zakeri-nasrabadiSystematicLiteratureReview2023b}   & 0.91 [0.88, 0.93] & 0.38 [0.21, 0.54]  & 0.60 [0.41, 0.77] & 0.46 [0.28, 0.60] & 0.43 [0.26, 0.57]   & 40\% [23, 59]          \\
			\citet{shevtsovControlTheoreticalSoftwareAdaptation2018a} & 0.85 [0.79, 0.91] & 0.58 [0.34, 0.76]  & 0.49 [0.30, 0.68] & 0.53 [0.34, 0.68] & 0.45 [0.24, 0.61]   & 51\% [32, 70]          \\
			\citet{linOpinionMiningSoftware2022a}                     & 0.91 [0.86, 0.95] & 0.67 [0.38, 0.84]  & 0.70 [0.41, 0.98] & 0.69 [0.42, 0.84] & 0.63 [0.36, 0.81]   & 30\% [2, 59]           \\
			\midrule
			\textbf{Average}                                         & 0.74 [0.72, 0.75] & 0.60 [0.54, 0.64]  & 0.53 [0.43, 0.56] & 0.48 [0.42, 0.51] & 0.37 [0.32, 0.42]   & 47\% [43, 50]          \\
			\bottomrule
		\end{tabular}
	}%
\end{table*}
\begin{table*}[htbp]
	\caption{Logistic regression model predicting the correctness of a single screening decision with the LLM and the secondary study. Models below the double horizontal line are from \citet{huotalaSESREvalDatasetEvaluating2025}.} 
	\label{tab:logistic_regression}
	\resizebox{\linewidth}{!}{%
		\begin{tabular}{lccccc}
			\toprule
			                                                         & \textbf{Est.} & \textbf{std.err} & \textbf{$z$} & \textbf{$p$}          & \makecell{\textbf{Odds} \\\textbf{ratio (OR)}}   \\
			\midrule 
			(Intercept)                                              & 4.9956        & 0.2881           & 17.342       & \textless{} 2e-16 *** & 147.76                  \\
			\midrule 
			\multicolumn{6}{l}{\textbf{Reference - Claude Opus 4.7 }}                                     \\
			GPT-5.5                                                  & -0.2348       & 0.1035           & -2.269       & 0.02326 *             & 0.79                    \\
			Grok 4.20 Beta                                           & -0.2747       & 0.1030           & -2.666       & 0.00766 **            & 0.76                    \\
			Kimi K2.6                                                & -0.1941       & 0.1040           & -1.867       & 0.06186 .             & 0.82                    \\
			Qwen 3.6 Plus                                            & -0.2747       & 0.1030           & -2.666       & 0.00766 **            & 0.76                    \\
			Gemini 3.1 Pro                                           & -0.1102       & 0.1050           & -1.050       & 0.29393               & 0.90                    \\
			DeepSeek V4 Pro                                          & -0.1890       & 0.1040           & -1.817       & 0.06926 .             & 0.83                    \\
			GLM 5.1                                                  & -0.1209       & 0.1049           & -1.153       & 0.24899               & 0.89                    \\
			\addlinespace
			\addlinespace
			\hline
			\hline
			\addlinespace
			\addlinespace
			Claude 3.7 Sonnet                                        & -0.2348       & 0.1035           & -2.269       & 0.02326 *             & 0.79                    \\
			DeepSeek R1                                              & -0.1890       & 0.1040           & -1.817       & 0.06926 .             & 0.83                    \\
			GPT-4.1                                                  & -0.4177       & 0.1014           & -4.118       & 3.83e-05 ***          & 0.66                    \\
			GPT-4.1 mini                                             & -0.0559       & 0.1057           & -0.529       & 0.59705               & 0.95                    \\
			GPT-4o                                                   & -0.4361       & 0.1013           & -4.307       & 1.66e-05 ***          & 0.65                    \\
			Llama 4 Maverick                                         & -0.2499       & 0.1033           & -2.419       & 0.01558 *             & 0.78                    \\
			o3-mini                                                  & -0.1683       & 0.1043           & -1.614       & 0.10657               & 0.85                    \\
			\addlinespace
			\midrule
			\addlinespace
			\multicolumn{6}{l}{\textbf{Reference - \citet{raniDecadeCodeComment2023a} }} \\
			\addlinespace
			\citet{casalaroModeldrivenEngineeringMobile2022a}         & -2.0906       & 0.2964           & -7.053       & 1.75e-12 ***          & 0.124                   \\
			\citet{sharbafConflictManagementTechniques2023a}          & -3.6242       & 0.2942           & -12.320      & < 2e-16 ***           & 0.027                   \\
			\citet{tambonHowCertifyMachine2022a}                      & -3.2768       & 0.2894           & -11.324      & < 2e-16 ***           & 0.038                   \\
			\citet{rodriguez-perezPerceivedDiversitySoftware2021a}    & -1.9675       & 0.2966           & -6.634       & 3.28e-11 ***          & 0.140                   \\
			\citet{bjarnasonEmpiricallyBasedModel2023a}               & -3.1666       & 0.2847           & -11.123      & < 2e-16 ***           & 0.042                   \\
			\citet{kuutilaTimePressureSoftware2020b}                  & -1.8817       & 0.2944           & -6.392       & 1.64e-10 ***          & 0.152                   \\
			\citet{geraldiSoftwareProductLine2020a}                   & -3.7823       & 0.2982           & -12.685      & < 2e-16 ***           & 0.023                   \\
			\citet{zhangTestingVerificationNeuralnetworkbased2020a}   & -6.3278       & 0.3123           & -20.262      & < 2e-16 ***           & 0.002                   \\
			\citet{tebesAnalyzingDocumentingSystematic2020a}          & -5.5829       & 0.2893           & -19.298      & < 2e-16 ***           & 0.004                   \\
			\citet{ebrahimiMobileAppPrivacy2021a}                     & -4.3087       & 0.2989           & -14.413      & < 2e-16 ***           & 0.013                   \\
			\citet{lewowskiHowFarAre2022a}                            & -2.2296       & 0.2971           & -7.505       & 6.14e-14 ***          & 0.108                   \\
			\citet{vandinterPredictiveMaintenanceUsing2022a}          & -3.1357       & 0.2944           & -10.650      & < 2e-16 ***           & 0.043                   \\
			\citet{stradowskiMachineLearningSoftware2023a}            & -3.0547       & 0.2901           & -10.529      & < 2e-16 ***           & 0.047                   \\
			\citet{alonsoSystematicMappingStudy2023b}                 & -6.0230       & 0.2980           & -20.208      & < 2e-16 ***           & 0.002                   \\
			\citet{somersDigitaltwinbasedTestingCyber2023b}           & -3.7030       & 0.2920           & -12.683      & < 2e-16 ***           & 0.025                   \\
			\citet{teixeiraModelingAutomaticCode2017a}                & -4.7432       & 0.2900           & -16.355      & < 2e-16 ***           & 0.009                   \\
			\citet{lenarduzziSystematicLiteratureReview2021a}         & -3.4868       & 0.2958           & -11.790      & < 2e-16 ***           & 0.031                   \\
			\citet{pereiraLearningSoftwareConfiguration2021a}         & -3.8763       & 0.3078           & -12.594      & < 2e-16 ***           & 0.021                   \\
			\citet{barisicMultiparadigmModelingCyber2022a}            & -4.3035       & 0.2907           & -14.806      & < 2e-16 ***           & 0.014                   \\
			\citet{deckersSystematicLiteratureReview2022a}            & -3.7793       & 0.2930           & -12.896      & < 2e-16 ***           & 0.023                   \\
			\citet{zakeri-nasrabadiSystematicLiteratureReview2023b}   & -2.4996       & 0.2848           & -8.777       & < 2e-16 ***           & 0.082                   \\
			\citet{shevtsovControlTheoreticalSoftwareAdaptation2018a} & -3.0345       & 0.2891           & -10.496      & < 2e-16 ***           & 0.048                   \\
			\citet{linOpinionMiningSoftware2022a}                     & -2.4916       & 0.3004           & -8.294       & < 2e-16 ***           & 0.083                   \\
			\bottomrule 
		\end{tabular}
	}%
\end{table*}

Here, we use the 5\% sampled dataset (SESR-Eval-Mini) for all the models we selected for RQ1.2 (Table~\ref{tab:test_llms_by_rq}). The LLMs benchmarked on the SESR-Eval-Mini show similar screening performance figures as on the full dataset\footnote{We can see that 95\% confidence intervals are wider in sampled than in full dataset.}. This result was somewhat expected, as the 5\% sample was drawn using power-allocated sampling, which attempted to sample by balancing the proportional and equal allocation. However, if we look closer at the proportional and equal allocation of the sample (Table~\ref{tab:sample_sizes}), we see that the power-allocated sampling had an I:E ratio of 21:79, while the original dataset had an I:E ratio of 12:88, as power-allocated sampling increased the proportional weight of smaller studies that had higher include ratio.

Tables \ref{tab:rq12_model_metrics_ci_primary} and \ref{tab:rq12_model_metrics_ci_secondary} show the accuracy, precision, recall, MCC, F1, and lost evidence scores of the models for primary studies and secondary studies, respectively. Table \ref{tab:logistic_regression} shows the logistic regression model, which predicts the correctness of a single screening decision with the LLM and the secondary study. The regression model was computed for RQ1.2 only, as we wanted to compare all new LLMs with the old LLMs from our prior work \citep{huotalaSESREvalDatasetEvaluating2025}.

From the regression model we can see that Claude Opus 4.7 has the highest odds for a correct screening decision, which surpasses the old GPT-4.1 mini by a slight margin. The statistically significant intercept (log-odds = 4.9956, p < 2e-16) corresponds to high baseline odds of a correct screening decision prediction, when using the reference model (Claude Opus 4.7) with the reference study \citep{raniDecadeCodeComment2023a} (odds ratio $\approx$ 147.76). Gemini 3.1 Pro, GLM 5.1, Kimi K2.6, DeepSeek R1, o3-mini and GPT-4.1 have roughly similar performance in correctness as the reference category, while the estimate is negative it is not statistically significant. The largest negative effects go for GPT-4o ($\beta$ = -0.4361, p < 1.66e-05, Odd's ratio (OR) $\approx$ 0.65) and GPT-4.1 ($\beta$ = -0.4177, p < 3.83e-05, OR $\approx$ 0.66), which suggests that these two old models have substantially poorer screening performance in terms of correctness compared to the newer LLMs (and also some older LLMs). For secondary studies, LLMs predicted the correct screening decision most on the study by \citet{raniDecadeCodeComment2023a}, with significantly worse performance on all other secondary studies.

\textbf{Screening performance across secondary studies.} The top-performing model with respect to the MCC score over secondary studies ($n=24$, see Table~\ref{tab:rq12_model_metrics_ci_secondary}) is Claude Opus 4.7 (new model) with MCC=$0.42$, followed by o3-mini (old model) with MCC=$0.40$, and in third place is DeepSeek V4 Pro (new model) with MCC=$0.38$. Then there is a bunch of models as all remaining models score between MCC=$0.34$ and MCC=$0.37$, with the exception of CLaude 3.7 Sonnet (MCC = $0.27$). With respect to OpenAI, we can state that previously their models -- GPT-4.1, GPT-4o, and o3-mini fared well when comparing across model providers. However, we now notice that GPT-5.5, which was their latest model, actually scores the same as GPT-4.1 mini and GPT-4o at MCC=$0.35$, indicating that those models have not progressed at all. As a coincidence, DeepSeek R1 (old) and DeepSeek V4 Pro (new) had the identical accuracy, precision, recall, F1, lost evidence scores and MCC (on primary study level), but a different MCC score on secondary study level. On average, we can see that the new LLMs perform marginally better (avg. new MCC = 0.365 vs. avg. old MCC = 0.347).

Then for the F1 score across secondary studies, we can see that the old GPT family still dominates, sharing the top-3 place with a mean F1 score of $0.53 ^1$ shared between GPT-4.1, o3-mini, and GPT-4o. The best new models are DeepSeek V4 Pro and Qwen 3.6 Plus, which are tied for fifth place with an F1 score of 0.50, followed by Claude Opus 4.7 with F1=$0.49^1$. Looking at the extended statistics from Online resource 3~\citep{zenodo_research_artifact}, screening performance between secondary studies varies significantly, where MCC ranges from $-0.27$ to $1.00$, and F1 from $0.00$ to $1.00$ for the new models.

Looking at the average cost per secondary study, the cost is under a dollar for 11 LLMs. Claude Opus 4.7 is the most expensive model with a cost per secondary study of 2.75 dollars. The second most expensive new model is GPT-5.5, which costs 1.83 dollars per secondary study. The cost per secondary study for DeepSeek V4 Pro, GLM 5.1, Grok 4.20 Beta, Kimi K2.6, and Qwen 3.6 Plus are on the 0.60 dollar area. By far the cheapest model is Llama 4 Maverick, which costs 7 cents (0.07 dollars) per secondary study. Looking at the paper screening speed - Kimi K2.6 is the slowest (225.6s) with DeepSeek V4 Pro in second place (103.0s). Third is Qwen 3.6 Plus (94.7s). Claude Opus 4.7 (15.6s), Gemini 3.1 Pro (14.3s), and GPT-5.5 (17.6s) are fairly fast in paper screening, and are almost as fast as the old GPT-4o (14.1s), but no LLM was faster than the old GPT-4.1 mini (10.5s).
 
\textbf{Screening performance across primary studies.} If we look at the primary-study averaged scores (Table~\ref{tab:rq12_model_metrics_ci_primary}), Claude Opus 4.7 still leads with MCC=$0.46$, followed by the old Llama 4 Maverick (old) (MCC=$0.45$) and GPT-4.1 mini (old) with MCC=$0.45$ in the shared second place. On the shared third place are o3-mini (old) and GPT-4o (old) with MCC of $0.44$ for both. GPT-5.5 and Grok 4.20 Beta are the worst-performing in terms of primary-study averaged MCC, with $0.39$ and $0.38$, respectively. On average, we see the opposite result than with the secondary studies: the old LLMs perform marginally better (avg. old MCC = 0.433 vs. avg. new MCC = 0.412).

However, the amount of lost evidence (Table~\ref{tab:rq12_model_metrics_ci_primary} rightmost column), differs significantly from the old LLMs. Claude Opus 4.7 loses 221 studies, which is substantially more than GPT-4o (128), GPT-4.1 (141), and Llama 4 Maverick (152). A similar pattern is observed for DeepSeek V4 Pro (200), GLM 5.1 (239), GPT-5.5 (213) and Grok 4.20 Beta (214). Looking at the per-study metrics averaged over all LLMs (Table~\ref{tab:rq12_per_study_metrics}), MCC ranges between $-0.20$ and $0.72$ and F1 score between $0.15$ and $0.87$, so differences between secondary studies are still quite significant.

\rqbox{
RQ1.2 results show that overall, the secondary studies have a larger effect than LLMs on screening performance. Model differences are modest: across secondary studies, the new models perform marginally better than old ones (avg. new MCC = 0.365 vs. avg. old MCC = 0.347), but when compared across primary studies, old models are still better (avg. new MCC = 0.412, avg. old MCC = 0.433). The new model Claude Opus 4.7 leads in the MCC metric by a narrow margin, and in logistic regression predicting the correctness of a screening decision. However, the old GPT-family still dominates when we look at the F1 score. New models tend to also lose more evidence (59\% vs. 49\%) than old models. Benchmarking on the SESR-Eval-Mini is cost-effective.
}
\subsection{RQ2: Computing the overall decision from criterion-level decisions}

\begin{sidewaystable*}[htbp]
	\centering
	\caption{Per-model accuracy, precision and recall, 5\% sample (RQ2).}
	\label{tab:rq2_v1_v2_comparison_ci_acc_pre_rec}
	\resizebox{\linewidth}{!}{%
		\begin{tabular}{lrrrrrrrrr}
			\toprule
			\textbf{Model} & \multicolumn{3}{c}{\textbf{Accuracy}} & \multicolumn{3}{c}{\textbf{Precision}} & \multicolumn{3}{c}{\textbf{Recall}} \\
			\cmidrule(lr){2-4}\cmidrule(lr){5-7}\cmidrule(lr){8-10}
			                         & \textit{V1}       & \textit{V2}       & $\Delta$                     & \textit{V1}       & \textit{V2}       & $\Delta$                      & \textit{V1}       & \textit{V2}       & $\Delta$                      \\
			\midrule
			\textbf{GPT-5.5}         & 0.82 [0.80, 0.84] & 0.83 [0.81, 0.84] & 0.00 [-0.01, +0.02]          & 0.60 [0.54, 0.66] & 0.64 [0.58, 0.71] & +0.04 [0.00, +0.09]           & 0.41 [0.36, 0.45] & 0.35 [0.31, 0.40] & \textbf{-0.05 [-0.09, -0.02]} \\
			\textbf{Claude Opus 4.7} & 0.85 [0.83, 0.86] & 0.84 [0.83, 0.85] & -0.01 [-0.01, 0.00]          & 0.75 [0.69, 0.81] & 0.75 [0.68, 0.81] & -0.01 [-0.05, +0.03]          & 0.38 [0.34, 0.43] & 0.34 [0.30, 0.39] & \textbf{-0.04 [-0.07, -0.01]} \\
			\textbf{Gemini 3.1 Pro}  & 0.83 [0.82, 0.85] & 0.82 [0.81, 0.84] & \textbf{-0.01 [-0.02, 0.00]} & 0.67 [0.61, 0.73] & 0.62 [0.56, 0.68] & \textbf{-0.05 [-0.09, -0.02]} & 0.39 [0.35, 0.44] & 0.38 [0.34, 0.43] & -0.01 [-0.04, +0.02]          \\
			\textbf{GLM 5.1}         & 0.83 [0.82, 0.85] & 0.82 [0.81, 0.84] & \textbf{-0.01 [-0.02, 0.00]} & 0.71 [0.65, 0.77] & 0.66 [0.59, 0.73] & \textbf{-0.05 [-0.09, -0.01]} & 0.33 [0.29, 0.38] & 0.31 [0.27, 0.35] & \textbf{-0.03 [-0.05, 0.00]}  \\
			\textbf{Kimi K2.6}       & 0.83 [0.81, 0.84] & 0.82 [0.81, 0.84] & 0.00 [-0.02, +0.01]          & 0.62 [0.56, 0.67] & 0.60 [0.54, 0.66] & -0.02 [-0.06, +0.03]          & 0.42 [0.37, 0.46] & 0.43 [0.38, 0.47] & +0.01 [-0.03, +0.05]          \\
			\textbf{Grok 4.20 Beta}  & 0.82 [0.80, 0.83] & 0.80 [0.79, 0.82] & \textbf{-0.01 [-0.02, 0.00]} & 0.58 [0.53, 0.63] & 0.54 [0.48, 0.59] & \textbf{-0.04 [-0.08, -0.01]} & 0.40 [0.36, 0.45] & 0.37 [0.32, 0.41] & -0.04 [-0.07, 0.00]           \\
			\textbf{Qwen 3.6 Plus}   & 0.82 [0.80, 0.83] & 0.82 [0.80, 0.83] & 0.00 [-0.01, +0.01]          & 0.57 [0.51, 0.62] & 0.58 [0.53, 0.64] & +0.02 [-0.01, +0.05]          & 0.48 [0.44, 0.53] & 0.45 [0.40, 0.49] & \textbf{-0.03 [-0.06, -0.01]} \\
			\textbf{DeepSeek V4 Pro} & 0.83 [0.81, 0.84] & 0.82 [0.80, 0.84] & -0.01 [-0.02, 0.00]          & 0.61 [0.56, 0.67] & 0.59 [0.54, 0.65] & -0.02 [-0.05, +0.02]          & 0.44 [0.40, 0.49] & 0.41 [0.37, 0.46] & -0.03 [-0.06, 0.00]           \\
			\midrule
			\textbf{Avg.}            & 0.83 [0.81, 0.84] & 0.82 [0.81, 0.84] & \textbf{-0.01 [-0.01, 0.00]} & 0.64 [0.59, 0.69] & 0.62 [0.57, 0.67] & -0.02 [-0.03, 0.00]           & 0.41 [0.37, 0.45] & 0.38 [0.34, 0.42] & \textbf{-0.03 [-0.04, -0.01]} \\
			\bottomrule
		\end{tabular}%
	}
	\bigskip
	\bigskip
	\caption{Per-model F1, MCC and lost evidence, 5\% sample (RQ2).}
	\label{tab:rq2_v1_v2_comparison_ci_f1_mcc_fn}
	\resizebox{\linewidth}{!}{%
		\begin{tabular}{lrrrrrrrrr}
			\toprule
			\textbf{Model} & \multicolumn{3}{c}{\textbf{F1}} & \multicolumn{3}{c}{\textbf{MCC}} & \multicolumn{3}{c}{\textbf{Lost Evidence (FN)}} \\
			\cmidrule(lr){2-4}\cmidrule(lr){5-7}\cmidrule(lr){8-10}
			                         & \textit{V1}       & \textit{V2}       & $\Delta$                      & \textit{V1}       & \textit{V2}       & $\Delta$                      & \textit{V1}    & \textit{V2}    & $\Delta$               \\
			\midrule
			\textbf{GPT-5.5}         & 0.48 [0.44, 0.53] & 0.45 [0.41, 0.50] & -0.03 [-0.07, +0.01]          & 0.39 [0.34, 0.44] & 0.38 [0.33, 0.44] & -0.01 [-0.05, +0.03]          & 213 [192, 234] & 232 [209, 256] & \textbf{+19 [+6, +33]} \\
			\textbf{Claude Opus 4.7} & 0.51 [0.46, 0.55] & 0.47 [0.42, 0.52] & \textbf{-0.04 [-0.07, -0.01]} & 0.46 [0.41, 0.51] & 0.43 [0.38, 0.48] & -0.03 [-0.06, 0.00]           & 221 [201, 242] & 235 [213, 257] & \textbf{+14 [+4, +25]} \\
			\textbf{Gemini 3.1 Pro}  & 0.50 [0.45, 0.54] & 0.47 [0.43, 0.52] & -0.03 [-0.05, 0.00]           & 0.43 [0.38, 0.48] & 0.39 [0.34, 0.44] & \textbf{-0.04 [-0.07, -0.01]} & 217 [196, 239] & 222 [201, 244] & +5 [-6, +15]           \\
			\textbf{GLM 5.1}         & 0.45 [0.41, 0.50] & 0.42 [0.37, 0.47] & \textbf{-0.03 [-0.06, -0.01]} & 0.41 [0.36, 0.46] & 0.36 [0.31, 0.42] & \textbf{-0.04 [-0.07, -0.01]} & 239 [216, 261] & 248 [225, 271] & \textbf{+9 [+1, +18]}  \\
			\textbf{Kimi K2.6}       & 0.50 [0.45, 0.54] & 0.50 [0.45, 0.54] & 0.00 [-0.04, +0.04]           & 0.41 [0.36, 0.46] & 0.40 [0.35, 0.45] & 0.00 [-0.05, +0.04]           & 209 [188, 230] & 205 [183, 226] & -4 [-19, +11]          \\
			\textbf{Grok 4.20 Beta}  & 0.48 [0.43, 0.52] & 0.44 [0.39, 0.48] & \textbf{-0.04 [-0.08, -0.01]} & 0.38 [0.33, 0.43] & 0.33 [0.28, 0.38] & \textbf{-0.05 [-0.09, -0.01]} & 214 [193, 235] & 227 [205, 250] & +13 [0, +27]           \\
			\textbf{Qwen 3.6 Plus}   & 0.52 [0.48, 0.56] & 0.51 [0.47, 0.55] & -0.01 [-0.04, +0.01]          & 0.41 [0.36, 0.46] & 0.40 [0.36, 0.45] & -0.01 [-0.04, +0.03]          & 185 [165, 206] & 197 [176, 218] & \textbf{+12 [+2, +23]} \\
			\textbf{DeepSeek V4 Pro} & 0.51 [0.47, 0.55] & 0.49 [0.44, 0.53] & -0.02 [-0.05, 0.00]           & 0.42 [0.37, 0.47] & 0.39 [0.34, 0.44] & -0.03 [-0.06, +0.01]          & 200 [180, 220] & 210 [190, 232] & +10 [-1, +21]          \\
			\midrule
			\textbf{Avg.}            & 0.49 [0.45, 0.53] & 0.47 [0.43, 0.51] & \textbf{-0.03 [-0.04, -0.01]} & 0.41 [0.37, 0.45] & 0.39 [0.35, 0.43] & \textbf{-0.02 [-0.04, -0.01]} & 212 [193, 232] & 222 [203, 242] & \textbf{+10 [+4, +16]} \\
			\bottomrule
		\end{tabular}%
	}
\end{sidewaystable*}

Table~\ref{tab:rq2_v1_v2_comparison_ci_acc_pre_rec} shows per-model accuracy, precision and recall, and Table~\ref{tab:rq2_v1_v2_comparison_ci_f1_mcc_fn} the F1, MCC, and lost evidence on whether computing the overall decision from individual criteria is better than letting the LLM make the overall decision. Technically speaking computing the decision from the individual criteria is a more correct approach. However, deviating from that can allow weighting of the criteria to occur in the LLM that then makes the overall decision. We can see that computing the overall screening decision from criterion-level decisions decreases the average accuracy, precision, recall, F1, and MCC. For GPT-5.5, the accuracy increased by one percentage point (from 0.82 to 0.83). Lost evidence increases by +10 on average, and the only LLM that had less lost evidence was Kimi K2.6 (-4). 

Bootstrapping demonstrates that the reduction in accuracy, precision, and MCC for Gemini 3.1 Pro, GLM 5.1, and Grok 4.20 Beta are statistically significant as zero is not part of the 95\% confidence interval. Recall reduction for GPT-5.5, Claude Opus 4.7, GLM 5.1, and Qwen 3.6 Plus are also statistically significant. The F1 score reduction is statistically significant for Claude Opus 4.7, GLM 5.1, and Grok 4.20 Beta. In terms of lost evidence, the increase in lost evidence is significant for most of the LLMs. 

\rqbox{
RQ2 shows that computing overall decisions from individual criteria makes screening stricter but not more precise: recall fell (0.41 to 0.38) and there were more false negatives (+10 on average). Given that false negatives (lost evidence) are worse than false positives, this change cannot be recommended.
}
\subsection{RQ3.1: LLM agreement - Overall decision}

\begin{table*}[htbp]
	\centering
	\caption{Gwet's AC1 between LLMs on binary include/exclude decisions on the full dataset and on the 5\% sample, with bootstrap confidence intervals (RQ3.1). 95\% percentile bootstrap confidence intervals in brackets ($B=2000$; papers resampled with replacement within each secondary study).}
	\label{tab:ac1_binary_ci}
	\resizebox{\linewidth}{!}{%
		\begin{tabular}{lll}
			\toprule
			& \multicolumn{2}{c}{\textbf{AC1}} \\
			\cmidrule(lr){2-3}
			\textbf{LLM pair}                         & \textbf{Full dataset} & \textbf{5\% sample}  \\
			\midrule
			\textbf{Claude 3.7 -- DeepSeek R1}        & 0.911 [0.908, 0.915]  & 0.861 [0.840, 0.881] \\
			\textbf{Claude 3.7 -- DeepSeek V4 Pro}    & 0.891 [0.887, 0.895]  & 0.847 [0.826, 0.868] \\
			\textbf{Claude 3.7 -- GPT-4.1}            & 0.813 [0.807, 0.818]  & 0.768 [0.742, 0.793] \\
			\textbf{Claude 3.7 -- GPT-4.1 mini}       & 0.909 [0.906, 0.913]  & 0.853 [0.833, 0.872] \\
			\textbf{Claude 3.7 -- GPT-4o}             & 0.826 [0.821, 0.831]  & 0.771 [0.744, 0.797] \\
			\textbf{Claude 3.7 -- Llama 4 Mav.}       & 0.874 [0.869, 0.878]  & 0.822 [0.797, 0.844] \\
			\textbf{Claude 3.7 -- o3-mini}            & 0.862 [0.858, 0.866]  & 0.817 [0.794, 0.839] \\
			\textbf{Claude 3.7 -- Qwen 3.6 Plus}      & 0.878 [0.873, 0.881]  & 0.833 [0.811, 0.854] \\
			\textbf{DeepSeek R1 -- DeepSeek V4 Pro}   & 0.890 [0.886, 0.894]  & 0.860 [0.839, 0.879] \\
			\textbf{DeepSeek R1 -- GPT-4.1}           & 0.814 [0.809, 0.820]  & 0.806 [0.781, 0.829] \\
			\textbf{DeepSeek R1 -- GPT-4.1 mini}      & 0.925 [0.922, 0.928]  & 0.887 [0.868, 0.904] \\
			\textbf{DeepSeek R1 -- GPT-4o}            & 0.829 [0.824, 0.834]  & 0.792 [0.766, 0.818] \\
			\textbf{DeepSeek R1 -- Llama 4 Mav.}      & 0.871 [0.867, 0.876]  & 0.838 [0.815, 0.860] \\
			\textbf{DeepSeek R1 -- o3-mini}           & 0.861 [0.856, 0.865]  & 0.832 [0.808, 0.853] \\
			\textbf{DeepSeek R1 -- Qwen 3.6 Plus}     & 0.873 [0.869, 0.877]  & 0.849 [0.826, 0.870] \\
			\textbf{DeepSeek V4 Pro -- GPT-4.1}       & 0.826 [0.821, 0.831]  & 0.799 [0.773, 0.823] \\
			\textbf{DeepSeek V4 Pro -- GPT-4.1 mini}  & 0.896 [0.892, 0.899]  & 0.873 [0.854, 0.891] \\
			\textbf{DeepSeek V4 Pro -- GPT-4o}        & 0.833 [0.828, 0.838]  & 0.775 [0.747, 0.801] \\
			\textbf{DeepSeek V4 Pro -- Llama 4 Mav.}  & 0.864 [0.860, 0.869]  & 0.824 [0.800, 0.846] \\
			\textbf{DeepSeek V4 Pro -- o3-mini}       & 0.918 [0.914, 0.921]  & 0.897 [0.879, 0.914] \\
			\textbf{DeepSeek V4 Pro -- Qwen 3.6 Plus} & 0.938 [0.935, 0.941]  & 0.921 [0.905, 0.936] \\
			\textbf{GPT-4.1 -- GPT-4.1 mini}          & 0.823 [0.817, 0.828]  & 0.794 [0.769, 0.819] \\
			\textbf{GPT-4.1 -- GPT-4o}                & 0.872 [0.867, 0.877]  & 0.853 [0.828, 0.875] \\
			\textbf{GPT-4.1 -- Llama 4 Mav.}          & 0.840 [0.835, 0.845]  & 0.825 [0.800, 0.848] \\
			\textbf{GPT-4.1 -- o3-mini}               & 0.838 [0.833, 0.843]  & 0.812 [0.786, 0.836] \\
			\textbf{GPT-4.1 -- Qwen 3.6 Plus}         & 0.826 [0.820, 0.831]  & 0.814 [0.789, 0.838] \\
			\textbf{GPT-4.1 mini -- GPT-4o}           & 0.826 [0.821, 0.831]  & 0.776 [0.750, 0.802] \\
			\textbf{GPT-4.1 mini -- Llama 4 Mav.}     & 0.877 [0.873, 0.881]  & 0.832 [0.810, 0.853] \\
			\textbf{GPT-4.1 mini -- o3-mini}          & 0.870 [0.866, 0.874]  & 0.852 [0.831, 0.872] \\
			\textbf{GPT-4.1 mini -- Qwen 3.6 Plus}    & 0.879 [0.875, 0.882]  & 0.855 [0.835, 0.876] \\
			\textbf{GPT-4o -- Llama 4 Mav.}           & 0.859 [0.855, 0.864]  & 0.832 [0.807, 0.855] \\
			\textbf{GPT-4o -- o3-mini}                & 0.837 [0.832, 0.842]  & 0.794 [0.768, 0.819] \\
			\textbf{GPT-4o -- Qwen 3.6 Plus}          & 0.829 [0.824, 0.834]  & 0.786 [0.759, 0.813] \\
			\textbf{Llama 4 Mav. -- o3-mini}          & 0.850 [0.845, 0.854]  & 0.815 [0.790, 0.839] \\
			\textbf{Llama 4 Mav. -- Qwen 3.6 Plus}    & 0.857 [0.852, 0.862]  & 0.813 [0.788, 0.837] \\
			\textbf{o3-mini -- Qwen 3.6 Plus}         & 0.919 [0.916, 0.923]  & 0.900 [0.882, 0.917] \\
			\midrule
			\textbf{Mean over pairs}                  & 0.864 [0.861, 0.867]  & 0.830 [0.815, 0.844] \\
			\bottomrule
		\end{tabular}%
	}
\end{table*}

Table~\ref{tab:ac1_binary_ci} shows Gwet's AC1 score on the binary decision on the full dataset -- the agreement between LLMs is generally high. The mean Gwet's AC1 score over pairs is 0.864. When looking at the range, the lowest Gwet's AC1 between any two LLMs is 0.813 (Claude 3.7 Sonnet vs. GPT-4.1), and the highest one is 0.938 between DeepSeek V4 Pro and Qwen 3.6 Plus. Looking at the 5\% sampled data (Table~\ref{tab:ac1_binary_ci}), the agreement between LLMs is slightly lower, where the mean AC1 score decreases to 0.830 and the range of agreement is 0.768 - 0.921. Out of the 36 pairs in the table, nine of them have an agreement less than 0.80, and contain either GPT-4.1 or GPT-4o as the other model. 
\rqbox{
RQ3.1 shows that LLMs make mostly the same decisions in paper screening, and the most disagreement comes from pairs where GPT-4.1 and GPT-4o are part of.
}
\subsection{RQ3.2: LLM agreement - Individual criteria}


\begin{table}[htbp]
	\centering
	\caption{Mean inter-model agreement (Gwet's AC1) on binary screening decision.}
	\begin{tabular}{lrl}
		\toprule
		\textbf{Type}      & \textbf{N} & \textbf{Mean AC1}                \\
		\midrule
		\textbf{IC}        & 93         & 0.778 [0.773, 0.783]             \\
		\textbf{EC}        & 110        & 0.868 [0.865, 0.870]             \\
		\textbf{IC $-$ EC} &            & \textbf{-0.090 [-0.095, -0.086]} \\
		\bottomrule
	\end{tabular}
	\label{tab:ic_ec_agreement_ci}
\end{table}

In order to get a better insight of the agreements and disagreements, we decided to look into individual criteria in RQ3.2. 
Table~\ref{tab:ic_ec_agreement_ci} shows the mean pairwise inter-model agreement grouped by criterion type. Exclusion criteria show slightly higher mean agreement than inclusion criteria (0.868 vs 0.778), which is statistically significant. 

As shown in Online resource 3 ~\citep{zenodo_research_artifact}, agreement levels for inclusion criteria vary considerably across the secondary studies. For example, in study \citet{stradowskiMachineLearningSoftware2023a}, agreement ranges from $0.709$ to $0.965$ across the inclusion criteria, whereas in study \citet{alonsoSystematicMappingStudy2023b}, it ranges from $0.065$ to $0.834$. 

A similar pattern is also observed for studies \citet{casalaroModeldrivenEngineeringMobile2022a,zakeri-nasrabadiSystematicLiteratureReview2023b,deckersSystematicLiteratureReview2022a,tambonHowCertifyMachine2022a,pereiraLearningSoftwareConfiguration2021a,rodriguez-perezPerceivedDiversitySoftware2021a,barisicMultiparadigmModelingCyber2022a,lenarduzziSystematicLiteratureReview2021a,linOpinionMiningSoftware2022a}, where substantial variation in agreement levels exists between individual inclusion criteria. The same pattern is observed partially for the exclusion criteria: agreement levels vary considerably across the secondary studies, and about half of the studies exhibit substantial variance between individual exclusion criteria.

Despite the variability, some individual inclusion and exclusion criteria show consistently high agreement across the secondary studies. For example, exclusion criteria related to the paper's language being other than English \citep{stradowskiMachineLearningSoftware2023a,alonsoSystematicMappingStudy2023b,somersDigitaltwinbasedTestingCyber2023b,zakeri-nasrabadiSystematicLiteratureReview2023b,lewowskiHowFarAre2022a,vandinterPredictiveMaintenanceUsing2022a,casalaroModeldrivenEngineeringMobile2022a,deckersSystematicLiteratureReview2022a,tambonHowCertifyMachine2022a,raniDecadeCodeComment2023a,pereiraLearningSoftwareConfiguration2021a,rodriguez-perezPerceivedDiversitySoftware2021a,barisicMultiparadigmModelingCyber2022a,ebrahimiMobileAppPrivacy2021a,tebesAnalyzingDocumentingSystematic2020a} generally exhibited very high levels of agreement in AC1 from 0.911 to 1.000. Similarly, criteria based on simple and objective characteristics of a study tended to yield higher agreement than criteria requiring greater subjective interpretation. To illustrate criteria where LLMs have high agreement: 

\begin{itemize}
    \item \citet{stradowskiMachineLearningSoftware2023a} "The paper is a primary study" (AC1 $0.965$)
    \item \citet{lewowskiHowFarAre2022a}  "Title or abstract of the paper indicates that at least one code smell/anti-pattern plays an important part of the study." (AC1 $0.876$)
    \item \citet{raniDecadeCodeComment2023a} "The topic of the paper is about code
comment quality" (AC1 $0.967$)
\end{itemize}

When looking at the criteria where LLMs disagree, we find two groups. Criteria that look simple and straightforward but whose agreement is still low, such as: 

\begin{itemize}
    \item \citet{tambonHowCertifyMachine2022a} "Papers that thrive to solely improve a
prediction metric such as accuracy,
precision, or recall" (AC1 $0.443$)
\item \citet{ebrahimiMobileAppPrivacy2021a} "Studies which explicitly addresses privacy
aspects of apps" (AC1 $0.618$)
\item  \citet{lenarduzziSystematicLiteratureReview2021a} "Publications that only mention prioritization
of TD in an introductory statement and do
not fully or partly focus on it" (AC1 $0.259$)
\item \citet{alonsoSystematicMappingStudy2023b} "Do not mainly comprise software
development-related MVPs (e.g., mainly
hardware or IoT related MVPs)" (AC1 $0.065$)

\end{itemize}

We must recognize that agreement is also influenced by the data, i.e. papers that are screened. Many criteria in the two bullet lists above have a good deal of subjectivity, e.g. "important part of the study"~\citep{lewowskiHowFarAre2022a} ($0.876$), while "do not fully or partly focus on it" ($0.259$) both have vagueness in them yet they score totally differently.

Then finally we have criteria that try to say too much and therefore do not perform very well: 

\begin{itemize}
\item \citet{rodriguez-perezPerceivedDiversitySoftware2021a} "We removed papers that analyze perceived
diversity from a software management
perspective (i.e., planning, scheduling,
resource allocation, execution, delivery) and
a team management perspective (i.e.,
communication) because these papers were
not related to perceived diversity in the
software development." (AC1 $0.530$)
\item \citet{bjarnasonEmpiricallyBasedModel2023a} "Articles that merely describe the use of
prototypes without considering methodological
perspectives such as principles and procedures
for the prototyping, articles that are not peer
reviewed or written in English, and duplicates of
studies already included in the systematic map" (AC1 $0.663$)
\end{itemize}

\subsection{RQ4: Refining the inclusion and exclusion criteria}

\begin{table*}[htbp]
	\centering
	\caption{Gwet's AC1 across the GPT-5.5, Claude Opus 4.7, Gemini 3.1 Pro, GLM 5.1, Kimi K2.6, Grok 4.20 Beta, Qwen 3.6 Plus, and DeepSeek V4 Pro LLMs for the refined criteria (RQ4).}
	\label{table:rq5_ac1_before_after_v3}
	\begin{adjustbox}{max size={\linewidth}{\textheight}}
		\begin{tabular}{lllrrr}
			\toprule
			& \multicolumn{2}{c}{\textbf{ID}} & \multicolumn{2}{c}{\textbf{AC1}} & \\
			\cmidrule(lr){2-3} \cmidrule(lr){4-5}
			\textbf{Study}                                         & \textbf{Old}         & \textbf{New} & \textbf{Old}           & \textbf{New} & \textbf{$\Delta$}                \\
			\midrule
			\citet{rodriguez-perezPerceivedDiversitySoftware2021a}  & IC1                  & IC1          & 0.902                  & 0.898        & -0.003 [-0.022, +0.016]          \\
			                                                       & IC2                  & IC2          & 0.985                  & 0.984        & -0.001 [-0.010, +0.009]          \\
			                                                       & IC3                  & IC3          & 1.000                  & 1.000        & 0.000 [0.000, 0.000]             \\
			                                                       & IC4                  & \textemdash  & 0.775                  & \textemdash  & \textemdash                      \\
			                                                       & IC5                  & \textemdash  & 0.716                  & \textemdash  & \textemdash                      \\
			                                                       & IC6                  & \textemdash  & 0.318                  & \textemdash  & \textemdash                      \\
			                                                       & IC7                  & \textemdash  & 0.759                  & \textemdash  & \textemdash                      \\
			                                                       & EC1                  & EC1          & 0.996                  & 0.998        & +0.002 [-0.004, +0.011]          \\
			                                                       & EC2                  & \textemdash  & 1.000                  & \textemdash  & \textemdash                      \\
			                                                       & EC3                  & EC2          & 0.941                  & 0.972        & \textbf{+0.031 [+0.011, +0.056]} \\
			                                                       & EC4                  & EC3          & 0.975                  & 0.991        & \textbf{+0.016 [+0.001, +0.032]} \\
			                                                       & EC5                  & EC4          & 0.876                  & 0.890        & +0.014 [-0.008, +0.038]          \\
			                                                       & EC6                  & EC5          & 0.987                  & 0.994        & +0.007 [-0.004, +0.018]          \\
			                                                       & EC7                  & EC6          & 0.992                  & 0.987        & -0.004 [-0.012, 0.000]           \\
			\midrule
			\citet{bjarnasonEmpiricallyBasedModel2023a}             & IC1                  & IC1          & 0.731                  & 0.734        & +0.003 [-0.031, +0.041]          \\
			                                                       & \multirow{2}{*}{EC1} & IC2          & \multirow{2}{*}{0.475} & 1.000        & \textbf{+0.525 [+0.443, +0.608]} \\
			                                                       &                      & EC1          &                        & 0.416        & -0.059 [-0.118, +0.005]          \\
			\midrule
			\citet{zhangTestingVerificationNeuralnetworkbased2020a} & IC1                  & IC1          & 0.759                  & 0.920        & \textbf{+0.162 [+0.079, +0.256]} \\
			                                                       & IC2                  & IC2          & 0.848                  & 0.875        & +0.027 [-0.060, +0.135]          \\
			                                                       & IC3                  & IC3          & 0.955                  & 0.944        & -0.011 [-0.039, 0.000]           \\
			                                                       & EC1                  & \textemdash  & 0.988                  & \textemdash  & \textemdash                      \\
			                                                       & EC2                  & IC4          & 1.000                  & 1.000        & 0.000 [0.000, 0.000]             \\
			                                                       & EC3                  & \textemdash  & 1.000                  & \textemdash  & \textemdash                      \\
			                                                       & EC4                  & IC5          & 0.851                  & 0.944        & +0.093 [-0.001, +0.200]          \\
			\midrule
			\citet{tebesAnalyzingDocumentingSystematic2020a}        & IC1                  & \textemdash  & 0.748                  & \textemdash  & \textemdash                      \\
			                                                       & IC2                  & IC1          & 0.948                  & 0.963        & +0.016 [-0.002, +0.038]          \\
			                                                       & IC3                  & IC2          & 0.868                  & 0.704        & \textbf{-0.163 [-0.251, -0.079]} \\
			                                                       & EC1                  & EC1          & 0.945                  & 0.948        & +0.003 [-0.028, +0.038]          \\
			                                                       & EC2                  & IC3          & 0.919                  & 0.955        & +0.036 [-0.008, +0.088]          \\
			                                                       & EC3                  & IC4          & 1.000                  & 1.000        & 0.000 [0.000, 0.000]             \\
			                                                       & EC4                  & IC5          & 0.889                  & 0.829        & \textbf{-0.059 [-0.115, -0.011]} \\
			                                                       & EC5                  & \textemdash  & 1.000                  & \textemdash  & \textemdash                      \\
			                                                       & EC6                  & \textemdash  & 1.000                  & \textemdash  & \textemdash                      \\
			                                                       & EC7                  & \textemdash  & 0.995                  & \textemdash  & \textemdash                      \\
			\midrule
			\citet{alonsoSystematicMappingStudy2023b}               & IC1                  & IC1          & 0.870                  & 0.848        & -0.022 [-0.083, +0.023]          \\
			                                                       & IC2                  & EC2          & 0.741                  & 0.130        & \textbf{-0.610 [-0.769, -0.424]} \\
			                                                       & IC3                  & EC1          & 0.138                  & 0.750        & \textbf{+0.612 [+0.384, +0.778]} \\
			                                                       & EC1                  & IC2          & 1.000                  & 1.000        & 0.000 [0.000, 0.000]             \\
			                                                       & EC2                  & \textemdash  & 0.949                  & \textemdash  & \textemdash                      \\
			                                                       & EC3                  & \textemdash  & 0.962                  & \textemdash  & \textemdash                      \\
			\midrule
			\citet{lenarduzziSystematicLiteratureReview2021a}       & IC1                  & IC1          & 0.883                  & 0.800        & \textbf{-0.083 [-0.146, -0.028]} \\
			                                                       & IC2                  & IC2          & 0.841                  & 0.817        & -0.024 [-0.075, +0.023]          \\
			                                                       & IC3                  & IC3          & 0.873                  & 0.833        & -0.040 [-0.097, +0.007]          \\
			                                                       & IC4                  & IC4          & 0.921                  & 0.881        & -0.041 [-0.086, +0.002]          \\
			                                                       & EC1                  & IC5          & 0.993                  & 1.000        & +0.007 [0.000, +0.021]           \\
			                                                       & EC2                  & \textemdash  & 0.989                  & \textemdash  & \textemdash                      \\
			                                                       & EC3                  & \textemdash  & 1.000                  & \textemdash  & \textemdash                      \\
			                                                       & EC4                  & EC1          & 0.921                  & 0.892        & -0.030 [-0.104, +0.039]          \\
			                                                       & EC5                  & \textemdash  & 0.994                  & \textemdash  & \textemdash                      \\
			                                                       & EC6                  & EC2          & 0.148                  & 0.215        & +0.067 [-0.011, +0.136]          \\
			                                                       & EC7                  & \textemdash  & 1.000                  & \textemdash  & \textemdash                      \\
			\midrule
			\citet{deckersSystematicLiteratureReview2022a}          & IC1                  & \textemdash  & 0.934                  & \textemdash  & \textemdash                      \\
			                                                       & IC2                  & IC1          & 1.000                  & 1.000        & 0.000 [0.000, 0.000]             \\
			                                                       & IC3                  & \textemdash  & 0.688                  & \textemdash  & \textemdash                      \\
			                                                       & EC1                  & IC2          & 0.674                  & 0.798        & \textbf{+0.124 [+0.043, +0.208]} \\
			                                                       & EC2                  & IC3          & 0.956                  & 0.769        & \textbf{-0.187 [-0.323, -0.064]} \\
			                                                       & EC3                  & EC3          & 0.994                  & 1.000        & +0.006 [0.000, +0.019]           \\
			                                                       & EC4                  & EC4          & 0.986                  & 0.987        & +0.001 [0.000, +0.002]           \\
			                                                       & EC5                  & EC5          & 0.596                  & 0.889        & \textbf{+0.293 [+0.167, +0.424]} \\
			                                                       & EC6                  & EC6          & 0.885                  & 0.939        & \textbf{+0.054 [+0.008, +0.117]} \\
			\midrule
			\textbf{Mean}                                          &                      &              & 0.847                  & 0.866        & \textbf{+0.019 [+0.008, +0.030]} \\
			\bottomrule
		\end{tabular}
	\end{adjustbox}
\end{table*}

\begin{table*}[htbp]
	\centering
	\caption{Lost evidence (false negatives) per model: RQ2 vs.\ RQ4 (refined criteria), restricted to the 7 secondary studies with refined criteria, 5\% sample. Percentages are relative to the total relevant papers in the sample.}
	\label{tab:rq5_lost_evidence}
	\begin{tabular}{lrrr}
		\toprule
		& \multicolumn{3}{c}{\textbf{Lost Evidence (FN)}} \\
		\cmidrule(lr){2-4}
		\textbf{Model}           & \textit{RQ2}        & \textit{RQ4}        & $\Delta$\textit{(RQ4$-$RQ2)} \\
		\midrule
		\textbf{GPT-5.5}         & 95 (89\%) [84, 106] & 87 (81\%) [76, 98]  & \textbf{-8 [-15, -2]}        \\
		\textbf{Claude Opus 4.7} & 99 (93\%) [89, 109] & 95 (89\%) [84, 105] & \textbf{-4 [-8, -1]}         \\
		\textbf{Gemini 3.1 Pro}  & 92 (86\%) [81, 103] & 89 (83\%) [78, 100] & -3 [-8, +1]                  \\
		\textbf{GLM 5.1}         & 99 (93\%) [89, 110] & 98 (92\%) [87, 109] & -1 [-6, +4]                  \\
		\textbf{Kimi K2.6}       & 94 (88\%) [83, 105] & 92 (86\%) [81, 102] & -2 [-8, +3]                  \\
		\textbf{Grok 4.20 Beta}  & 94 (88\%) [84, 105] & 92 (86\%) [81, 102] & -2 [-7, +3]                  \\
		\textbf{Qwen 3.6 Plus}   & 89 (83\%) [78, 100] & 86 (80\%) [74, 97]  & -3 [-8, +2]                  \\
		\textbf{DeepSeek V4 Pro} & 95 (89\%) [84, 105] & 74 (69\%) [63, 86]  & \textbf{-21 [-30, -13]}      \\
		\midrule
		\textbf{Avg.}            & 95 (88\%) [85, 105] & 89 (83\%) [79, 99]  & \textbf{-6 [-9, -2]}         \\
		\bottomrule
	\end{tabular}%
\end{table*}

\begin{sidewaystable*}[htbp]
	\centering
	\caption{RQ2 vs.\ RQ4 screening metrics.}
	\label{tab:rq5_vs_rq2}
	\begin{adjustbox}{max size={\linewidth}{\textheight}}
		\begin{tabular}{lrrrrrrrrrrrr}
			\toprule
			\textbf{Model} & \multicolumn{3}{c}{\textbf{F1$^1$}} & \multicolumn{3}{c}{\textbf{MCC$^1$}} & \multicolumn{3}{c}{\textbf{Recall}} & \multicolumn{3}{c}{\textbf{Lost Evidence (FN)}} \\
			\cmidrule(lr){2-4}\cmidrule(lr){5-7}\cmidrule(lr){8-10}\cmidrule(lr){11-13}
			                         & \textit{RQ2}      & \textit{RQ4}      & $\Delta$                      & \textit{RQ2}       & \textit{RQ4}       & $\Delta$             & \textit{RQ2}      & \textit{RQ4}      & $\Delta$                      & \textit{RQ2}           & \textit{RQ4}      & $\Delta$                \\
			\midrule
			\textbf{GPT-5.5}         & 0.18 [0.07, 0.27] & 0.23 [0.13, 0.31] & +0.05 [-0.03, +0.13]          & 0.12 [0.00, 0.22]  & 0.12 [0.01, 0.20]  & -0.01 [-0.10, +0.09] & 0.11 [0.06, 0.17] & 0.19 [0.11, 0.26] & \textbf{+0.07 [+0.02, +0.14]} & 95 [84, 106]           & 87 [76, 98]       & \textbf{-8 [-15, -2]}   \\
			\textbf{Claude Opus 4.7} & 0.15 [0.05, 0.23] & 0.16 [0.07, 0.24] & +0.01 [-0.03, +0.06]          & 0.16 [0.03, 0.23]  & 0.15 [0.03, 0.22]  & -0.01 [-0.03, +0.03] & 0.07 [0.03, 0.13] & 0.11 [0.06, 0.17] & \textbf{+0.04 [+0.01, +0.08]} & 99 [89, 109]           & 95 [84, 105]      & \textbf{-4 [-8, -1]}    \\
			\textbf{Gemini 3.1 Pro}  & 0.19 [0.09, 0.27] & 0.21 [0.11, 0.29] & +0.02 [-0.03, +0.08]          & 0.12 [0.01, 0.21]  & 0.15 [0.04, 0.23]  & +0.03 [-0.02, +0.08] & 0.14 [0.08, 0.21] & 0.17 [0.10, 0.24] & +0.03 [-0.01, +0.07]          & 92 [81, 103]           & 89 [78, 100]      & -3 [-8, +1]             \\
			\textbf{GLM 5.1}         & 0.13 [0.04, 0.21] & 0.14 [0.05, 0.21] & +0.01 [-0.03, +0.06]          & 0.12 [-0.01, 0.19] & 0.11 [-0.02, 0.19] & -0.01 [-0.07, +0.06] & 0.07 [0.03, 0.13] & 0.08 [0.04, 0.14] & +0.01 [-0.04, +0.06]          & 99 [89, 110]           & 98 [87, 109]      & -1 [-6, +4]             \\
			\textbf{Kimi K2.6}       & 0.15 [0.07, 0.22] & 0.17 [0.08, 0.25] & +0.02 [-0.03, +0.08]          & 0.09 [-0.01, 0.17] & 0.10 [-0.02, 0.19] & +0.01 [-0.08, +0.10] & 0.12 [0.06, 0.18] & 0.14 [0.08, 0.21] & +0.02 [-0.03, +0.07]          & 94 [83, 105]           & 92 [81, 102]      & -2 [-8, +3]             \\
			\textbf{Grok 4.20 Beta}  & 0.17 [0.07, 0.25] & 0.17 [0.08, 0.25] & +0.01 [-0.06, +0.07]          & 0.11 [-0.01, 0.21] & 0.06 [-0.06, 0.14] & -0.05 [-0.15, +0.05] & 0.12 [0.07, 0.18] & 0.14 [0.08, 0.21] & +0.02 [-0.03, +0.07]          & 94 [84, 105]           & 92 [81, 102]      & -2 [-7, +3]             \\
			\textbf{Qwen 3.6 Plus}   & 0.18 [0.09, 0.26] & 0.19 [0.11, 0.28] & +0.02 [-0.03, +0.06]          & 0.13 [0.02, 0.20]  & 0.09 [-0.02, 0.18] & -0.03 [-0.09, +0.02] & 0.17 [0.10, 0.24] & 0.20 [0.13, 0.27] & +0.03 [-0.02, +0.08]          & 89 [78, 100]           & 86 [74, 97]       & -3 [-8, +2]             \\
			\textbf{DeepSeek V4 Pro} & 0.17 [0.07, 0.24] & 0.27 [0.18, 0.34] & \textbf{+0.10 [+0.04, +0.16]} & 0.14 [0.03, 0.21]  & 0.08 [-0.02, 0.17] & -0.05 [-0.12, +0.02] & 0.11 [0.06, 0.17] & 0.31 [0.22, 0.40] & \textbf{+0.20 [+0.12, +0.28]} & 95 [84, 105]           & 74 [63, 86]       & \textbf{-21 [-30, -13]} \\
			\midrule
			\textbf{Avg.}            & 0.16 [0.08, 0.23] & 0.19 [0.11, 0.26] & +0.03 [0.00, +0.07]           & 0.12 [0.02, 0.19]  & 0.11 [0.01, 0.18]  & -0.01 [-0.05, +0.02] & 0.12 [0.07, 0.17] & 0.17 [0.11, 0.23] & \textbf{+0.05 [+0.02, +0.08]} & 95 [85, 105]           & 89 [79, 99]       & \textbf{-6 [-9, -2]}    \\
			\bottomrule
		\end{tabular}%
	\end{adjustbox}
	\footnotesize $^1$ Mean of the per-secondary-study scores (undefined MCC counts as 0). Recall and Lost Evidence are pooled over papers. Avg.\ is the mean over LLMs.
\end{sidewaystable*}

\begin{table*}[htbp]
	\centering
	\caption{Per-study RQ2 vs.\ RQ4 screening metrics (V2 prompt, 5\% sample, updated-criteria subset, averaged across LLMs).}
	\label{tab:rq5_study_comparison}
	\resizebox{\textwidth}{!}{%
		\begin{tabular}{lrrrrrrrrrr}
			\toprule
			\textbf{Study} & \multicolumn{2}{c}{\textbf{Accuracy}} & \multicolumn{2}{c}{\textbf{Precision}} & \multicolumn{2}{c}{\textbf{Recall}} & \multicolumn{2}{c}{\textbf{F1}} & \multicolumn{2}{c}{\textbf{MCC}} \\
			\cmidrule(lr){2-3}\cmidrule(lr){4-5}\cmidrule(lr){6-7}\cmidrule(lr){8-9}\cmidrule(lr){10-11}
			                                                       & \textit{RQ2} & \textit{RQ4} & \textit{RQ2} & \textit{RQ4} & \textit{RQ2} & \textit{RQ4} & \textit{RQ2} & \textit{RQ4} & \textit{RQ2} & \textit{RQ4}                                          \\
			\midrule
			\citet{lenarduzziSystematicLiteratureReview2021a}       & 0.87         & 0.86         & 0.73         & 0.81         & 0.23         & 0.23         & 0.33         & 0.33         & 0.34         & 0.36                                                  \\
			\citet{deckersSystematicLiteratureReview2022a}          & 0.74         & 0.68         & 0.20         & 0.19         & 0.75         & 0.96         & 0.31         & 0.31         & 0.29         & 0.34                                                  \\
			\citet{tebesAnalyzingDocumentingSystematic2020a}        & 0.30         & 0.34         & 0.80         & 0.79         & 0.13         & 0.19         & 0.23         & 0.30         & 0.02         & 0.04                                                  \\
			\citet{zhangTestingVerificationNeuralnetworkbased2020a} & 0.09         & 0.11         & 0.88         & 1.00         & 0.09         & 0.11         & 0.16         & 0.20         & 0.00         & 0.00                                                  \\
			\citet{alonsoSystematicMappingStudy2023b}               & 0.21         & 0.24         & 0.13         & 0.71         & 0.01         & 0.09         & 0.02         & 0.15         & 0.02         & -0.11                                                 \\
			\citet{bjarnasonEmpiricallyBasedModel2023a}             & 0.78         & 0.74         & 0.04         & 0.03         & 1.00         & 1.00         & 0.08         & 0.06         & 0.17         & 0.15                                                  \\
			\citet{rodriguez-perezPerceivedDiversitySoftware2021a}  & 0.94         & 0.94         & 0.06         & 0.00         & 0.02         & 0.00         & 0.03         & 0.00         & 0.02         & -0.01                                                 \\
			\midrule
			\textbf{Avg.}                                          & 0.56         & 0.56         & 0.40         & 0.50         & 0.32         & 0.37         & 0.16         & 0.19         & 0.12         & 0.11                                                  \\
			\bottomrule
		\end{tabular}%
	}
\end{table*}

In analyzing RQ3.2, we noticed that some criteria could have benefited from refinement, and hypothesized that such refinement might lead to improvements in screening accuracy. 
Therefore, in RQ4, we selected studies for criteria refinement based on two factors: low screening accuracy (RQ1) and low agreement in screening criteria (RQ3.1 and RQ3.2). This rests on the assumption that an increase in agreement between LLMs would lead to an increase in accuracy. Next we go over the results in two separate sections. See Tables \ref{table:rq5_ac1_before_after_v3}, \ref{tab:rq5_lost_evidence}, \ref{tab:rq5_vs_rq2}, and \ref{tab:rq5_study_comparison} for the results. The detailed criteria refinement results are available as Online resource 4~\citep{zenodo_research_artifact}.

\subsubsection{Agreement -- Gwet's AC1}

On average, Gwet's AC1 score increased slightly from $0.847$ to $0.866$ after the criteria refinement. However, Gwet's AC1 comparison (Table~\ref{table:rq5_ac1_before_after_v3}) shows that when criteria were reworded, the AC1 score for the reworded criteria increased largely, or had statistical significance in some studies. For example, see the following changes in Table \ref{table:rq5_ac1_before_after_v3}:

\begin{itemize}
    \item ~\citet{rodriguez-perezPerceivedDiversitySoftware2021a}: $EC3 \rightarrow EC2$ and $EC4 \rightarrow EC3$
    \item ~\citet{bjarnasonEmpiricallyBasedModel2023a}: $EC1 \rightarrow IC2$
    \item ~\citet{zhangTestingVerificationNeuralnetworkbased2020a} $IC1$
    \item ~\citet{alonsoSystematicMappingStudy2023b}: $IC3 \rightarrow EC1$
    \item ~\citet{deckersSystematicLiteratureReview2022a}: $EC5$
\end{itemize}

Conversely, the AC1 score decreased notably in~\citet{alonsoSystematicMappingStudy2023b} for $IC2 \rightarrow EC2$, \citet{deckersSystematicLiteratureReview2022a} ($EC2 \rightarrow IC3$), and \citet{tebesAnalyzingDocumentingSystematic2020a} ($IC3 \rightarrow IC2$). The rest of the refined criteria exhibited similar levels of agreement, staying within a tenth of the original AC1. Based on bootstrapping, the change in AC1 score is statistically significant for some criteria -- refining those criteria did either improve or degrade AC1.

For more detailed analysis of study-wise changes, we analyze three studies in more detail~\citep{rodriguez-perezPerceivedDiversitySoftware2021a,bjarnasonEmpiricallyBasedModel2023a,alonsoSystematicMappingStudy2023b}. 

In study \citet{rodriguez-perezPerceivedDiversitySoftware2021a}, criteria IC4, IC5, IC6, IC7, and EC2 were deleted, as they were related to the paper's metadata, not the paper's title and abstract. Most of the reworded criteria exhibited similar levels of performance, e.g. $EC3 \rightarrow EC2$ and $EC4 \rightarrow EC3$ had an improvement in Gwet's AC1 of $+0.031$ and $+0.016$, respectively. The exclusion criterion \textit{EC3} was originally written as: \begin{quote}"We removed discussion papers, i.e., those papers studying the perceived diversity in a psychological, social, or philosophical context because they were not related to SE activities."~\citep{rodriguez-perezPerceivedDiversitySoftware2021a}\end{quote} Then EC3 was reworded as: \begin{quote}"Discussion papers, i.e., those papers studying the perceived diversity in a psychological, social, or philosophical context."\end{quote} This rewording was statistically significant, given the rewording process attempted to preserve the original wording while improving the clarity of the criteria.

In study \citet{bjarnasonEmpiricallyBasedModel2023a}, criteria EC1 was split into IC2 and EC1. The new IC2 was simply written as: "Written in English" (AC1 = $1.000$), which was assumed to have a high level of agreement among LLMs. The other split criterion (EC1) states that \begin{quote}"Articles that merely describe the use of prototypes without considering methodological perspectives such as principles and procedures for the prototyping."\end{quote} has no more ambiguity. As the original Gwet's AC1 score for criterion EC1 was $0.475$, this is clear evidence that criteria that contain multiple conditions benefit significantly by splitting them into parts. While the paper was written in English, it would have skewed the original LLM's reliability.

In study \citet{alonsoSystematicMappingStudy2023b}, EC2 and EC3 were deleted because they were related to the paper's publication year and the published journal, which yet again are impossible to be determined from the title and abstract without having access to the paper's metadata. The rewording exhibited similar behavior compared to study \citet{rodriguez-perezPerceivedDiversitySoftware2021a} and \citet{bjarnasonEmpiricallyBasedModel2023a} in rewording, and moving the criteria from inclusion to exclusion criteria for $IC3 \rightarrow EC1$. IC3 that was previously written as \begin{quote}"Do not mainly comprise software development-related MVPs (e.g., mainly hardware or IoT related MVPs)"~\citep{alonsoSystematicMappingStudy2023b} \end{quote} was reworded and flipped as EC1: \begin{quote}"Comprise other than software development-related Minimum Viable Products (e.g., hardware or IoT related Minimum Viable Products)"\end{quote} Here, expanding acronyms and flipping the inclusion criterion into an exclusion criterion increased Gwet's AC1 from $0.138$ to $0.750$.

\subsubsection{Screening accuracy}

Lost evidence (Table~\ref{tab:rq5_lost_evidence}) indicates that refining the criteria led to a small increase in retaining evidence, i.e. lost evidence reduction (-6), when compared to RQ2. Per-model differences follow the same pattern more or less, where GLM 5.1 exhibited the smallest reduction in lost evidence (-1) compared to DeepSeek V4 Pro (-21), GPT-5.5 (-8) and Claude Opus 4.7 (-4).

F1 and MCC (Table~\ref{tab:rq5_vs_rq2}) show that the average F1 score of the tested LLMs improved slightly from 0.16 to 0.19, whereas MCC decreased from 0.12 to 0.11. Looking at the model-level results, DeepSeek V4 Pro had a statistically significant increase in F1 score from 0.17 to 0.27. The remaining seven LLMs exhibited only a slight improvement in F1 score. MCC increased only for Gemini 3.1 Pro (+0.03) and Kimi K2.6 (+0.01). Recall was observed to improve with every LLM, and DeepSeek V4 Pro having a statistically significant increase of +0.20 along with GPT-5.5 (+0.07) and Claude Opus 4.7 (+0.04). The same was observed for lost evidence, where the three LLMs above had also a reduction in lost evidence that was statistically significant.

The per-study results (Table~\ref{tab:rq5_study_comparison}) were surprising, although the average F1 score improved slightly from 0.16 to 0.19. Average accuracy did not improve, while recall increased from 0.32 to 0.37. Precision also increased from 0.40 to 0.50, which indicates that refining the criteria helped identify more relevant papers and also reduced false positives. MCC degraded by 0.01 on average. At the study level, study \citet{deckersSystematicLiteratureReview2022a} showed the strongest improvement in recall, while study~\citet{lenarduzziSystematicLiteratureReview2021a} showed an increase only in precision and MCC.

\rqbox{
RQ4 showed modest improvement in LLM agreement, in retaining evidence and recall when original screening criteria were refined.
}

\section{Discussion}\label{sec:discussion}

\subsection{LLM performance (RQ1)}


Despite the 18-month gap between the first execution of SESR-Eval and the re-run in this paper, the newer models on the full dataset (RQ1.1) are at best marginally more accurate than the seven earlier ones. On both MCC and F1 metrics, DeepSeek V4 Pro (F1=0.48, MCC=0.41) and Qwen 3.6 Plus (F1=0.48, MCC=0.40) did not reach the F1 and MCC of older LLMs, such as GPT-4.1 mini (F1=0.50, MCC=0.46). Given that these two new models are approximately five times more expensive than GPT-4.1 mini (the best-performing old LLM in terms of MCC alongside Llama 4 Maverick), conducting paper screening on these expensive models is not recommended. In addition, the new models were among the slowest ones tested. However, the results on the SESR-Eval-Mini dataset (RQ1.2) diverged: Claude Opus 4.7 (MCC=0.46) claimed the crown as the best model for paper screening by a narrow margin over GPT-4.1 mini and Llama 4 Maverick (MCC=0.45). As we did not test Claude Opus 4.7 on the full dataset, we cannot rule out that it is genuinely better. However, its per-token costs make screening thousands of papers expensive, so this gain must be weighed.

Taken together, the results suggest that we may be approaching a ceiling, where latency and costs grow faster than screening performance. Claude Opus 4.7 demonstrates that performance improvements are still possible. Part of this latency might be accounted for by the larger models, where additional inference time is required, as was observed for the old LLMs (DeepSeek R1 and o3-mini) in the prior work ~\citep{huotalaSESREvalDatasetEvaluating2025}. The inflated token costs across the field (Table~\ref{tab:companies_and_models}) might also be accounted for by the larger models, increased demand of LLMs or reasoning tokens, which all necessitate additional compute power. All the new LLMs were tested with reasoning turned on, so this itself inflates costs, even if token prices stay at the same level. 

Since the new models do not clearly perform better in paper screening, we might have to look somewhere else for improvement. One possibility is that the abstracts may not provide enough information for LLMs to accurately make the screening decision, which has been previously discussed in \citet{Felizardo_Deizepe_Coutinho_Gomes_Meireles_Gerosa_Steinmacher_2025}. Human reviewers perform paper screening using the same information available, so we would assume that LLMs should be able to complete the same task with acceptable accuracy. However, human reviewers bring domain expertise and knowledge of the SR's goals, which LLMs might have difficulties in. Human reviewers also tend to disagree with one another at this stage, which may cap LLM screening performance regardless of model capability. Providing LLMs with additional context, that may be implicit to human reviewers, about the screening task may therefore be one potential direction for improving screening performance. This would help in keeping costs under control, as larger and newer models tend to have higher per-token costs. 

Agent-based approaches are a second direction worth studying, as the design paradigm for generative AI has gradually been shifting towards agentic AI. In particular, task decomposition could split the screening process into smaller tasks, with each AI agent evaluating a specific inclusion or exclusion criterion. This could improve screening performance by letting each LLM focus more on a single criteria, and aggregating the results in the end. However, multiple AI agents multiply LLM calls per paper, which could further inflate costs. In addition, LLMs could be provided more context about the SR being conducted, so their context aligns better with the interests of the inclusion and exclusion criteria.

\subsection{Effects of prompt design and output structure on screening performance (RQ2)}

Computing the overall screening decision from the criterion-level decision decreased precision, recall, F1, MCC, and accuracy metrics. In addition, the V2 prompt and output structure made LLMs lose more relevant evidence. \citet{adamPromptEngineeringLarge2026} tested 515 different prompts, and had an F1 score of 0.276, recall of 0.836 and precision of 0.185. While we tested two versions of prompts and output schemas (V1 and V2), our prompts achieved a better F1 score of 0.47, but precision and recall (0.62 and 0.38) did not have the same pattern. As previously mentioned, targeting a good precision while maximizing recall is the goal for LLM-based paper screening~\citep{huotalaSESREvalDatasetEvaluating2025}, which our V2 prompt and structure did not yet achieve. 

This would indicate that the user prompt would need even bigger changes compared to the baseline, to reach the goal of good precision with maximized recall. \citet{dennstadtTitleAbstractScreening2024} mentioned that even small changes to the user prompt significantly impacted the screening performance. Our results show that the changes we implemented to the V2 user prompt impacted screening performance ever so slightly. This would indicate that LLMs behave very differently on certain changes made to the prompt, and more radical changes would have been needed, or modifying the system prompt instead of the user prompt, as LLMs tend to respect and follow the system prompt more strictly than the user prompt. 

\citet{matsuiHumanComparableSensitivityLarge2024a} and \citet{shameemExploratoryStudyEffectiveness2026} observed that LLMs classified papers incorrectly because of insufficient domain knowledge, misinterpreted inclusion criteria, or domain-specific literature. While our changes to the V2 included clarifications to the interpretation of inclusion and exclusion criteria, the amount of lost evidence increased slightly and the overall performance degrades. This result aligns with Matsui et al. and Shameem et al., which would indicate that the LLMs had trouble understanding the criteria, even after our changes were meant to clarify their application and improve understanding. \citet{jozwiakREFLECTIVETIABCosteffectivePrompt2026} observed that prompt optimization (which is an area of prompt design) improved recall by +3.7\% to +37.1\% for the LLMs -- in our results we observed an average reduction of -0.03 and between -0.06 and +0.01 for the tested LLMs. Our result is the opposite, and we believe that more radical changes are needed to observe a larger difference.

\subsection{LLM agreement with each other and across inclusion and exclusion criteria (RQ3.1 \& RQ3.2)}

In our experiments, LLMs generally had an acceptable level of agreement on the overall binary include/exclude decisions (AC1=0.864), and on per-criterion decisions (include AC1=0.778, exclude AC1=0.868). In one prior work, \citet{guoAutomatedPaperScreening2024b} agreement between models was also acceptable, with a quite high kappa = 0.96. 
Paradoxically, getting benefits from the Condorcet Jury Theorem, see Section~\ref{sec:background}, would require us to have high correctness (MCC) in individual LLM decisions but low agreement. Such a setup would provide the biggest benefits, as the mistakes by one LLM would not be repeated by another LLM. Unfortunately, it seems that we have low correctness (MCC) and high agreement between LLMs, making the possible gains of combining decisions limited. In another prior work, \citet{thodeExploringUseLLMs2025} used multiple LLMs and included a paper if either LLM would include it. Obviously, this increased recall while decreasing precision, as each paper gets multiple tries to be included.

\subsection{Inclusion and exclusion criteria refinement (RQ4)}

Refining the inclusion and exclusion criteria produced encouraging and informative results. We hypothesized that ambiguous criteria in the secondary studies were degrading screening performance. The mean F1 score increased only by a margin, from 0.16 to 0.19, showing that rewording the criteria does not alone explain the low precision. However, the Gwet's AC1 scoring shows that several criteria became considerably easier for the LLMs to interpret. The criteria that combined multiple conditions made the LLMs' judgment unable to distinguish between the two conditions. For instance, as checking two conditions that a paper was written in English and that it described prototypes without a methodological perspective in a single criterion, the LLM lost the language check outcome completely. Therefore, we strongly recommend that inclusion and exclusion criteria that are used with LLMs to screen papers should be singular, with acronyms explained and written in a form that is easy to understand by the LLMs. Using paper metadata in inclusion and exclusion criteria for paper screening is also not advisable, as the LLMs cannot make that information up without being fed additional context or allowing the LLM to retrieve that metadata from a third party source.

\subsection{Limitations and Threats to Validity}

Next, the limitations and threats to the validity of our work are outlined.

\textbf{Generalization of results.} We benchmarked only SE-specific secondary studies and their primary studies in this work. This was because our prior work~\citep{huotalaSESREvalDatasetEvaluating2025}, which established the gold-standard dataset used, is itself limited to the SE domain. We suspect our findings might not generalize to other domains, such as medicine or health technology assessment, as the conventions of inclusion and exclusion criteria, and the structure of abstracts can differ.

\textbf{Reliability of screening results.} There is always the risk that LLMs can hallucinate and therefore have an effect on the screening results. We have attempted to mitigate this threat by benchmarking multiple LLMs, to minimize the effect of possible hallucinations the LLMs might have in their token generation chains.

\textbf{Only zero-shot prompting technique was tested.} We tested only the zero-shot prompting technique, as the additional prompting strategies would have introduced an additional variable that could disrupt the differences between LLMs. However, our prior work evaluated one-shot (OS), few-shot (FS), and few-shot with chain-of-thought (FS-CoT) prompting techniques, which showed that using prompting strategies improves LLM performance~\citep{huotalaPromiseChallengesUsing2024a}. Our evaluation in this work therefore reflects the "baseline" result, which is assumed to improve with OS, FS and FS-CoT prompting techniques.

\textbf{One LLM did not support structured outputs via native or tool calling.} In this case, we resorted to prompted output for Qwen 3.6 Plus, where the LLM was instructed in-prompt to output JSON directly rather than by tool-calling or outputting natively. This inflated costs, as the schema had to be included in every request, which also might have increased the possibility of malformed outputs. However, we configured Pydantic AI to retry the requests if the output did not match the expected schema, to minimize this threat. Fortunately, most LLMs supported either tool calling or native output.

\textbf{Undefined MCC scores.} We noticed that the scripts we used to compute MCC scores for Tables \ref{tab:rq11_model_metrics_ci_secondary} and \ref{tab:rq12_model_metrics_ci_secondary} (and the per-study Tables \ref{tab:rq11_per_study_metrics} and \ref{tab:rq12_per_study_metrics}) warned us about undefined MCC scores, because two of the studies, \citet{zhangTestingVerificationNeuralnetworkbased2020a} and \citet{pereiraLearningSoftwareConfiguration2021a}, contained no excluded papers, in both the full dataset and the sampled dataset. This makes MCC's denominator zero regardless of the model's prediction. We believe this might have negatively impacted the validity of the MCC if averaged over secondary studies, as the zero values could have dragged down the MCC score by approximately 2/24 (8.3\%). We decided not to filter out the two studies, so that the average is still taken from the n=24 secondary studies. As the possible reduction of the MCC score here is relative for every LLM, this does not change how the LLMs compare against each other.

\textbf{Most new models were run at low reasoning effort.} Six of the eight models in the experiments were run with reasoning effort set to low, with Qwen 3.6 Plus at medium and DeepSeek V4 Pro at high. The choice was motivated by prior evidence that the reasoning setting brings negligible gains in screening performance~\citep{huotalaSESREvalDatasetEvaluating2025}. However, the evidence predates the models evaluated in this paper, and reasoning effort may have different effects for newer models. Our results for the low-effort models should be read as a lower bound rather than their peak performance. Notably, the two models (Qwen 3.6 Plus and DeepSeek V4 Pro) run at different reasoning settings did not have meaningful improvement in screening performance.

\textbf{Possibility of training data contamination.} The SESR-Eval dataset was released in October 2025, so models released after that date may have been trained on the dataset. However, the data from the secondary studies might be in the training data, as they were released years before the dataset itself. As we observed only marginal performance improvements for the newer models, we believe this threat is unlikely to have affected our benchmark results. It cannot be fully ruled out, however, as most LLM vendors do not disclose what data the models have been trained on.

\textbf{Incorrect ground truths in SESR-Eval.}
It is likely that some ground truth labels in SESR-Eval are incorrect, as the original secondary study authors may have also made mistakes. Furthermore, it is also possible that not all decisions in the SESR-Eval dataset were made based solely on the title and abstract, even though SESR-Eval purposefully sourced studies where such decision points were visible. However, we think it is highly unlikely that such issues alone would explain the rather limited LLM screening performance. Further support that LLM mistakes in screening software engineering papers are real -- and not an artifact of poor ground truth -- is provided by \citet{mantyla2026understanding}, where the original secondary study authors screened new papers manually, compared their results with LLM decisions, and analyzed the reasons for LLM screening mistakes. In that study, as in this paper, LLM performance in the screening task was modest.

\textbf{Refinement of inclusion and exclusion criteria.} The refined inclusion and exclusion criteria are our interpretation of the original authors' intent. There is a risk that the rewritten or split criteria might not match the original author's intent. We mitigated this threat by limiting the changes to: 1) expanding abbreviations and acronyms, which does not affect the original meaning of the criterion; 2) splitting compound criteria into separate rules, if there is a distinct separation between them; 3) removing criteria that cannot be verified from the title or abstract, since we do not want the LLMs to make up the information and guess; 4) converting negated criteria to affirmative form, which LLMs generally interpret more reliably; 5) preserving the original wording as much as possible; and 6) clarifying vague terms in the criteria using definitions straight from the paper. In addition, all deleted, reworded, and added rules were reviewed by two of the authors. Nevertheless, as the refinement process was a manual and laborious process, human error cannot be ruled out.

\section{Conclusions and future work}

\label{sec:conclusions}

In this paper, we extended the SESR-Eval~\citep{huotalaSESREvalDatasetEvaluating2025} study considerably by benchmarking multiple new LLMs for paper screening, creating the new SESR-Eval-Mini dataset, investigating if prompt design and output structuring affected screening performance, analyzing agreement in screening decisions across LLMs and across inclusion and exclusion criteria, and finally, examining whether refining the inclusion and exclusion criteria had any effect on screening performance.

Despite the 18-month gap between \citet{huotalaSESREvalDatasetEvaluating2025} and this work, we found at most marginal improvements in LLM screening performance. This suggests that the publicly reported advancements of LLMs do not translate into improved screening performance for software engineering papers. Thus, the evidence indicates we must state that LLM progress has largely stalled in this respect.

Differences in screening performance are larger between secondary studies than between LLMs, and also with the new LLMs, meaning that the performance varies more based on the study than the used model. This means that while benchmarking new LLMs in screening tasks, it is important to test a broad selection of secondary studies.

Following the finding that differences between secondary studies are large, we improved the screening rules for certain secondary studies that performed poorly in screening, but found only modest improvements in screening performance. As improving screening rules makes little difference, this means that secondary study differences are likely to stem from other sources. One such source to be investigated in future studies is the primary studies that each secondary study screens. It is quite possible that some datasets (sets of primary studies) are more difficult to screen correctly than others.

Taken together, LLMs are still not reliable enough for replacing humans in the paper screening process of SRs. However, we feel that LLMs can still be a useful companion in paper screening. The current bottleneck of paper screening of LLMs is not the raw compute and model capabilities, but how much context paper abstracts and the prompts contain, and in what kind of prompt the screening task is provided to the LLMs. 

As future work, we are specifically looking into designing a distributed screening architecture, where inclusion and exclusion criteria are split between multiple small language model (SLM) agents. We believe this approach has high potential in increasing screening accuracy, as each SLM agent would focus on evaluating a single criterion rather than juggling between varying amounts of criteria simultaneously while trying to arrive at a decision from a single piece of text. As another future work, we are looking into ways that would maximize recall with the usage of probability thresholds, which would open the possibility of using LLMs to filter out studies before human screening. Filtering would be especially valuable in review contexts where the terminology used is not settled, as database searches often yield thousands of results for human screening resulting in high amounts of manual work and excluded papers. Lastly, we are looking at investigating the effect of other prompting techniques in a more robust manner, and the combination of few-shot (FS) and chain-of-thought (CoT) prompting together with multi-agent systems.

\section*{Data availability}

The scripts, data, and other materials used to conduct this work are available in the research artifact \citep{zenodo_research_artifact}.

\backmatter

\bmhead{Acknowledgments}

The first and third authors have been supported by the Strategic Research Council of Research Council of Finland
(Grant ID 358471).

\section*{Declarations}

The authors declare that they have no conflict of interest.

\bibliography{main}


\begin{thebibliography}{71}
\ifx \bisbn   \undefined \def \bisbn  #1{ISBN #1}\fi
\ifx \binits  \undefined \def \binits#1{#1}\fi
\ifx \bauthor  \undefined \def \bauthor#1{#1}\fi
\ifx \batitle  \undefined \def \batitle#1{#1}\fi
\ifx \bjtitle  \undefined \def \bjtitle#1{#1}\fi
\ifx \bvolume  \undefined \def \bvolume#1{\textbf{#1}}\fi
\ifx \byear  \undefined \def \byear#1{#1}\fi
\ifx \bissue  \undefined \def \bissue#1{#1}\fi
\ifx \bfpage  \undefined \def \bfpage#1{#1}\fi
\ifx \blpage  \undefined \def \blpage #1{#1}\fi
\ifx \burl  \undefined \def \burl#1{\textsf{#1}}\fi
\ifx \doiurl  \undefined \def \doiurl#1{\url{https://doi.org/#1}}\fi
\ifx \betal  \undefined \def \betal{\textit{et al.}}\fi
\ifx \binstitute  \undefined \def \binstitute#1{#1}\fi
\ifx \binstitutionaled  \undefined \def \binstitutionaled#1{#1}\fi
\ifx \bctitle  \undefined \def \bctitle#1{#1}\fi
\ifx \beditor  \undefined \def \beditor#1{#1}\fi
\ifx \bpublisher  \undefined \def \bpublisher#1{#1}\fi
\ifx \bbtitle  \undefined \def \bbtitle#1{#1}\fi
\ifx \bedition  \undefined \def \bedition#1{#1}\fi
\ifx \bseriesno  \undefined \def \bseriesno#1{#1}\fi
\ifx \blocation  \undefined \def \blocation#1{#1}\fi
\ifx \bsertitle  \undefined \def \bsertitle#1{#1}\fi
\ifx \bsnm \undefined \def \bsnm#1{#1}\fi
\ifx \bsuffix \undefined \def \bsuffix#1{#1}\fi
\ifx \bparticle \undefined \def \bparticle#1{#1}\fi
\ifx \barticle \undefined \def \barticle#1{#1}\fi
\bibcommenthead
\ifx \bconfdate \undefined \def \bconfdate #1{#1}\fi
\ifx \botherref \undefined \def \botherref #1{#1}\fi
\ifx \url \undefined \def \url#1{\textsf{#1}}\fi
\ifx \bchapter \undefined \def \bchapter#1{#1}\fi
\ifx \bbook \undefined \def \bbook#1{#1}\fi
\ifx \bcomment \undefined \def \bcomment#1{#1}\fi
\ifx \oauthor \undefined \def \oauthor#1{#1}\fi
\ifx \citeauthoryear \undefined \def \citeauthoryear#1{#1}\fi
\ifx \endbibitem  \undefined \def \endbibitem {}\fi
\ifx \bconflocation  \undefined \def \bconflocation#1{#1}\fi
\ifx \arxivurl  \undefined \def \arxivurl#1{\textsf{#1}}\fi
\csname PreBibitemsHook\endcsname

\bibitem[\protect\citeauthoryear{Aubakirova et~al.}{2026}]{openrouter}
\begin{botherref}
\oauthor{\bsnm{Aubakirova}, \binits{M.}},
\oauthor{\bsnm{Atallah}, \binits{A.}},
\oauthor{\bsnm{Clark}, \binits{C.}},
\oauthor{\bsnm{Summerville}, \binits{J.}},
\oauthor{\bsnm{Midha}, \binits{A.}}:
State of AI: An Empirical 100 Trillion Token Study with OpenRouter.
arXiv
(2026).
\doiurl{10.48550/ARXIV.2601.10088}
\end{botherref}
\endbibitem

\bibitem[\protect\citeauthoryear{Adam et~al.}{2026}]{adamPromptEngineeringLarge2026}
\begin{botherref}
\oauthor{\bsnm{Adam}, \binits{T.J.}},
\oauthor{\bsnm{Abosabie}, \binits{S.A.S.}},
\oauthor{\bsnm{Dittmer}, \binits{M.}},
\oauthor{\bsnm{Wolf}, \binits{E.}},
\oauthor{\bsnm{Abosabie}, \binits{S.A.}},
\oauthor{\bsnm{Behnke}, \binits{C.}},
\oauthor{\bsnm{Baier}, \binits{F.}},
\oauthor{\bsnm{Weickmann}, \binits{A.}},
\oauthor{\bsnm{K\"{o}ser}, \binits{L.}},
\oauthor{\bsnm{Correll}, \binits{C.U.}},
\oauthor{\bsnm{Rutsch}, \binits{N.}}:
Prompt engineering of large language models for paper screening in medical meta-analyses and systematic reviews: {A} prospective comparative study.
Research Synthesis Methods,
1--18
(2026)
\doiurl{10.1017/rsm.2026.10093}
\end{botherref}
\endbibitem

\bibitem[\protect\citeauthoryear{Alonso et~al.}{2023}]{alonsoSystematicMappingStudy2023b}
\begin{barticle}
\bauthor{\bsnm{Alonso}, \binits{S.}},
\bauthor{\bsnm{Kalinowski}, \binits{M.}},
\bauthor{\bsnm{Ferreira}, \binits{B.}},
\bauthor{\bsnm{Barbosa}, \binits{S.D.J.}},
\bauthor{\bsnm{Lopes}, \binits{H.}}:
\batitle{A systematic mapping study and practitioner insights on the use of software engineering practices to develop {{MVPs}}}.
\bjtitle{Information and Software Technology}
\bvolume{156},
\bfpage{107144}
(\byear{2023})
\doiurl{10.1016/j.infsof.2022.107144}
\end{barticle}
\endbibitem

\bibitem[\protect\citeauthoryear{Abramowitz et~al.}{2025}]{abramowitzWhoReviewsReviewers2025}
\begin{bchapter}
\bauthor{\bsnm{Abramowitz}, \binits{B.}},
\bauthor{\bsnm{Lev}, \binits{O.}},
\bauthor{\bsnm{Mattei}, \binits{N.}}:
\bctitle{Who {Reviews} {The} {Reviewers}? {A} {Multi}-{Level} {Jury} {Problem}}.
In: \bbtitle{Proceedings of the 24th {International} {Conference} on {Autonomous} {Agents} and {Multiagent} {Systems}}.
\bsertitle{{AAMAS} '25},
pp. \bfpage{14}--\blpage{22}.
\bpublisher{International Foundation for Autonomous Agents and Multiagent Systems},
\blocation{Richland, SC}
(\byear{2025}).
\doiurl{10.5555/3709347.3743512}
\end{bchapter}
\endbibitem

\bibitem[\protect\citeauthoryear{Afonso et~al.}{2026}]{afonso2026profolaf}
\begin{bchapter}
\bauthor{\bsnm{Afonso}, \binits{M.}},
\bauthor{\bsnm{Saavedra}, \binits{N.}},
\bauthor{\bsnm{Louren{\c{c}}o}, \binits{B.}},
\bauthor{\bsnm{Mendes}, \binits{A.}},
\bauthor{\bsnm{Ferreira}, \binits{J.F.}}:
\bctitle{{{ProfOlaf}}: {{Semi-automated}} {{Tool}} for {{Systematic}} {{Literature}} {{Reviews}}}.
In: \bbtitle{Proceedings of the 34th ACM International Conference on the Foundations of Software Engineering}.
\bsertitle{{{FSE}} {{Companion}} '26},
pp. \bfpage{262}--\blpage{266}.
\bpublisher{ACM},
\blocation{New York, NY, USA}
(\byear{2026}).
\doiurl{10.1145/3803437.3806432}
\end{bchapter}
\endbibitem

\bibitem[\protect\citeauthoryear{Bankier}{1988}]{bankierPowerAllocationsDetermining1988}
\begin{barticle}
\bauthor{\bsnm{Bankier}, \binits{M.D.}}:
\batitle{Power {Allocations}: {Determining} {Sample} {Sizes} for {Subnational} {Areas}}.
\bjtitle{The American Statistician}
\bvolume{42}(\bissue{3}),
\bfpage{174}--\blpage{177}
(\byear{1988})
\doiurl{10.2307/2684995}
\end{barticle}
\endbibitem

\bibitem[\protect\citeauthoryear{Bjarnason et~al.}{2023}]{bjarnasonEmpiricallyBasedModel2023a}
\begin{barticle}
\bauthor{\bsnm{Bjarnason}, \binits{E.}},
\bauthor{\bsnm{Lang}, \binits{F.}},
\bauthor{\bsnm{Mj{\"o}berg}, \binits{A.}}:
\batitle{An empirically based model of software prototyping: A mapping study and a multi-case study}.
\bjtitle{Empirical Software Engineering}
\bvolume{28}(\bissue{5}),
\bfpage{115}
(\byear{2023})
\doiurl{10.1007/s10664-023-10331-w}
\end{barticle}
\endbibitem

\bibitem[\protect\citeauthoryear{Behrouzian et~al.}{2026}]{behrouzianLinguisticPolarityDecision2026}
\begin{botherref}
\oauthor{\bsnm{Behrouzian}, \binits{A.M.}},
\oauthor{\bsnm{Meleti}, \binits{M.}},
\oauthor{\bsnm{Colangelo}, \binits{M.T.}},
\oauthor{\bsnm{Calciolari}, \binits{E.}},
\oauthor{\bsnm{Galli}, \binits{C.}}:
Linguistic {Polarity} and {Decision} {Architecture} in {LLM}-{Based} {Abstract} {Screening} for {Systematic} {Reviews}.
Information
\textbf{17}(5)
(2026)
\doiurl{10.3390/info17050449}
\end{botherref}
\endbibitem

\bibitem[\protect\citeauthoryear{Bari{\v s}i{\'c} et~al.}{2022}]{barisicMultiparadigmModelingCyber2022a}
\begin{barticle}
\bauthor{\bsnm{Bari{\v s}i{\'c}}, \binits{A.}},
\bauthor{\bsnm{Ruchkin}, \binits{I.}},
\bauthor{\bsnm{Savi{\'c}}, \binits{D.}},
\bauthor{\bsnm{Mohamed}, \binits{M.A.}},
\bauthor{\bsnm{{Al-Ali}}, \binits{R.}},
\bauthor{\bsnm{Li}, \binits{L.W.}},
\bauthor{\bsnm{Mkaouar}, \binits{H.}},
\bauthor{\bsnm{Eslampanah}, \binits{R.}},
\bauthor{\bsnm{Challenger}, \binits{M.}},
\bauthor{\bsnm{Blouin}, \binits{D.}},
\bauthor{\bsnm{Nikiforova}, \binits{O.}},
\bauthor{\bsnm{Cicchetti}, \binits{A.}}:
\batitle{Multi-paradigm modeling for cyber--physical systems: {{A}} systematic mapping review}.
\bjtitle{Journal of Systems and Software}
\bvolume{183},
\bfpage{111081}
(\byear{2022})
\doiurl{10.1016/j.jss.2021.111081}
\end{barticle}
\endbibitem

\bibitem[\protect\citeauthoryear{Casalaro et~al.}{2022}]{casalaroModeldrivenEngineeringMobile2022a}
\begin{barticle}
\bauthor{\bsnm{Casalaro}, \binits{G.L.}},
\bauthor{\bsnm{Cattivera}, \binits{G.}},
\bauthor{\bsnm{Ciccozzi}, \binits{F.}},
\bauthor{\bsnm{Malavolta}, \binits{I.}},
\bauthor{\bsnm{Wortmann}, \binits{A.}},
\bauthor{\bsnm{Pelliccione}, \binits{P.}}:
\batitle{Model-driven engineering for mobile robotic systems: A systematic mapping study}.
\bjtitle{Software and Systems Modeling}
\bvolume{21}(\bissue{1}),
\bfpage{19}--\blpage{49}
(\byear{2022})
\doiurl{10.1007/s10270-021-00908-8}
\end{barticle}
\endbibitem

\bibitem[\protect\citeauthoryear{Chiang and Lee}{2023}]{Chiang_Lee_2023}
\begin{botherref}
\oauthor{\bsnm{Chiang}, \binits{C.-H.}},
\oauthor{\bsnm{Lee}, \binits{H.-y.}}:
Can Large Language Models Be an Alternative to Human Evaluations?
Association for Computational Linguistics
(2023).
\doiurl{10.18653/v1/2023.acl-long.870}
\end{botherref}
\endbibitem

\bibitem[\protect\citeauthoryear{Deizepe et~al.}{2026}]{deizepe2026ailitreview}
\begin{bchapter}
\bauthor{\bsnm{Deizepe}, \binits{A.}},
\bauthor{\bsnm{Felizardo}, \binits{K.R.}},
\bauthor{\bsnm{Pereira}, \binits{R.}}:
\bctitle{{AI LitReview}: A web-based tool for {LLM}-assisted systematic literature reviews}.
In: \bbtitle{Proceedings of the XL Simp{\'{o}}sio Brasileiro de Engenharia de Software (SBES 2026)}
(\byear{2026})
\end{bchapter}
\endbibitem

\bibitem[\protect\citeauthoryear{{Delgado-Chaves} et~al.}{2025}]{delgado-chavesTransformingLiteratureScreening2025}
\begin{barticle}
\bauthor{\bsnm{{Delgado-Chaves}}, \binits{F.M.}},
\bauthor{\bsnm{Jennings}, \binits{M.J.}},
\bauthor{\bsnm{Atalaia}, \binits{A.}},
\bauthor{\bsnm{Wolff}, \binits{J.}},
\bauthor{\bsnm{Horvath}, \binits{R.}},
\bauthor{\bsnm{Mamdouh}, \binits{Z.M.}},
\bauthor{\bsnm{Baumbach}, \binits{J.}},
\bauthor{\bsnm{Baumbach}, \binits{L.}}:
\batitle{Transforming literature screening: {{The}} emerging role of large language models in systematic reviews}.
\bjtitle{Proceedings of the National Academy of Sciences}
\bvolume{122}(\bissue{2}),
\bfpage{2411962122}
(\byear{2025})
\doiurl{10.1073/pnas.2411962122}
\end{barticle}
\endbibitem

\bibitem[\protect\citeauthoryear{Deckers and Lago}{2022}]{deckersSystematicLiteratureReview2022a}
\begin{barticle}
\bauthor{\bsnm{Deckers}, \binits{R.}},
\bauthor{\bsnm{Lago}, \binits{P.}}:
\batitle{Systematic literature review of domain-oriented specification techniques}.
\bjtitle{Journal of Systems and Software}
\bvolume{192},
\bfpage{111415}
(\byear{2022})
\doiurl{10.1016/j.jss.2022.111415}
\end{barticle}
\endbibitem

\bibitem[\protect\citeauthoryear{Du et~al.}{2024}]{duMachineLearningModels2024}
\begin{barticle}
\bauthor{\bsnm{Du}, \binits{J.}},
\bauthor{\bsnm{Soysal}, \binits{E.}},
\bauthor{\bsnm{Wang}, \binits{D.}},
\bauthor{\bsnm{He}, \binits{L.}},
\bauthor{\bsnm{Lin}, \binits{B.}},
\bauthor{\bsnm{Wang}, \binits{J.}},
\bauthor{\bsnm{Manion}, \binits{F.J.}},
\bauthor{\bsnm{Li}, \binits{Y.}},
\bauthor{\bsnm{Wu}, \binits{E.}},
\bauthor{\bsnm{Yao}, \binits{L.}}:
\batitle{Machine learning models for abstract screening task - {A} systematic literature review application for health economics and outcome research}.
\bjtitle{BMC Medical Research Methodology}
\bvolume{24},
\bfpage{108}
(\byear{2024})
\doiurl{10.1186/s12874-024-02224-3}
\end{barticle}
\endbibitem

\bibitem[\protect\citeauthoryear{Dennst{\"a}dt et~al.}{2024}]{dennstadtTitleAbstractScreening2024}
\begin{barticle}
\bauthor{\bsnm{Dennst{\"a}dt}, \binits{F.}},
\bauthor{\bsnm{Zink}, \binits{J.}},
\bauthor{\bsnm{Putora}, \binits{P.M.}},
\bauthor{\bsnm{Hastings}, \binits{J.}},
\bauthor{\bsnm{Cihoric}, \binits{N.}}:
\batitle{Title and abstract screening for literature reviews using large language models: An exploratory study in the biomedical domain}.
\bjtitle{Systematic Reviews}
\bvolume{13}(\bissue{1}),
\bfpage{158}
(\byear{2024})
\doiurl{10.1186/s13643-024-02575-4}
\end{barticle}
\endbibitem

\bibitem[\protect\citeauthoryear{Efron and Tibshirani}{1994}]{Efron_Tibshirani_1994}
\begin{bbook}
\bauthor{\bsnm{Efron}, \binits{B.}},
\bauthor{\bsnm{Tibshirani}, \binits{R.J.}}:
\bbtitle{An Introduction to the Bootstrap},
\bedition{1}st edn.,
pp. \bfpage{1}--\blpage{456}.
\bpublisher{Chapman and Hall/CRC},
\blocation{New York, NY, USA}
(\byear{1994}).
\doiurl{10.1201/9780429246593}
\end{bbook}
\endbibitem

\bibitem[\protect\citeauthoryear{Ebrahimi et~al.}{2021}]{ebrahimiMobileAppPrivacy2021a}
\begin{barticle}
\bauthor{\bsnm{Ebrahimi}, \binits{F.}},
\bauthor{\bsnm{Tushev}, \binits{M.}},
\bauthor{\bsnm{Mahmoud}, \binits{A.}}:
\batitle{Mobile app privacy in software engineering research: {{A}} systematic mapping study}.
\bjtitle{Information and Software Technology}
\bvolume{133},
\bfpage{106466}
(\byear{2021})
\doiurl{10.1016/j.infsof.2020.106466}
\end{barticle}
\endbibitem

\bibitem[\protect\citeauthoryear{Felizardo et~al.}{2025}]{Felizardo_Deizepe_Coutinho_Gomes_Meireles_Gerosa_Steinmacher_2025}
\begin{bchapter}
\bauthor{\bsnm{Felizardo}, \binits{K.R.}},
\bauthor{\bsnm{Deizepe}, \binits{A.}},
\bauthor{\bsnm{Coutinho}, \binits{D.}},
\bauthor{\bsnm{Gomes}, \binits{G.}},
\bauthor{\bsnm{Meireles}, \binits{M.}},
\bauthor{\bsnm{Gerosa}, \binits{M.}},
\bauthor{\bsnm{Steinmacher}, \binits{I.}}:
\bctitle{{{On}} the {{Difficulties}} of {{Conducting}} and {{Replicating}} {{Systematic}} {{Literature}} {{Reviews}} {{Studies}} {{Using}} {{LLMs}} in {{Software}} {{Engineering}}}.
In: \bbtitle{2025 IEEE/ACM International Workshop on Methodological Issues with Empirical Studies in Software Engineering (WSESE)},
pp. \bfpage{20}--\blpage{23}.
\bpublisher{IEEE},
\blocation{Ottawa, ON, Canada}
(\byear{2025}).
\doiurl{10.1109/WSESE66602.2025.00010}
\end{bchapter}
\endbibitem

\bibitem[\protect\citeauthoryear{Felizardo et~al.}{2024}]{felizardoChatGPTApplicationSystematic2024}
\begin{bchapter}
\bauthor{\bsnm{Felizardo}, \binits{K.R.}},
\bauthor{\bsnm{Lima}, \binits{M.S.}},
\bauthor{\bsnm{Deizepe}, \binits{A.}},
\bauthor{\bsnm{Conte}, \binits{T.U.}},
\bauthor{\bsnm{Steinmacher}, \binits{I.}}:
\bctitle{{{ChatGPT}} application in {{Systematic Literature Reviews}} in {{Software Engineering}}: An evaluation of its accuracy to support the selection activity}.
In: \bbtitle{Proceedings of the 18th {{ACM}}/{{IEEE}} {{International}} {{Symposium}} on {{Empirical}} {{Software}} {{Engineering}} and {{Measurement}}}.
\bsertitle{{{ESEM}} '24},
pp. \bfpage{25}--\blpage{36}.
\bpublisher{Association for Computing Machinery},
\blocation{New York, NY, USA}
(\byear{2024}).
\doiurl{10.1145/3674805.3686666}
\end{bchapter}
\endbibitem

\bibitem[\protect\citeauthoryear{Guo et~al.}{2024}]{guoAutomatedPaperScreening2024b}
\begin{barticle}
\bauthor{\bsnm{Guo}, \binits{E.}},
\bauthor{\bsnm{Gupta}, \binits{M.}},
\bauthor{\bsnm{Deng}, \binits{J.}},
\bauthor{\bsnm{Park}, \binits{Y.-J.}},
\bauthor{\bsnm{Paget}, \binits{M.}},
\bauthor{\bsnm{Naugler}, \binits{C.}}:
\batitle{Automated {{Paper Screening}} for {{Clinical Reviews Using Large Language Models}}: {{Data Analysis Study}}}.
\bjtitle{Journal of Medical Internet Research}
\bvolume{26}(\bissue{1}),
\bfpage{48996}
(\byear{2024})
\doiurl{10.2196/48996}
\end{barticle}
\endbibitem

\bibitem[\protect\citeauthoryear{Geraldi et~al.}{2020}]{geraldiSoftwareProductLine2020a}
\begin{barticle}
\bauthor{\bsnm{Geraldi}, \binits{R.T.}},
\bauthor{\bsnm{Reinehr}, \binits{S.}},
\bauthor{\bsnm{Malucelli}, \binits{A.}}:
\batitle{Software product line applied to the internet of things: {{A}} systematic literature review}.
\bjtitle{Information and Software Technology}
\bvolume{124},
\bfpage{106293}
(\byear{2020})
\doiurl{10.1016/j.infsof.2020.106293}
\end{barticle}
\endbibitem

\bibitem[\protect\citeauthoryear{Gwet}{2008}]{gwet2008computing}
\begin{barticle}
\bauthor{\bsnm{Gwet}, \binits{K.L.}}:
\batitle{Computing inter-rater reliability and its variance in the presence of high agreement}.
\bjtitle{British Journal of Mathematical and Statistical Psychology}
\bvolume{61}(\bissue{1}),
\bfpage{29}--\blpage{48}
(\byear{2008})
\doiurl{10.1348/000711006X126600}
\end{barticle}
\endbibitem

\bibitem[\protect\citeauthoryear{Huotala et~al.}{2025a}]{huotalaSESREvalDatasetEvaluating2025}
\begin{bchapter}
\bauthor{\bsnm{Huotala}, \binits{A.}},
\bauthor{\bsnm{Kuutila}, \binits{M.}},
\bauthor{\bsnm{M{\"a}ntyl{\"a}}, \binits{M.}}:
\bctitle{{{SESR-Eval}}: {{Dataset}} for {{Evaluating LLMs}} in the {{Title-Abstract Screening}} of {{Systematic Reviews}}}.
In: \bbtitle{Proceedings of the 19th {{ACM}}/{{IEEE}} {{International}} {{Symposium}} on {{Empirical}} {{Software}} {{Engineering}} and {{Measurement}}}.
\bsertitle{{{ESEM}} '25},
pp. \bfpage{1}--\blpage{12}.
\bpublisher{IEEE},
\blocation{Piscataway, NJ, USA}
(\byear{2025}).
\doiurl{10.48550/arXiv.2507.19027}
\end{bchapter}
\endbibitem

\bibitem[\protect\citeauthoryear{Huotala et~al.}{2025b}]{huotala2025research}
\begin{barticle}
\bauthor{\bsnm{Huotala}, \binits{A.}},
\bauthor{\bsnm{Kuutila}, \binits{M.}},
\bauthor{\bsnm{M{\"a}ntyl{\"a}}, \binits{M.}}:
\batitle{Research artifacts in secondary studies: A systematic mapping in software engineering}.
\bjtitle{Information and Software Technology}
\bvolume{187},
\bfpage{107830}
(\byear{2025})
\doiurl{10.1016/j.infsof.2025.107830}
\end{barticle}
\endbibitem

\bibitem[\protect\citeauthoryear{Huotala et~al.}{2026b}]{zenodo_research_artifact}
\begin{botherref}
\oauthor{\bsnm{Huotala}, \binits{A.}},
\oauthor{\bsnm{Kuutila}, \binits{M.}},
\oauthor{\bsnm{Mäntylä}, \binits{M.}}:
{{Dataset}} for: {{Has}} {{LLM}} {{Screening}} {{Performance}} {{Stalled}} in {{Systematic}} {{Reviews?}}
Zenodo
(2026).
\doiurl{10.5281/zenodo.23117229}
\end{botherref}
\endbibitem

\bibitem[\protect\citeauthoryear{Huotala et~al.}{2026c}]{sesr_eval_research_artifact}
\begin{botherref}
\oauthor{\bsnm{Huotala}, \binits{A.}},
\oauthor{\bsnm{Kuutila}, \binits{M.}},
\oauthor{\bsnm{M{\"a}ntyl{\"a}}, \binits{M.}}:
{{Dataset}} for: {{SESR-Eval}}: {{Dataset}} for {{Evaluating}} {{LLMs}} in the {{Title-Abstract}} {{Screening}} of {{Systematic}} {{Reviews}}.
Zenodo
(2026).
\doiurl{10.5281/zenodo.23205534}
\end{botherref}
\endbibitem

\bibitem[\protect\citeauthoryear{Huotala et~al.}{2024}]{huotalaPromiseChallengesUsing2024a}
\begin{bchapter}
\bauthor{\bsnm{Huotala}, \binits{A.}},
\bauthor{\bsnm{Kuutila}, \binits{M.}},
\bauthor{\bsnm{Ralph}, \binits{P.}},
\bauthor{\bsnm{M{\"a}ntyl{\"a}}, \binits{M.}}:
\bctitle{The {{Promise}} and {{Challenges}} of {{Using LLMs}} to {{Accelerate}} the {{Screening Process}} of {{Systematic Reviews}}}.
In: \bbtitle{Proceedings of the 28th {{International}} {{Conference}} on {{Evaluation}} and {{Assessment}} in {{Software Engineering}}}.
\bsertitle{{{EASE}} '24},
pp. \bfpage{262}--\blpage{271}.
\bpublisher{Association for Computing Machinery},
\blocation{New York, NY, USA}
(\byear{2024}).
\doiurl{10.1145/3661167.3661172}
\end{bchapter}
\endbibitem

\bibitem[\protect\citeauthoryear{Huotala et~al.}{2026a}]{huotalaFSE2026}
\begin{bchapter}
\bauthor{\bsnm{Huotala}, \binits{A.}},
\bauthor{\bsnm{Kuutila}, \binits{M.}},
\bauthor{\bsnm{Turtio}, \binits{O.-P.}},
\bauthor{\bsnm{Sipil{\"a}}, \binits{S.}},
\bauthor{\bsnm{M{\"a}ntyl{\"a}}, \binits{M.}}:
\bctitle{{{AISysRev}} - {{LLM-based}} {{Tool}} for {{Title-Abstract Screening}}}.
In: \bbtitle{Proceedings of the 34th ACM International Conference on the Foundations of Software Engineering}.
\bsertitle{{{FSE}} {{Companion}} '26},
pp. \bfpage{142}--\blpage{146}.
\bpublisher{ACM},
\blocation{New York, NY, USA}
(\byear{2026}).
\doiurl{10.1145/3803437.3806408}
\end{bchapter}
\endbibitem

\bibitem[\protect\citeauthoryear{Hamel et~al.}{2020}]{hamelEvaluationDistillerSRsMachine2020}
\begin{barticle}
\bauthor{\bsnm{Hamel}, \binits{C.}},
\bauthor{\bsnm{Kelly}, \binits{S.E.}},
\bauthor{\bsnm{Thavorn}, \binits{K.}},
\bauthor{\bsnm{Rice}, \binits{D.B.}},
\bauthor{\bsnm{Wells}, \binits{G.A.}},
\bauthor{\bsnm{Hutton}, \binits{B.}}:
\batitle{An evaluation of {DistillerSR}'s machine learning-based prioritization tool for title/abstract screening -- impact on reviewer-relevant outcomes}.
\bjtitle{BMC Medical Research Methodology}
\bvolume{20}(\bissue{1}),
\bfpage{256}
(\byear{2020})
\doiurl{10.1186/s12874-020-01129-1}
\end{barticle}
\endbibitem

\bibitem[\protect\citeauthoryear{J\'ozwiak et~al.}{2026}]{jozwiakREFLECTIVETIABCosteffectivePrompt2026}
\begin{barticle}
\bauthor{\bsnm{J\'ozwiak}, \binits{A.}},
\bauthor{\bsnm{Imre}, \binits{A.}},
\bauthor{\bsnm{Hagym\'asy}, \binits{J.}},
\bauthor{\bsnm{Tittmann}, \binits{J.}},
\bauthor{\bsnm{Nagy}, \binits{A.}},
\bauthor{\bsnm{Kov\'acs}, \binits{S.}},
\bauthor{\bsnm{Kardas}, \binits{P.}},
\bauthor{\bsnm{{van Boven}}, \binits{J.F.}},
\bauthor{\bsnm{Mommers}, \binits{I.}},
\bauthor{\bsnm{\'Agh}, \binits{T.}}:
\batitle{{REFLECTIVE}-{TIAB}: cost-effective prompt optimization for large language model-based title and abstract screening in literature reviews}.
\bjtitle{Expert Review of Pharmacoeconomics \& Outcomes Research}
\bvolume{26}(\bissue{6}),
\bfpage{747}--\blpage{755}
(\byear{2026})
\doiurl{10.1080/14737167.2026.2691995}
\end{barticle}
\endbibitem

\bibitem[\protect\citeauthoryear{Kitchenham and Charters}{2007}]{kitchenhamGuidelinesPerformingSystematic2007}
\begin{botherref}
\oauthor{\bsnm{Kitchenham}, \binits{B.}},
\oauthor{\bsnm{Charters}, \binits{S.}}:
{{Guidelines}} for {{Performing}} {{Systematic}} {{Literature}} {{Reviews}} in {{Software Engineering}}.
Technical report,
Keele University
(January 2007)
\end{botherref}
\endbibitem

\bibitem[\protect\citeauthoryear{Kitchenham et~al.}{2004}]{kitchenhamEvidenceBasedSoftwareEngineering}
\begin{bchapter}
\bauthor{\bsnm{Kitchenham}, \binits{B.A.}},
\bauthor{\bsnm{Dyb{\aa}}, \binits{T.}},
\bauthor{\bsnm{J{\o}rgensen}, \binits{M.}}:
\bctitle{Evidence-{Based} {Software} {Engineering}}.
In: \bbtitle{Proceedings of the 26th {International} {Conference} on {Software} {Engineering}}.
\bsertitle{{ICSE} '04},
pp. \bfpage{273}--\blpage{281}.
\bpublisher{IEEE Computer Society},
\blocation{USA}
(\byear{2004}).
\doiurl{10.5555/998675.999432}
\end{bchapter}
\endbibitem

\bibitem[\protect\citeauthoryear{Kuutila et~al.}{2020}]{kuutilaTimePressureSoftware2020b}
\begin{barticle}
\bauthor{\bsnm{Kuutila}, \binits{M.}},
\bauthor{\bsnm{M{\"a}ntyl{\"a}}, \binits{M.}},
\bauthor{\bsnm{Farooq}, \binits{U.}},
\bauthor{\bsnm{Claes}, \binits{M.}}:
\batitle{Time pressure in software engineering: {{A}} systematic review}.
\bjtitle{Information and Software Technology}
\bvolume{121},
\bfpage{106257}
(\byear{2020})
\doiurl{10.1016/j.infsof.2020.106257}
\end{barticle}
\endbibitem

\bibitem[\protect\citeauthoryear{Lenarduzzi et~al.}{2021}]{lenarduzziSystematicLiteratureReview2021a}
\begin{barticle}
\bauthor{\bsnm{Lenarduzzi}, \binits{V.}},
\bauthor{\bsnm{Besker}, \binits{T.}},
\bauthor{\bsnm{Taibi}, \binits{D.}},
\bauthor{\bsnm{Martini}, \binits{A.}},
\bauthor{\bsnm{Arcelli~Fontana}, \binits{F.}}:
\batitle{A systematic literature review on {{Technical Debt}} prioritization: {{Strategies}}, processes, factors, and tools}.
\bjtitle{Journal of Systems and Software}
\bvolume{171},
\bfpage{110827}
(\byear{2021})
\doiurl{10.1016/j.jss.2020.110827}
\end{barticle}
\endbibitem

\bibitem[\protect\citeauthoryear{Lin et~al.}{2022}]{linOpinionMiningSoftware2022a}
\begin{barticle}
\bauthor{\bsnm{Lin}, \binits{B.}},
\bauthor{\bsnm{Cassee}, \binits{N.}},
\bauthor{\bsnm{Serebrenik}, \binits{A.}},
\bauthor{\bsnm{Bavota}, \binits{G.}},
\bauthor{\bsnm{Novielli}, \binits{N.}},
\bauthor{\bsnm{Lanza}, \binits{M.}}:
\batitle{Opinion {{Mining}} for {{Software Development}}: {{A Systematic Literature Review}}}.
\bjtitle{ACM Trans. Softw. Eng. Methodol.}
\bvolume{31}(\bissue{3}),
\bfpage{38}--\blpage{13841}
(\byear{2022})
\doiurl{10.1145/3490388}
\end{barticle}
\endbibitem

\bibitem[\protect\citeauthoryear{List and Goodin}{2001}]{listEpistemicDemocracyGeneralizing2001}
\begin{barticle}
\bauthor{\bsnm{List}, \binits{C.}},
\bauthor{\bsnm{Goodin}, \binits{R.E.}}:
\batitle{Epistemic {Democracy}: {Generalizing} the {Condorcet} {Jury} {Theorem}}.
\bjtitle{Journal of Political Philosophy}
\bvolume{9}(\bissue{3}),
\bfpage{277}--\blpage{306}
(\byear{2001})
\doiurl{10.1111/1467-9760.00128}
\end{barticle}
\endbibitem

\bibitem[\protect\citeauthoryear{Lewowski and Madeyski}{2022}]{lewowskiHowFarAre2022a}
\begin{barticle}
\bauthor{\bsnm{Lewowski}, \binits{T.}},
\bauthor{\bsnm{Madeyski}, \binits{L.}}:
\batitle{How far are we from reproducible research on code smell detection? {{A}} systematic literature review}.
\bjtitle{Information and Software Technology}
\bvolume{144},
\bfpage{106783}
(\byear{2022})
\doiurl{10.1016/j.infsof.2021.106783}
\end{barticle}
\endbibitem

\bibitem[\protect\citeauthoryear{Li et~al.}{2024}]{liEvaluatingEffectivenessLarge2024}
\begin{barticle}
\bauthor{\bsnm{Li}, \binits{M.}},
\bauthor{\bsnm{Sun}, \binits{J.}},
\bauthor{\bsnm{Tan}, \binits{X.}}:
\batitle{Evaluating the effectiveness of large language models in abstract screening: A comparative analysis}.
\bjtitle{Systematic Reviews}
\bvolume{13}(\bissue{1}),
\bfpage{219}
(\byear{2024})
\doiurl{10.1186/s13643-024-02609-x}
\end{barticle}
\endbibitem

\bibitem[\protect\citeauthoryear{Madeyski et~al.}{2026}]{madeyskiLLM4SCREENLITRecommendationsAssessing2026}
\begin{barticle}
\bauthor{\bsnm{Madeyski}, \binits{L.}},
\bauthor{\bsnm{Kitchenham}, \binits{B.}},
\bauthor{\bsnm{Shepperd}, \binits{M.}}:
\batitle{{LLM4SCREENLIT}: {Recommendations} on assessing the performance of large language models for screening literature in systematic reviews}.
\bjtitle{Information and Software Technology}
\bvolume{198},
\bfpage{108204}
(\byear{2026})
\doiurl{10.1016/j.infsof.2026.108204}
\end{barticle}
\endbibitem

\bibitem[\protect\citeauthoryear{M{\"a}ntyl{\"a} et~al.}{2026}]{mantyla2026understanding}
\begin{bchapter}
\bauthor{\bsnm{M{\"a}ntyl{\"a}}, \binits{M.}},
\bauthor{\bsnm{Matsubara}, \binits{P.}},
\bauthor{\bsnm{Felizardo}, \binits{K.R.}},
\bauthor{\bsnm{Kuutila}, \binits{M.}},
\bauthor{\bsnm{Gerosa}, \binits{M.}},
\bauthor{\bsnm{{de Sousa}~Sampaio}, \binits{S.}},
\bauthor{\bsnm{Conte}, \binits{T.}},
\bauthor{\bsnm{Steinmacher}, \binits{I.}}:
\bctitle{Understanding {{LLMs}} in {{Title-Abstract Screening}}: From disagreements to recommendations}.
In: \bbtitle{Euromicro Conference on Software Engineering and Advanced Applications},
pp. \bfpage{226}--\blpage{242}.
\bpublisher{Springer},
\blocation{Cham, Switzerland}
(\byear{2026}).
\doiurl{10.1007/978-3-032-36587-3_16}
\end{bchapter}
\endbibitem

\bibitem[\protect\citeauthoryear{Manning et~al.}{2008}]{Manning_Raghavan_Schutze_2008}
\begin{bbook}
\bauthor{\bsnm{Manning}, \binits{C.D.}},
\bauthor{\bsnm{Raghavan}, \binits{P.}},
\bauthor{\bsnm{Sch\"{u}tze}, \binits{H.}}:
\bbtitle{Introduction to Information Retrieval}.
\bpublisher{Cambridge University Press},
\blocation{Cambridge}
(\byear{2008})
\end{bbook}
\endbibitem

\bibitem[\protect\citeauthoryear{Matsui et~al.}{2024}]{matsuiHumanComparableSensitivityLarge2024a}
\begin{barticle}
\bauthor{\bsnm{Matsui}, \binits{K.}},
\bauthor{\bsnm{Utsumi}, \binits{T.}},
\bauthor{\bsnm{Aoki}, \binits{Y.}},
\bauthor{\bsnm{Maruki}, \binits{T.}},
\bauthor{\bsnm{Takeshima}, \binits{M.}},
\bauthor{\bsnm{Takaesu}, \binits{Y.}}:
\batitle{Human-{{Comparable Sensitivity}} of {{Large Language Models}} in {{Identifying Eligible Studies Through Title}} and {{Abstract Screening}}: 3-{{Layer Strategy Using GPT-3}}.5 and {{GPT-4}} for {{Systematic Reviews}}}.
\bjtitle{Journal of Medical Internet Research}
\bvolume{26}(\bissue{1}),
\bfpage{52758}
(\byear{2024})
\doiurl{10.2196/52758}
\end{barticle}
\endbibitem

\bibitem[\protect\citeauthoryear{Marshall and Wallace}{2019}]{marshallSystematicReviewAutomation2019}
\begin{barticle}
\bauthor{\bsnm{Marshall}, \binits{I.J.}},
\bauthor{\bsnm{Wallace}, \binits{B.C.}}:
\batitle{Toward systematic review automation: a practical guide to using machine learning tools in research synthesis}.
\bjtitle{Systematic Reviews}
\bvolume{8}(\bissue{1}),
\bfpage{163}
(\byear{2019})
\doiurl{10.1186/s13643-019-1074-9}
\end{barticle}
\endbibitem

\bibitem[\protect\citeauthoryear{Ouzzani et~al.}{2016}]{ouzzani2016rayyan}
\begin{barticle}
\bauthor{\bsnm{Ouzzani}, \binits{M.}},
\bauthor{\bsnm{Hammady}, \binits{H.}},
\bauthor{\bsnm{Fedorowicz}, \binits{Z.}},
\bauthor{\bsnm{Elmagarmid}, \binits{A.}}:
\batitle{Rayyan---a web and mobile app for systematic reviews}.
\bjtitle{Systematic reviews}
\bvolume{5}(\bissue{1}),
\bfpage{210}
(\byear{2016})
\doiurl{10.1186/s13643-016-0384-4}
\end{barticle}
\endbibitem

\bibitem[\protect\citeauthoryear{Pereira et~al.}{2021}]{pereiraLearningSoftwareConfiguration2021a}
\begin{barticle}
\bauthor{\bsnm{Pereira}, \binits{J.A.}},
\bauthor{\bsnm{Acher}, \binits{M.}},
\bauthor{\bsnm{Martin}, \binits{H.}},
\bauthor{\bsnm{J{\'e}z{\'e}quel}, \binits{J.-M.}},
\bauthor{\bsnm{Botterweck}, \binits{G.}},
\bauthor{\bsnm{Ventresque}, \binits{A.}}:
\batitle{Learning software configuration spaces: {{A}} systematic literature review}.
\bjtitle{Journal of Systems and Software}
\bvolume{182},
\bfpage{111044}
(\byear{2021})
\doiurl{10.1016/j.jss.2021.111044}
\end{barticle}
\endbibitem

\bibitem[\protect\citeauthoryear{Qiu et~al.}{2026}]{qiuPromptBasedTutorialLarge2026}
\begin{botherref}
\oauthor{\bsnm{Qiu}, \binits{T.}},
\oauthor{\bsnm{Jiang}, \binits{H.}},
\oauthor{\bsnm{Chen}, \binits{M.S.}},
\oauthor{\bsnm{Lin}, \binits{T.}}:
A {Prompt}-{Based} {Tutorial} for {Large} {Language} {Model}--{Assisted} {Screening} in {Systematic} {Reviews} and {Meta}-{Analyses}.
International Journal of Technology Assessment in Health Care,
1--31
(2026)
\doiurl{10.1017/S0266462326104073}
\end{botherref}
\endbibitem

\bibitem[\protect\citeauthoryear{Ralph et~al.}{2020}]{ralph2020empirical}
\begin{botherref}
\oauthor{\bsnm{Ralph}, \binits{P.}},
\oauthor{\bsnm{Ali}, \binits{N.}},
\oauthor{\bsnm{Baltes}, \binits{S.}},
\oauthor{\bsnm{Bianculli}, \binits{D.}},
\oauthor{\bsnm{Diaz}, \binits{J.}},
\oauthor{\bsnm{Dittrich}, \binits{Y.}},
\oauthor{\bsnm{Ernst}, \binits{N.}},
\oauthor{\bsnm{Felderer}, \binits{M.}},
\oauthor{\bsnm{Feldt}, \binits{R.}},
\oauthor{\bsnm{Filieri}, \binits{A.}},
\oauthor{\bsnm{{de Fran\c{c}a}}, \binits{B.B.N.}},
\oauthor{\bsnm{Furia}, \binits{C.A.}},
\oauthor{\bsnm{Gay}, \binits{G.}},
\oauthor{\bsnm{Gold}, \binits{N.}},
\oauthor{\bsnm{Graziotin}, \binits{D.}},
\oauthor{\bsnm{He}, \binits{P.}},
\oauthor{\bsnm{Hoda}, \binits{R.}},
\oauthor{\bsnm{Juristo}, \binits{N.}},
\oauthor{\bsnm{Kitchenham}, \binits{B.}},
\oauthor{\bsnm{Lenarduzzi}, \binits{V.}},
\oauthor{\bsnm{Mart\'{\i}nez}, \binits{J.}},
\oauthor{\bsnm{Melegati}, \binits{J.}},
\oauthor{\bsnm{Mendez}, \binits{D.}},
\oauthor{\bsnm{Menzies}, \binits{T.}},
\oauthor{\bsnm{Molleri}, \binits{J.}},
\oauthor{\bsnm{Pfahl}, \binits{D.}},
\oauthor{\bsnm{Robbes}, \binits{R.}},
\oauthor{\bsnm{Russo}, \binits{D.}},
\oauthor{\bsnm{Saarim\"{a}ki}, \binits{N.}},
\oauthor{\bsnm{Sarro}, \binits{F.}},
\oauthor{\bsnm{Taibi}, \binits{D.}},
\oauthor{\bsnm{Siegmund}, \binits{J.}},
\oauthor{\bsnm{Spinellis}, \binits{D.}},
\oauthor{\bsnm{Staron}, \binits{M.}},
\oauthor{\bsnm{Stol}, \binits{K.}},
\oauthor{\bsnm{Storey}, \binits{M.-A.}},
\oauthor{\bsnm{Taibi}, \binits{D.}},
\oauthor{\bsnm{Tamburri}, \binits{D.}},
\oauthor{\bsnm{Torchiano}, \binits{M.}},
\oauthor{\bsnm{Treude}, \binits{C.}},
\oauthor{\bsnm{Turhan}, \binits{B.}},
\oauthor{\bsnm{Wang}, \binits{X.}},
\oauthor{\bsnm{Vegas}, \binits{S.}}:
Empirical Standards for Software Engineering Research
(2020).
\doiurl{10.48550/arXiv.2010.03525}
\end{botherref}
\endbibitem

\bibitem[\protect\citeauthoryear{Rani et~al.}{2023}]{raniDecadeCodeComment2023a}
\begin{barticle}
\bauthor{\bsnm{Rani}, \binits{P.}},
\bauthor{\bsnm{Blasi}, \binits{A.}},
\bauthor{\bsnm{Stulova}, \binits{N.}},
\bauthor{\bsnm{Panichella}, \binits{S.}},
\bauthor{\bsnm{Gorla}, \binits{A.}},
\bauthor{\bsnm{Nierstrasz}, \binits{O.}}:
\batitle{A decade of code comment quality assessment: {{A}} systematic literature review}.
\bjtitle{Journal of Systems and Software}
\bvolume{195},
\bfpage{111515}
(\byear{2023})
\doiurl{10.1016/j.jss.2022.111515}
\end{barticle}
\endbibitem

\bibitem[\protect\citeauthoryear{{Rodr{\'i}guez-P{\'e}rez} et~al.}{2021}]{rodriguez-perezPerceivedDiversitySoftware2021a}
\begin{barticle}
\bauthor{\bsnm{{Rodr{\'i}guez-P{\'e}rez}}, \binits{G.}},
\bauthor{\bsnm{Nadri}, \binits{R.}},
\bauthor{\bsnm{Nagappan}, \binits{M.}}:
\batitle{Perceived diversity in software engineering: A systematic literature review}.
\bjtitle{Empirical Software Engineering}
\bvolume{26}(\bissue{5}),
\bfpage{102}
(\byear{2021})
\doiurl{10.1007/s10664-021-09992-2}
\end{barticle}
\endbibitem

\bibitem[\protect\citeauthoryear{Shameem et~al.}{2026}]{shameemExploratoryStudyEffectiveness2026}
\begin{barticle}
\bauthor{\bsnm{Shameem}, \binits{M.}},
\bauthor{\bsnm{Alshayeb}, \binits{M.}},
\bauthor{\bsnm{Niazi}, \binits{M.}},
\bauthor{\bsnm{Mahmood}, \binits{S.}},
\bauthor{\bsnm{Ahmad}, \binits{A.}},
\bauthor{\bsnm{Al~Dallal}, \binits{J.}}:
\batitle{An exploratory study on effectiveness of {GPT}-4o in conducting sub-tasks of systematic literature reviews}.
\bjtitle{Automated Software Engineering}
\bvolume{33}(\bissue{3}),
\bfpage{81}
(\byear{2026})
\doiurl{10.1007/s10515-026-00629-6}
\end{barticle}
\endbibitem

\bibitem[\protect\citeauthoryear{Shevtsov et~al.}{2018}]{shevtsovControlTheoreticalSoftwareAdaptation2018a}
\begin{barticle}
\bauthor{\bsnm{Shevtsov}, \binits{S.}},
\bauthor{\bsnm{Berekmeri}, \binits{M.}},
\bauthor{\bsnm{Weyns}, \binits{D.}},
\bauthor{\bsnm{Maggio}, \binits{M.}}:
\batitle{Control-{{Theoretical Software Adaptation}}: {{A Systematic Literature Review}}}.
\bjtitle{IEEE Transactions on Software Engineering}
\bvolume{44}(\bissue{8}),
\bfpage{784}--\blpage{810}
(\byear{2018})
\doiurl{10.1109/TSE.2017.2704579}
\end{barticle}
\endbibitem

\bibitem[\protect\citeauthoryear{Somers et~al.}{2023}]{somersDigitaltwinbasedTestingCyber2023b}
\begin{barticle}
\bauthor{\bsnm{Somers}, \binits{R.J.}},
\bauthor{\bsnm{Douthwaite}, \binits{J.A.}},
\bauthor{\bsnm{Wagg}, \binits{D.J.}},
\bauthor{\bsnm{Walkinshaw}, \binits{N.}},
\bauthor{\bsnm{Hierons}, \binits{R.M.}}:
\batitle{Digital-twin-based testing for cyber--physical systems: {{A}} systematic literature review}.
\bjtitle{Information and Software Technology}
\bvolume{156},
\bfpage{107145}
(\byear{2023})
\doiurl{10.1016/j.infsof.2022.107145}
\end{barticle}
\endbibitem

\bibitem[\protect\citeauthoryear{Serretti}{2026}]{serrettiBenchmarkingLocalSchemaConstrained2026}
\begin{botherref}
\oauthor{\bsnm{Serretti}, \binits{A.}}:
Benchmarking a {Local} {Schema}-{Constrained} {Large} {Language} {Model} {Pipeline} for {Abstract} {Screening} and {Evidence} {Mapping}.
Cureus
\textbf{18}(6)
(2026)
\doiurl{10.7759/cureus.111193}
\end{botherref}
\endbibitem

\bibitem[\protect\citeauthoryear{Stradowski and Madeyski}{2023}]{stradowskiMachineLearningSoftware2023a}
\begin{barticle}
\bauthor{\bsnm{Stradowski}, \binits{S.}},
\bauthor{\bsnm{Madeyski}, \binits{L.}}:
\batitle{Machine learning in software defect prediction: {{A}} business-driven systematic mapping study}.
\bjtitle{Information and Software Technology}
\bvolume{155},
\bfpage{107128}
(\byear{2023})
\doiurl{10.1016/j.infsof.2022.107128}
\end{barticle}
\endbibitem

\bibitem[\protect\citeauthoryear{Shteingart et~al.}{2020}]{shteingartMajorityVotingCondorcets2020}
\begin{botherref}
\oauthor{\bsnm{Shteingart}, \binits{H.}},
\oauthor{\bsnm{Marom}, \binits{E.}},
\oauthor{\bsnm{Itkin}, \binits{I.}},
\oauthor{\bsnm{Shabat}, \binits{G.}},
\oauthor{\bsnm{Kolomenkin}, \binits{M.}},
\oauthor{\bsnm{Salhov}, \binits{M.}},
\oauthor{\bsnm{Katzir}, \binits{L.}}:
Majority {Voting} and the {Condorcet}'s {Jury} {Theorem}.
arXiv
(2020).
\doiurl{10.48550/arXiv.2002.03153}
\end{botherref}
\endbibitem

\bibitem[\protect\citeauthoryear{Sujau et~al.}{2025}]{sujauAcceleratingDiseaseModel2025}
\begin{barticle}
\bauthor{\bsnm{Sujau}, \binits{M.}},
\bauthor{\bsnm{Wada}, \binits{M.}},
\bauthor{\bsnm{Vall\'{e}e}, \binits{E.}},
\bauthor{\bsnm{Hillis}, \binits{N.}},
\bauthor{\bsnm{Su\v{s}njak}, \binits{T.}}:
\batitle{Accelerating {Disease} {Model} {Parameter} {Extraction}: {An} {LLM}-{Based} {Ranking} {Approach} to {Select} {Initial} {Studies} for {Literature} {Review} {Automation}}.
\bjtitle{Machine Learning and Knowledge Extraction}
\bvolume{7}(\bissue{2}),
\bfpage{28}
(\byear{2025})
\doiurl{10.3390/make7020028}
\end{barticle}
\endbibitem

\bibitem[\protect\citeauthoryear{Sharbaf et~al.}{2023}]{sharbafConflictManagementTechniques2023a}
\begin{barticle}
\bauthor{\bsnm{Sharbaf}, \binits{M.}},
\bauthor{\bsnm{Zamani}, \binits{B.}},
\bauthor{\bsnm{Suny{\'e}}, \binits{G.}}:
\batitle{Conflict management techniques for model merging: A systematic mapping review}.
\bjtitle{Software and Systems Modeling}
\bvolume{22}(\bissue{3}),
\bfpage{1031}--\blpage{1079}
(\byear{2023})
\doiurl{10.1007/s10270-022-01050-9}
\end{barticle}
\endbibitem

\bibitem[\protect\citeauthoryear{Teixeira et~al.}{2017}]{teixeiraModelingAutomaticCode2017a}
\begin{barticle}
\bauthor{\bsnm{Teixeira}, \binits{S.}},
\bauthor{\bsnm{Agrizzi}, \binits{B.A.}},
\bauthor{\bsnm{Filho}, \binits{J.G.P.}},
\bauthor{\bsnm{Rossetto}, \binits{S.}},
\bauthor{\bsnm{Baldam}, \binits{R.D.L.}}:
\batitle{Modeling and automatic code generation for wireless sensor network applications using model-driven or business process approaches: {{A}} systematic mapping study}.
\bjtitle{Journal of Systems and Software}
\bvolume{132},
\bfpage{50}--\blpage{71}
(\byear{2017})
\doiurl{10.1016/j.jss.2017.06.024}
\end{barticle}
\endbibitem

\bibitem[\protect\citeauthoryear{Tran et~al.}{2023}]{tranSensitivitySpecificityAvoidable2023}
\begin{botherref}
\oauthor{\bsnm{Tran}, \binits{V.-T.}},
\oauthor{\bsnm{Gartlehner}, \binits{G.}},
\oauthor{\bsnm{Yaacoub}, \binits{S.}},
\oauthor{\bsnm{Boutron}, \binits{I.}},
\oauthor{\bsnm{Schwingshackl}, \binits{L.}},
\oauthor{\bsnm{Stadelmaier}, \binits{J.}},
\oauthor{\bsnm{Sommer}, \binits{I.}},
\oauthor{\bsnm{Aboulayeh}, \binits{F.}},
\oauthor{\bsnm{Afach}, \binits{S.}},
\oauthor{\bsnm{Meerpohl}, \binits{J.}},
\oauthor{\bsnm{Ravaud}, \binits{P.}}:
Sensitivity, Specificity and Avoidable Workload of Using a Large Language Models for Title and Abstract Screening in Systematic Reviews and Meta-Analyses.
medRxiv
(2023).
\doiurl{10.1101/2023.12.15.23300018}
\end{botherref}
\endbibitem

\bibitem[\protect\citeauthoryear{Thode et~al.}{2025}]{thodeExploringUseLLMs2025}
\begin{botherref}
\oauthor{\bsnm{Thode}, \binits{L.}},
\oauthor{\bsnm{Iftikhar}, \binits{U.}},
\oauthor{\bsnm{Mendez}, \binits{D.}}:
Exploring the use of {LLMs} for the selection phase in systematic literature studies.
Information and Software Technology
\textbf{184}
(2025)
\doiurl{10.1016/j.infsof.2025.107757}
\end{botherref}
\endbibitem

\bibitem[\protect\citeauthoryear{Tambon et~al.}{2022}]{tambonHowCertifyMachine2022a}
\begin{barticle}
\bauthor{\bsnm{Tambon}, \binits{F.}},
\bauthor{\bsnm{Laberge}, \binits{G.}},
\bauthor{\bsnm{An}, \binits{L.}},
\bauthor{\bsnm{Nikanjam}, \binits{A.}},
\bauthor{\bsnm{Mindom}, \binits{P.S.N.}},
\bauthor{\bsnm{Pequignot}, \binits{Y.}},
\bauthor{\bsnm{Khomh}, \binits{F.}},
\bauthor{\bsnm{Antoniol}, \binits{G.}},
\bauthor{\bsnm{Merlo}, \binits{E.}},
\bauthor{\bsnm{Laviolette}, \binits{F.}}:
\batitle{How to certify machine learning based safety-critical systems? {{A}} systematic literature review}.
\bjtitle{Automated Software Engineering}
\bvolume{29}(\bissue{2}),
\bfpage{38}
(\byear{2022})
\doiurl{10.1007/s10515-022-00337-x}
\end{barticle}
\endbibitem

\bibitem[\protect\citeauthoryear{Tebes et~al.}{2020}]{tebesAnalyzingDocumentingSystematic2020a}
\begin{barticle}
\bauthor{\bsnm{Tebes}, \binits{G.}},
\bauthor{\bsnm{Peppino}, \binits{D.}},
\bauthor{\bsnm{Becker}, \binits{P.}},
\bauthor{\bsnm{Matturro}, \binits{G.}},
\bauthor{\bsnm{Solari}, \binits{M.}},
\bauthor{\bsnm{Olsina}, \binits{L.}}:
\batitle{Analyzing and documenting the systematic review results of software testing ontologies}.
\bjtitle{Information and Software Technology}
\bvolume{123},
\bfpage{106298}
(\byear{2020})
\doiurl{10.1016/j.infsof.2020.106298}
\end{barticle}
\endbibitem

\bibitem[\protect\citeauthoryear{Trad et~al.}{2025}]{tradStreamliningSystematicReviews2025}
\begin{barticle}
\bauthor{\bsnm{Trad}, \binits{F.}},
\bauthor{\bsnm{Yammine}, \binits{R.}},
\bauthor{\bsnm{Charafeddine}, \binits{J.}},
\bauthor{\bsnm{Chakhtoura}, \binits{M.}},
\bauthor{\bsnm{Rahme}, \binits{M.}},
\bauthor{\bsnm{El-Hajj~Fuleihan}, \binits{G.}},
\bauthor{\bsnm{Chehab}, \binits{A.}}:
\batitle{Streamlining systematic reviews with large language models using prompt engineering and retrieval augmented generation}.
\bjtitle{BMC Medical Research Methodology}
\bvolume{25}(\bissue{1}),
\bfpage{130}
(\byear{2025})
\doiurl{10.1186/s12874-025-02583-5}
\end{barticle}
\endbibitem

\bibitem[\protect\citeauthoryear{{van de Schoot} et~al.}{2021}]{vandeschootOpenSourceMachine2021a}
\begin{barticle}
\bauthor{\bsnm{{van de Schoot}}, \binits{R.}},
\bauthor{\bsnm{{de Bruin}}, \binits{J.}},
\bauthor{\bsnm{Schram}, \binits{R.}},
\bauthor{\bsnm{Zahedi}, \binits{P.}},
\bauthor{\bsnm{{de Boer}}, \binits{J.}},
\bauthor{\bsnm{Weijdema}, \binits{F.}},
\bauthor{\bsnm{Kramer}, \binits{B.}},
\bauthor{\bsnm{Huijts}, \binits{M.}},
\bauthor{\bsnm{Hoogerwerf}, \binits{M.}},
\bauthor{\bsnm{Ferdinands}, \binits{G.}},
\bauthor{\bsnm{Harkema}, \binits{A.}},
\bauthor{\bsnm{Willemsen}, \binits{J.}},
\bauthor{\bsnm{Ma}, \binits{Y.}},
\bauthor{\bsnm{Fang}, \binits{Q.}},
\bauthor{\bsnm{Hindriks}, \binits{S.}},
\bauthor{\bsnm{Tummers}, \binits{L.}},
\bauthor{\bsnm{Oberski}, \binits{D.L.}}:
\batitle{An open source machine learning framework for efficient and transparent systematic reviews}.
\bjtitle{Nature Machine Intelligence}
\bvolume{3}(\bissue{2}),
\bfpage{125}--\blpage{133}
(\byear{2021})
\doiurl{10.1038/s42256-020-00287-7}
\end{barticle}
\endbibitem

\bibitem[\protect\citeauthoryear{Van~Dinter et~al.}{2022}]{vandinterPredictiveMaintenanceUsing2022a}
\begin{barticle}
\bauthor{\bsnm{Van~Dinter}, \binits{R.}},
\bauthor{\bsnm{Tekinerdogan}, \binits{B.}},
\bauthor{\bsnm{Catal}, \binits{C.}}:
\batitle{Predictive maintenance using digital twins: {{A}} systematic literature review}.
\bjtitle{Information and Software Technology}
\bvolume{151},
\bfpage{107008}
(\byear{2022})
\doiurl{10.1016/j.infsof.2022.107008}
\end{barticle}
\endbibitem

\bibitem[\protect\citeauthoryear{Vaswani et~al.}{2023}]{Vaswani_Shazeer_Parmar_Uszkoreit_Jones_Gomez_Kaiser_Polosukhin_2023}
\begin{botherref}
\oauthor{\bsnm{Vaswani}, \binits{A.}},
\oauthor{\bsnm{Shazeer}, \binits{N.}},
\oauthor{\bsnm{Parmar}, \binits{N.}},
\oauthor{\bsnm{Uszkoreit}, \binits{J.}},
\oauthor{\bsnm{Jones}, \binits{L.}},
\oauthor{\bsnm{Gomez}, \binits{A.N.}},
\oauthor{\bsnm{Kaiser}, \binits{L.}},
\oauthor{\bsnm{Polosukhin}, \binits{I.}}:
Attention is all you need
(arXiv:1706.03762)
(2023)
\doiurl{10.48550/arXiv.1706.03762}
\end{botherref}
\endbibitem

\bibitem[\protect\citeauthoryear{White et~al.}{2025}]{livebench}
\begin{bchapter}
\bauthor{\bsnm{White}, \binits{C.}},
\bauthor{\bsnm{Dooley}, \binits{S.}},
\bauthor{\bsnm{Roberts}, \binits{M.}},
\bauthor{\bsnm{Pal}, \binits{A.}},
\bauthor{\bsnm{Feuer}, \binits{B.}},
\bauthor{\bsnm{Jain}, \binits{S.}},
\bauthor{\bsnm{Shwartz-Ziv}, \binits{R.}},
\bauthor{\bsnm{Jain}, \binits{N.}},
\bauthor{\bsnm{Saifullah}, \binits{K.}},
\bauthor{\bsnm{Dey}, \binits{S.}},
\bauthor{\bsnm{Shubh-Agrawal}},
\bauthor{\bsnm{Sandha}, \binits{S.S.}},
\bauthor{\bsnm{Naidu}, \binits{S.V.}},
\bauthor{\bsnm{Hegde}, \binits{C.}},
\bauthor{\bsnm{LeCun}, \binits{Y.}},
\bauthor{\bsnm{Goldstein}, \binits{T.}},
\bauthor{\bsnm{Neiswanger}, \binits{W.}},
\bauthor{\bsnm{Goldblum}, \binits{M.}}:
\bctitle{{LiveBench}: A challenging, contamination-free {LLM} benchmark}.
In: \bbtitle{The Thirteenth International Conference on Learning Representations}
(\byear{2025}).
\doiurl{10.48550/arXiv.2406.19314}
\end{bchapter}
\endbibitem

\bibitem[\protect\citeauthoryear{Wongpakaran et~al.}{2013}]{wongpakaranComparisonCohensKappa2013}
\begin{botherref}
\oauthor{\bsnm{Wongpakaran}, \binits{N.}},
\oauthor{\bsnm{Wongpakaran}, \binits{T.}},
\oauthor{\bsnm{Gwet}, \binits{K.}}:
A comparison of {Cohen}'s {Kappa} and {Gwet}'s {AC1} when calculating inter-rater reliability coefficients: {A} study conducted with personality disorder samples.
BMC Medical Research Methodology
\textbf{13}
(2013)
\doiurl{10.1186/1471-2288-13-61}
\end{botherref}
\endbibitem

\bibitem[\protect\citeauthoryear{Zhang and Li}{2020}]{zhangTestingVerificationNeuralnetworkbased2020a}
\begin{barticle}
\bauthor{\bsnm{Zhang}, \binits{J.}},
\bauthor{\bsnm{Li}, \binits{J.}}:
\batitle{Testing and verification of neural-network-based safety-critical control software: {{A}} systematic literature review}.
\bjtitle{Information and Software Technology}
\bvolume{123},
\bfpage{106296}
(\byear{2020})
\doiurl{10.1016/j.infsof.2020.106296}
\end{barticle}
\endbibitem

\bibitem[\protect\citeauthoryear{{Zakeri-Nasrabadi} et~al.}{2023}]{zakeri-nasrabadiSystematicLiteratureReview2023b}
\begin{barticle}
\bauthor{\bsnm{{Zakeri-Nasrabadi}}, \binits{M.}},
\bauthor{\bsnm{Parsa}, \binits{S.}},
\bauthor{\bsnm{Ramezani}, \binits{M.}},
\bauthor{\bsnm{Roy}, \binits{C.}},
\bauthor{\bsnm{Ekhtiarzadeh}, \binits{M.}}:
\batitle{A systematic literature review on source code similarity measurement and clone detection: {{Techniques}}, applications, and challenges}.
\bjtitle{Journal of Systems and Software}
\bvolume{204},
\bfpage{111796}
(\byear{2023})
\doiurl{10.1016/j.jss.2023.111796}
\end{barticle}
\endbibitem

\end{thebibliography}

\end{document}